\documentclass[twocolumn,prb,10pt,aps,longbibliography,superscriptaddress]{revtex4-2}
\usepackage{graphicx}
\usepackage{amsmath}
\usepackage{amssymb}
\usepackage{amsfonts}
\usepackage{dsfont}
\usepackage{dcolumn}
\usepackage{bbm}
\usepackage{mathtools}
\usepackage{braket}
\usepackage{physics}
\usepackage{siunitx}
\usepackage[unicode]{hyperref}
\usepackage[nameinlink]{cleveref}
\Crefname{section}{Sec.}{Secs.}
\Crefname{equation}{Eq.}{Eqs.}
\Crefname{figure}{Fig.}{Figs.}
\Crefname{tabular}{Tab.}{Tabs.}

\usepackage{bbold} 
\usepackage{nicefrac}
\usepackage{tikz}
\usetikzlibrary{arrows.meta}
\usepackage{lipsum}  
\usepackage{xcolor}

\usepackage[makeroom]{cancel}

\usepackage{xfrac}
\usepackage{booktabs}
\usepackage{chngcntr}

\definecolor{tdo_green}{HTML}{83B818}
\definecolor{tdo_darkgreen}{HTML}{839A00}

\newcommand{\bes}{\begin{subequations}}
\newcommand{\ees}{\end{subequations}}
\newcommand{\be}{\begin{equation}}
\newcommand{\ee}{\end{equation}}

\makeatletter
  \newcommand\flausr{\@fleqntrue}
\makeatother

\usepackage{subfiles}

\begin{document}

    \title{Influence of quadrupolar interaction on NMR spectroscopy}


    \author{Alina Joch}
    \email{alina.joch@tu-dortmund.de}
    \affiliation{Condensed Matter Theory, 
    TU Dortmund University, Otto-Hahn-Stra\ss{}e 4, 44221 Dortmund, Germany}
    \affiliation{DLR, Germany}

    \author{G\"otz S.\ Uhrig}
    \email{goetz.uhrig@tu-dortmund.de}
    \affiliation{Condensed Matter Theory, 
    TU Dortmund University, Otto-Hahn-Stra\ss{}e 4, 44221 Dortmund, Germany}

    \date{\textrm{\today}}

    \begin{abstract}
Optically driven electronic spins coupled in quantum dots to nuclear spins
show a pre-pulse signal (revival amplitude) after having been trained by long periodic 
sequences of pulses. The size of this revival amplitude depends 
on the external magnetic field in a specific way 
due to the varying commensurability of the nuclear Larmor precession period
with the time $T_\text{rep}$ between two consecutive pulses. In theoretical simulations, sharp dips
occur at fields when an integer number of precessions fits in $T_\text{rep}$; this feature could be used
to identify nuclear isotopes spectroscopically. But these sharp and characteristic dips  have 
not (yet) been detected in experiment. We study whether the nuclear quadrupolar interaction
is the reason for this discrepancy because it perturbs the nuclear precessions. But our simulations show
that the absolute width of the dips and their relative depth are not changed by quadrupolar interactions.
Only the absolute depth decreases. We conclude that quadrupolar interaction alone cannot be the
reason for the absence of the characteristic dips in experiment.
		\end{abstract}
		
    \maketitle

    \section{Introduction}
    \label{s:introduction}
		
		Exploiting the quantum properties of an electronic spin for quantum information
		processing represents a very attractive route since more than two decades
		\cite{loss98,niels00,ladd10b}. One promising kind of realization uses
		localized electronic spins in nanostructures of semiconductors \cite{urbas13,gangl19,chekh20}.
		The manipulation by laser pulses has turned out to be a fruitful way to control
		these spins \cite{awsch02,dyako17,slavc10} on very long time scales and with high
		temporal resolution \cite{greil06b,greil09a,greil09b,belyk16,evers21}.
		
\begin{figure}[htb]
      \centering
      \includegraphics[width=\columnwidth]{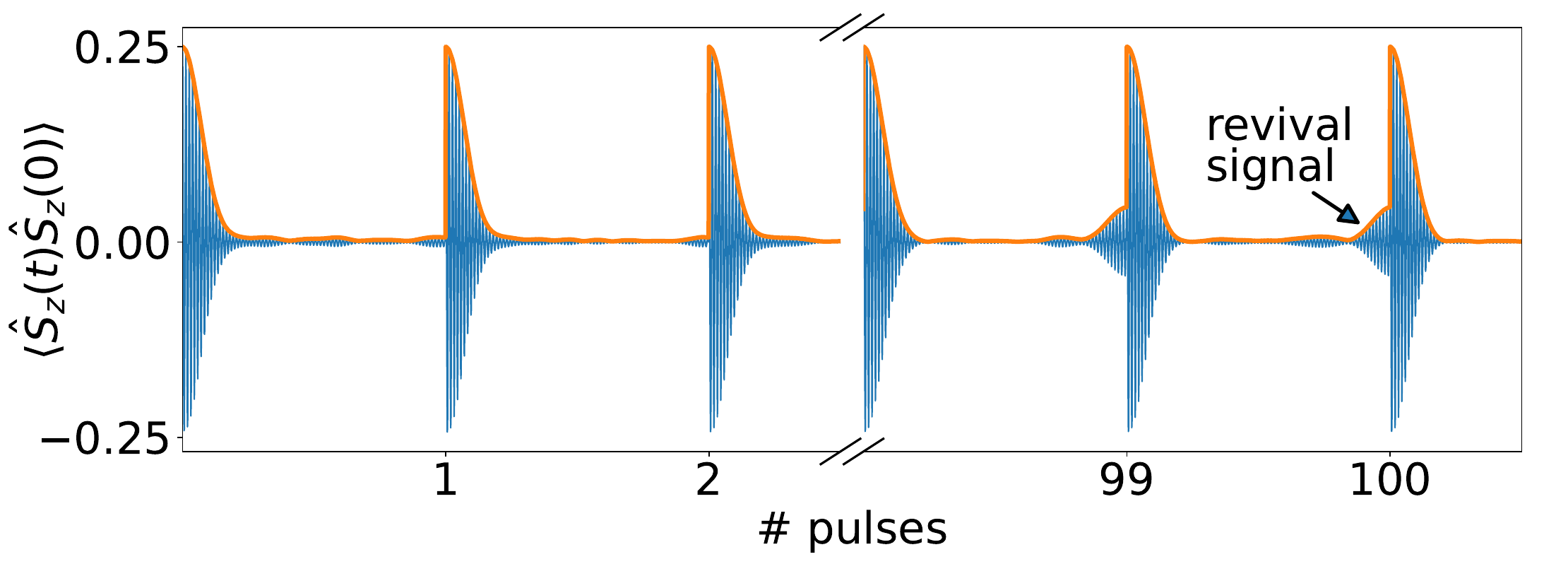}
      \caption[Visualization of revival amplitude]
			{Schematic illustration of the revival amplitude that occurs after a certain number of pulses prior to the next pulse. Blue curve: fast Larmor precessions; orange curve: envelope.}
      \label{intro}
   \end{figure}

		A particularly intriguing phenomenon is nuclei induced frequency focussing (NIFF)
		of the electronic spins in quantum dots
		\cite{greil06b,petro12,beuge16,jasch17,klein18,scher18,scher20}.
		This effect occurs when an electronic spin is repeatedly excited by long periodic
		trains of short laser pulses which orient the electronic spin. The latter is 
		generically coupled to many nuclear spins of the isotopes of the semiconductor.
		The long periodic train of pulses imprints a certain pattern on the distribution 
		of the overall magnetization of the nuclear spins (Overhauser field) 
		\cite{petro12,beuge16,jasch17,scher18}.  This comb-like
		pattern favors the Larmor precessions of the electronic spin which are commensurate with
		the repetition time $T_\text{rep}$ of the pulses. This means that an integer
		number of precessional revolutions takes place between two consecutive pulses 
		\cite{beuge16,jasch17,klein18}. 
		The patterned Overhauser field in turn acts similar to a spin Hahn
		echo in nuclear magnetic resonance (NMR).
	  Even an ensemble of slightly different quantum dots shows a collective response
		prior to the next pulse which can be detected, for instance by Faraday rotation. This collective response
		is called revival amplitude or pre-pulse signal because it seems to be revived from no signal very analogous
		to the spin Hahn echo. It is visualized in Fig.\ \ref{intro}. The revival amplitude is large for an integer number of nuclear Larmor precessions
		within $T_\text{rep}$ and small for an half-integer number of spin revolutions.
		
		This intriguing phenomenon depends on the strength of the applied magnetic field.
		Clearly, the magnetic field modified by the Overhauser field determines the 
		Larmor frequency of the electron or hole.
		In addition, a second commensurability plays a crucial role, namely the number of
		nuclear Larmor precessions
		fitting into the interval between two pulses. This commensurability influences the
		degree of revival. If the number of half-revolutions
		of the nuclear spins within the repetition period is odd, a broad minimum 
		of the revival amplitude is observed. If the number of half-revolutions is even, i.e.,
		the number of nuclear spin revolution is an integer, a very sharp minimum is observed.
		For small spin baths it could be shown that the interplay of the commensurabilities
		can be used to purify disordered quantum mixtures by reducing their entropy close
		to zero \cite{uhrig20}.
		
		In any case, a strongly non-monotonic dependence of the revival amplitude 
		on the applied field results. Shallow minima and sharp dips
		alternate upon increasing magnetic field \cite{klein18,scher20}. In experiment, however, only
		shallow minima occur \cite{klein18}. The absence of the sharp dips is not understood so far.
		First results from fully quantum mechanical calculations \cite{klein18} 
		did not reproduce the sharp dips suggesting that they only
		occur in semi-classical treatments based on truncated Wigner approximation
		\cite{david15,klein18,scher20}. Yet, recent numerical data from the quantum simulation of 
		significantly larger baths of nuclear spins revealed that the quantum mechanical results
		approach the semi-classical ones if the considered nuclear spin baths are large enough
		\cite{klein22}. 
		
	  If the bath of nuclear spins consists of various different isotopes with different
		nuclear Larmor frequencies the magnetic fields at which minima occur also vary
		\cite{beuge17,scher21a}. In particular, the sharp dips at integer nuclear precessions
		within $T_\text{rep}$ can be distinguished because each isotopes induces
		such dips. This has been simulated in Ref.\ \onlinecite{scher21a} for up to four
		different isotopes and it has been proposed
		to use these dips to analyze the content of these isotopes in the quantum dots by
		scanning the revival amplitudes. This idea was dubbed NMR spectroscopy.
		An experimental verification of this approach is still lacking due to the absence
		of sharp dips in the so far measured data \cite{klein18}. The question arises which mechanism
		may destroy the sharp dips. Since they stem from nuclear Larmor precessions any coupling of
		the nuclear spins is a candidate to smear out the resonances responsible for the
		sharp dips. The nuclear gyromagnetic ratios of the isotopes are a feature of the 
		nuclei themselves and thus very robust against external influences. Thus, we are 
		looking for further
		mechanisms prone to influence nuclear Larmor precessions, except the explicitly 
		considered hyperfine coupling to the electronic spins.
		
		Candidates for further detrimental couplings are (i) the quadrupolar interactions of 
		nuclear spins to the gradients of the electric field if the nuclear spins are larger
		than $I=1/2$ \cite{bulut12,bulut14,hackm15,glase16,glazo18} 
		and (ii) dipole-dipole interactions among the nuclear spins \cite{merku02,schli03}.
		It was shown that the measured spin noise can be understood by including the quadrupolar
		interaction \cite{hackm15,glase16}. It is found that the effects of quadrupolar
		couplings kick in at about 1$\mu$s while the dipolar effects are estimated
		to be relevant at the time scale of 10 to 100$\mu$s \cite{merku02,schli03}. 
		Thus, we will consider the quadrupolar coupling and its influence
		on the sharp dips in NMR spectroscopy. An investigation of the weaker 
		dipole-dipole interactions is left to future studies.
		
	  Since the quadrupolar interaction breaks the spin rotation invariance it is 
		a good candidate to perturb the Larmor precessions. Hence, the question we
		are posing and answering here is whether the inclusion of quadrupolar effects
		can explain the absence of sharp dips in experiment.
		
    The article is set up as follows. After this Introduction, the model for describing a 
		quantum dot is introduced in Sect.\  \ref{s:model}. Particularly, the quadrupolar
    interaction is discussed and two scenarios for the electric potential are 
    presented. Then, Sec.\  \ref{s:methods} briefly explains the truncated Wigner approach 
		(TWA) used to solve the equations of motion and an extension for quadratic terms. 
    This includes the effective treatment of the polarizing laser pulses
		as well as details of the employed numerics.
    Subsequently, the spin noise in absence and presence of the quadrupolar
		interaction is studied to make sure that our approach reproduces
		known results. In the subsequent Sec.\  \ref{s:revivalamplitude} we examine
		the magnetic field dependence of the revival amplitude for two scenarios of the 
		electrical potential. It turns out that they
			yield almost the same results. Finally, we analyze the shape of the
			dips at integer numbers of the nuclear spin revolutions. In the Conclusions
			\ref{s:conclusion}, we argue
			that the quadrupolar coupling reduces the absolute values of the depth
			and the width of the sharp dips. But the depth \emph{relative} to the total signal
			does not change	significantly. In other words, the actual shape does  not change very much
			upon inclusion of the quadrupolar coupling so that the absence of the
			sharp dips in experiment cannot be related to the quadrupolar interactions.
			At least, it cannot be the only reason.

    \section{Model}
    \label{s:model}

\subsection{Central Spin Model}
\label{ss:model}
Due to the strong localization of the electron in the quantum dot (QD), the hyperfine
interaction is the dominant interaction between the electron and the nuclear spins
\cite{merku02, schli03, lee05}. If $N$ nuclei are considered, we can describe 
this system by the Hamiltonian
\begin{equation}\label{CSMHamilton}
    \hat{H}_{\mathrm{CSM}} = \sum\limits_{k=1}^N A_k \hspace{0.2em} 
		\hat{\hspace{-0.2em}\pmb{S}} \cdot \hspace{0.2em} \hat{\hspace{-0.2em}\pmb{I}}_k  ,
\end{equation}
which is referred to as the central spin model (CSM) \cite{gaudi76,gaudi83}.
Operators are marked with a hat and vectors are printed in bold.
The spin quantum numbers are $S = 1/2$ for the electron spin and 
$I = 3/2$ for the nuclear spins of the relevant isotopes of Ga and As \cite{stone14}.

The coupling constants $A_k$ are proportional to the probability that the electron
is present at the spot where the nucleus resides. Assuming that the confining potential
of the quantum dots is roughly parabolic it is natural to conclude that the electronic wave
function is a Gaussian. Thus, the distribution of the couplings constants $A_k$
also follows a Gaussian dependence
\begin{equation}\label{wavefunction}
    A_k \propto \exp \left[ - \left( {r_k/l_0} \right)^2 \right] , 
\end{equation}
where $l_0$ is the characteristic oscillator length. Since there are 
about $10^4$ to $10^6$ nuclear spins \cite{merku02, schli03, lee05} coupled
sizeably to the electron a direct quantum mechanical calculation is out of the question.
Even a semi-classical simulation, see below, is beyond reach for realistic
values of $N$. Much smaller bath sizes are used in the simulations. 
In order to properly address the characteristic energy and thus time scales, we
normalize the coupling to capture the experimental time scales of the electronic spin.
Due to the disordered state of the nuclear spin ensemble the normalization 
needs to be done by the root mean square \cite{stane13, stane14b, grave16}
\begin{equation}\label{AQ}
    A_{\mathrm{Q}} \coloneqq \sqrt{\sum_{k=1}^N A_k^2 } .
\end{equation}
Keeping this energy scale fixed, it is possible to compare time scales for different sets of
nuclear spins differing in number and in the distribution of their couplings.

We focus on rather flat QDs so that we can assume them to be two-dimensional.
To describe a uniform distribution of nuclear spins one can draw random
numbers for the radii $r_k$ and random
angles $\phi_k$ for each given radius. 
We introduce a cutoff radius $r_{\mathrm{cut}}$ as an upper limit to keep the distribution
of coupling normalizable. If the exact position of the nucleus is not relevant for the simulations, 
the angles $\phi_k$ are not needed. Then, in two dimensions, the couplings are
distributed according to the exponential parametrization \cite{farib13a,farib13b,fause17a,scher18,scher21c}
\begin{equation}
\label{Akindex}
    A_k \propto \exp \left( -\gamma k \right)  
\end{equation}
with $\gamma \approx 2/N_{\mathrm{eff}}$. The effective number $N_\text{eff}$ of
coupled nuclear spins is the number of significantly coupled spins and can
be defined by \cite{fause17a}
\be
\label{eq:Neff-def}
N_\text{eff} \coloneqq \frac{\left(\sum_k A_k\right)^2}{A_\text{Q}^2} .
\ee

In our calculations we use a slightly different defintion including the radius $r_k$ in units of $l_0$
\begin{equation}\label{exponentialdistribution}
    A_k \propto \exp \left( -r_k^2 \right) \quad \mathrm{with} \quad r_k \in \left[\,0, r_{\mathrm{cut}}  \right],
\end{equation}
where the cutoff radius is chosen to be $\sqrt{N \gamma}$ to be able to compare with previously mentioned method \cite{scher18,scher21c}.

The Zeeman splitting stems from the coupling of the magnetic moment to an external magnetic field
$B_{\mathrm{ext}}$. For the electron spin, it reads
\begin{equation}
\label{ezHamilton}
    \hat{H}_{\mathrm{eZ}} = \gamma_{\mathrm{e}} B_{\mathrm{ext}} 
		\pmb{n}_{\mathrm{B}} \cdot \hspace{0.2em}\hat{\hspace{-0.2em}\pmb{S}} ,
\end{equation}
where $\pmb{n}_{\mathrm{B}} = \pmb{e}_x$ is the direction of the applied magnetic field 
and $\gamma_{\mathrm{e}} = g_{\mathrm{e}} \mu_{\mathrm{B}} \hbar^{-1}$ is the electronic gyromagnetic ratio. 
The $g$-factor $g_{\mathrm{e}}$ for an electron in a QD is approximately given by 
$|g_{\mathrm{e}}| \approx 0.555$ \cite{greil07a}.
For the nuclear spins we analogously obtain
\begin{equation}
\label{nzHamilton}
    \hat{H}_{\mathrm{nZ}} = 
		\sum\limits_{k=1}^{N} \gamma_{\mathrm{n}} B_{\mathrm{ext}} \pmb{n}_{\mathrm{B}} \cdot 
		\hspace{0.2em}\hat{\hspace{-0.2em}\pmb{I}}_k  .
\end{equation}
The gyromagnetic ratio $\gamma_{\mathrm{n}} = \gamma_{\mathrm{e}} / z$ 
is several orders of magnitude smaller due to the much larger nuclear mass in comparison to the 
electron mass. In GaAs $z$ takes a value of about 800 \cite{beuge17,jasch17}.

\subsection{Quadrupolar Interaction}
\label{ss:quadint}

A nuclear spin with a spin 
larger than $1/2$ interacts with the local electric field gradients 
caused by the strain fields present in QDs because of its quadrupolar moment \cite{bulut12, bulut14, glazo18}.
In GaAs quantum dots, all nuclear spins fulfill $I > 1/2$ \cite{stone14}. 
The interaction with the electric field gradient is
described by the Hamiltonian \cite{glazo18}
\begin{align}
    \hat{H}_{\mathrm{quad}} = \sum\limits_{\alpha, \beta \in \lbrace x,y,z \rbrace} 
		& V_{\alpha\beta} \frac{|e| Q}{12I (2I-1)} 
		\left[3 \left( \hat{I}^{\alpha} \hat{I}^{\beta}  
		+ \hat{I}^{\beta} \hat{I}^{\alpha} \right) \right. 
		\nonumber
		\\ & \left. - 2I (I+1) \right] 
    \label{eq:HamiltonQuadWW}
\end{align}
where we used the short hand 
\be
V_{\alpha\beta} \coloneqq \frac{\partial^2 V}{\partial r^{\alpha} \partial r^{\beta}} .
\ee
Furthermore, $e$ is the elementary charge and $Q$ the strength of the quadrupolar moment of the nucleus.
By diagonalizing the Hessian matrix $V_{\alpha \beta}$, the Hamilton operator \cite{abrag78, bulut12}
can be recast in the form
\begin{equation}
    \hat{H}^{\prime}_{\mathrm{quad}} = 
		q \left[ \left( 3\hat{I}_z^2 - \pmb{I}^2 \right) + \eta \left( \hat{I}_x^2 - \hat{I}_y^2 \right) \right] ,
\end{equation}
where $x$, $y$, and $z$ are the principal axes of the electric field 
gradient at the given site of a nucleus and all non-diagonal elements of the tensor $V_{\alpha \beta}$ are zero.
Here, $q$ embodies the prefactors in \eqref{eq:HamiltonQuadWW} and $\eta$ is given by
\begin{equation}
    \eta = \frac{V_{xx} - V_{yy}}{V_{zz}}  .
\end{equation}
For each of the $N$ nuclear spins we introduce the vectors $\pmb{n}^\alpha_k$ with $\alpha\in\{ x,y,z\}$
for the principal axes at site $k$. Eventually, we obtain \cite{hackm15, fisch22}
\begin{equation}
\begin{aligned}
    \hat{H}^{\prime}_{\mathrm{quad}} = \sum\limits_{k=1}^N \frac{q_k}{3} &\left[ \left(3\left(\hat{\hspace{-0.2em}\pmb{I}}_k \cdot \pmb{n}_k^z \right)^2 
		- \pmb{I}^2 \right) \right. \\ 
		&+  \left. \eta_k \left(  \left(\hat{\hspace{-0.2em}\pmb{I}}_k \cdot \pmb{n}_k^x \right)^2 
		-  \left(\hat{\hspace{-0.2em}\pmb{I}}_k \cdot \pmb{n}_k^y \right)^2 \right) \right] .
    \label{Hquadnk}
\end{aligned}
\end{equation}
The direction $\pmb{n}_k^z$ refers to the normalized principal axis
 corresponding to the largest eigenvalue of the quadrupolar Hesse matrix \cite{hackm15}.

We do not have precise knowledge of the form of $V$ at each nuclear site. Furthermore, 
calculations with an exact $V$ would be extremely demanding. Hence, we 
consider two opposing scenarios for $V$:
\begin{itemize}
    \item[(A)] Simple model:\\
              The way quantum dots are produced and the resulting strain \cite{urbas13, warbu13} 
              suggest a quadratic behavior in a coarse-grained average increasing from the center of the
							flat QD in the $xy$ plane to its boundary. This suggests 
							the assumption of a quadratic dependence
              \begin{equation}
							\label{quadrchoiceofV}
                  V = ax^2 + by^2 +cz^2 
              \end{equation}
              with $c = 0$ and $a = b$. 
							In this approach, all $q_k$ have the same value and all nuclear sites experience the 
							same Hesse matrix with the same principal axes since the Hesse matrix is formed from 
							the second derivatives of $V$.
    \item[(B)] Stochastic model:\\
              To construct a set of potentials $V$ that is close to the situation in 
							real QDs we use the results from first-principle calculations \cite{bulut12,bulut14} .
							The distributions of the parameters $q_k$, $\eta_k$, and of the angle $\theta_k$ computed in 
              Ref.\  \cite{bulut14} are used to randomly draw numbers from them and
							to assign these numbers to the spin at $\pmb{r_k}$. The angle $\theta_k$ measures the angle
							of $\pmb{n}_k^z$ to the normal of the substrate surface on which the QD is formed.
							This is done for all spins. Thereby, the overall distribution of the quadrupolar
							parameters it taken into account. Possible correlations between the distributions
							of the parameters are neglected. But we find evidence, see below, that they hardly
							matter.
\end{itemize}

Before simulating the quadrupolar interaction, we first fix its
energy scale relative to the hyperfine couplings $A_k$.
In Ref.\  \onlinecite{hackm15}
\begin{equation}
    Q_{\mathrm{r}} = \frac{\sum_k q_k}{\sum_k A_k}
\end{equation}
was introduced. We will use this parameter for comparisons with the results in Ref.\  \onlinecite{hackm15}.
But since in other studies, such as the one in Refs.\  \cite{bulut12, bulut14}, 
individual values are given instead of sums, we also introduce
\begin{equation}
    Q = \frac{\sqrt{\overline{q_k^2}}}{A_k^{\mathrm{min}}}  ,
\end{equation}
where $A_k^{\mathrm{min}}$ is the minimum value of the $A_k$ used in the simulation 
and $\sqrt{\overline{q_k^2}}$ is the root mean square of $q_k$. 
For given values of $Q$ and $Q_{\mathrm{r}}$, respectively, the value
$a=b$ in the simple model for $V$ follows.

The individual couplings $q_k$ in experiment are of the order $\mathrm{neV}$ \cite{bulut12} and 
the hyperfine couplings $A_k$ are of the order $0.1 - 1 \, \mathrm{\mu eV}$ \cite{merku02, lee05}. 
This means that the values of $Q$ range between 0.01 and 0.1.
As stated before, due to the constraints on runtime it is no possible to use the experimental
parameters. Hence, we reduce the effective number of nuclear spins by a factor $\lambda =100$, but
the energy scale $A_Q$ has to stay the same which means that the $A_k$ 
need to be increased by the factor $\sqrt{\lambda}=10$. 
To be concrete, we treat $N_\text{eff}=200$ instead of $10^4$ to $10^5$.
The relative parameters
$Q$ and $Q_\text{r}$ are kept constant so that the quadrupolar coupling are
also scaled up by $\sqrt{\lambda}$. 

In addition, the quadratic scaling of the runtime with the strength of the 
external magnetic field makes the simulations extremely tedious \cite{scher18, scher21c}.
In order to observe the intricate interplay between the electronic and the
nuclear Larmor precessions at lower values of  the magnetic fields,
one has to re-scale the $z$ factor down as well.
This means that one increases
the nuclear gyromagnetic ratio so that lower values of the magnetic fields
yield the desired odd and even number of half-turns of the nuclear spins.
If we use $z\to z_\text{eff}=z/\sqrt{\lambda}$ we have the additional advantage
that the ratio of the nuclear Zeeman energy to the hyperfine coupling
and the quadrupolar interactions does not change. But we will also
study larger values of $z_\text{eff}$.

In summary, there is a large parameter range to be covered:
The number of nuclear spins $N$ and the number of nuclear spins effectively 
coupled to the central spin $N_{\mathrm{eff}} \approx 2 / \gamma$. 
The parameter $z$ defining the ratio between the gyromagnetic 
ratio of the central spin and of the nuclear spins. 
The parameter $Q$ introduced above for the quadrupolar
couplings and the magnetic field $B$.
All these parameters are varied, only $N=100$ (unless denoted otherwise), 
$A_{\mathrm{Q}} = 1.19$ and $T_{\mathrm{rep}} = 13.2 \, \mathrm{ns}$ 
will be kept fixed in the following. The values for $A_{\mathrm{Q}}$
and $T_{\mathrm{rep}}$ are chosen because they are the ones 
corresponding to generic experiments.

    \section{Methods}
    \label{s:methods}
		
		We briefly introduce the approaches that we are using in the sequel
		to unveil the spin dynamics and the revival amplitudes in particular.
		
\subsection{Truncated Wigner Approximation}
	\label{ss:twa}	
		
  We use the  truncated Wigner approximation (TWA) which is a semiclassical approach \cite{polko10, stane14b}. 
  In this approach, the spin operators are expressed by classical vectors
	which obey classical equations of motion. The quantum character is captured
	by averaging over distributions of the initial conditions. These are given
	by Wigner functions which are only quasi-distributions because they
	also take negative values. In leading order for the quantum dynamics 
	of angular momenta \cite{david15}, it is sufficient to assume the initial
	distributions to be Gaussian. Then the mean values and the variance suffice
	to fully determine the initial distribution. Mean value and variance 
	are computed from the quantum mechanical expectation values of the 
	corresponding operators.
	Concretely, we solved the equations of motion for 
  $10^4 - 10^6$ different initial conditions drawn from the thus determined 
	normal distributions. The obtained time-dependent averages approximate the
	quantum expectation values. Note that any TWA is exact for systems which are bilinear in  bosonic variables and linear in spin
	variables \cite{polko10} as is implied by Ehrenfest's theorem. In all other
	cases, the TWA represents an approximation which only  captures leading quantum effects.
	By construction, its use is best justified if the dynamics is close to a classical one.

  The classical equations of motion are given by \cite{david15}
  \begin{equation}\label{Bewgl}
      \frac{\partial}{\partial t} S_\alpha^{\mathrm{cl}} = 
			\varepsilon_{\alpha \beta \gamma} 
			\frac{\partial H_{\mathrm{W}}}{\partial S_\beta^{\mathrm{cl}}} S_\gamma^{\mathrm{cl}} \, 
  \end{equation}
  with $\varepsilon_{\alpha \beta \gamma}$ being the tensor elements of the Levi-Civita tensor
  and $H_{\mathrm{W}}$ the Weyl-symbol of $H$.
	Note that only the three spin operators occur; we refer to this approach as the SU(2) TWA
	for short. Below, we will also consider extensions of the 
	SU(2) TWA denoted SU(n) TWA. Their advantage is to capture the dynamics 
	induced by non-linear Hamiltonians exactly as well.
  
  The  classical equations of motion in SU(2) TWA for the Hamilton function 
  \begin{equation}
      H = H_{\mathrm{CSM}} + H_{\mathrm{eZ}} + H_{\mathrm{nZ}} \, 
  \end{equation}
  read
  \begin{subequations}
  \begin{align}
      \frac{\mathrm{d}}{\mathrm{d}t} \pmb{S} &= \left( \sum_{k=1}^N  A_k \pmb{I}_k + \pmb{h} \right) 
			\times \pmb{S}  ,\label{bewglen1}
\\			
      \frac{\mathrm{d}}{\mathrm{d}t} \pmb{I}_k &= \left( A_k \pmb{S} + \pmb{h}_n \right) 
			\times \pmb{I}_k \, , \quad k \in {1,..,N}  .\label{bewglen2}
  \end{align}
  \end{subequations}
  The vector $\pmb{h}$ is given by
  \begin{equation}
      \pmb{h} = (-h,0,0)^T \qquad \mathrm{with} \qquad h = \gamma_{\mathrm{e}} B_{\mathrm{ext}}
  \end{equation}
  and its nuclear counterpart $\pmb{h}_{\mathrm{n}} $ by $\pmb{h}/z$.
  The initial conditions are drawn from Gaussian distributions resulting from 
  the quantum mechanical expectation values. The applied method is only an 
  approximation for our discussed model due to the hyperfine spin-spin interactions.

\subsection{Extended Truncated Wigner Approximation}	
	\label{ss:etwa}	
	
  For the quadrupolar interaction, even the local part of the Hamilton operator
	at a given nucleus $k$ comprises quadratic terms in \eqref{Hquadnk} so that SU(2) TWA is not exact.
	We consider this a serious caveat. It is known that non-local terms tend to
	average out because the influence of many interaction partners justifies this
	averaging. Generically, mean-field approaches become exact for large coordination
	numbers for this reason. Thus, the quantum fluctuations of non-local
	term tend to be small and a semi-classical approach is appropriate.
	In contrast, the local dynamics does not experience averaging effects for given
	spin. For this reason, we aim at an extension of SU(2) TWA such that the local
	dynamics is represented faithfully. Indeed, such an extension is possible
	at the expense of increasing the number of relevant observables \cite{david15}.
	
	The idea is to represent the
  quadratic terms as linear combinations of the generators of the respective SU(n) group 
	where $n=2I+1$ is the dimension of the local spin Hilbert space. For spin $I=1/2$
	the SU(2) TWA is locally exact. For spin $I=1$ we would need to consider
	SU(3) TWA and for the relevant case of $I=3/2$ we should use SU(4) TWA.
	In this way, one can express the local Hamiltonian as linear sum of an 
	extended set of operators. For SU(3) one needs 8 operators in total while
	SU(4) needs 15 operators. In general, $n^2-1$ operators are sufficient because
	any matrix of dimension $n$ can be represented as linear sum of $n^2$ operators
	including the identity. Hence, the extended SU($n$) will be locally exact.
	We emphasize, however, that the interactions between different spins are
	not exactly represented. Here, we again refer to the justification that a large number
	of interaction partners reduces the influence of fluctuations
	and thus renders the dynamics more classical and less quantum.
	The reduction of relative fluctuations allows one to derive a dynamic mean-field
	theory of dense spin systems \cite{grass21}.
	
  The equations of motion in the extended SU(3) version read \cite{david15}
  \begin{equation}
	\label{Bewgl2}
      \frac{\partial}{\partial t} X_\alpha^{\mathrm{cl}} = 
			f_{\alpha \beta \gamma} \frac{\partial H_{\mathrm{W}}}{\partial X_\beta^{\mathrm{cl}}} 
			X_\gamma^{\mathrm{cl}}  ,
  \end{equation}
  where $X$ are the generators of the respective SU(n) group.
  The structure constants $f_{\alpha \beta \gamma}$ can be calculated to fulfil
	the relations
  \begin{equation}
      [X_\alpha^{\mathrm{cl}},X_\beta^{\mathrm{cl}}] = i f_{\alpha \beta \gamma} X_\gamma^{\mathrm{cl}} .
  \end{equation}
	
  In the comparisons of various approaches, we also include results obtained from
	the equations of motion derived  in Ref.\  \cite{fisch22} using a path 
  integral formalism. This derivation yields an additional prefactor $\left(1-1/(2I)\right)$
	in front of the classical equations of motion of the SU(2) TWA which result from local bilinear
	interaction terms such as the quadrupolar interaction.
	The prefactor nicely captures the fact that no local bilinear terms occur for $I=1/2$.
  
\begin{figure}[htb]
      \centering
      \includegraphics[width=\columnwidth]{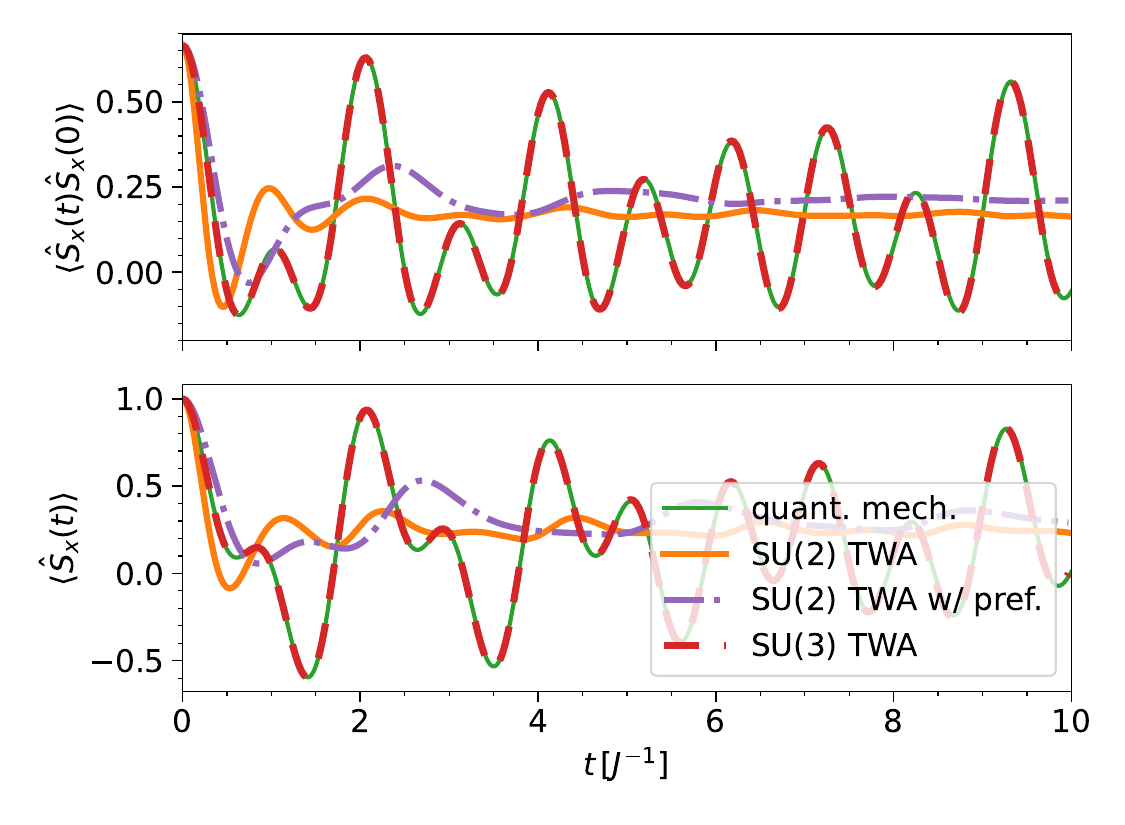}
      \caption[Comparison of the SU(2) and SU(3) TWA with the exact quantum mechanical result 
			for the quadrupolar Hamiltonian for a single spin $I=1$]
			{Comparison of the SU(2) TWA, the SU(3) TWA, the SU(2) TWA including the additional prefactor (denoted 
			``w/ pref.''), and the quantum mechanical result (denoted ``quant.\  mech.'') 
			for the quadrupolar Hamiltonian \eqref{Hquadtwa3} with $I = 1$
       considering the correlation (top) and the expectation value (bottom) of the $x$-component. 
       The parameters are set to $q=1 J$, $\eta = 0.5$, and $h = \sqrt{6} J$. 
			The TWA results are averaged over $10^6$ configurations drawn from normal distributions.}
      \label{SU3_Hquad}
   \end{figure}

  As a generic example, we apply the two SU(2) TWA approaches and the SU(3) TWA to the quadrupolar Hamiltonian
  \begin{equation}
	\label{Hquadtwa3}
      \hat{H}_{\mathrm{quad}} = q \left[ \left( 3\hat{I}_z^2 - \pmb{I}^2 \right) 
			+ \eta \left( \hat{I}_x^2 - \hat{I}_y^2 \right) \right] 
			+ \pmb{h} \cdot \hspace{0.2em} \hat{\hspace{-0.2em}\pmb{I}}  
  \end{equation}
  with $\pmb{h} = \frac{h}{\sqrt{3}} (1, 1, 1)^T$ and $I=1$. The specific equations of motion 
	based on Ref.\  \cite{david15} are given in Appendix \ref{a:eomHquad}.
	The results are shown in Fig.\  \ref{SU3_Hquad} for the parameters $q=1J$, $\eta = 0.5$, and $h = \sqrt{6} J$.
	The energy scale is $J$ and thus the time scale $1/J$ since we set $\hbar$ to unity for simplicity.
  One clearly sees that the SU(2) TWA without and with prefactor rather quickly deviate
	from the exact quantum mechanical results for the autocorrelation $\langle \hat{S}^x(t) \hat{S}^x(0) \rangle$ 
	as well as for the expectation value $\langle \hat{S}^x(t) \rangle$.
  The prefactor derived by Fischer et al.\ yields
	some improvement, but only for very short times. But the extended TWA, here SU(3) TWA, yields a perfect
	match to the quantum mechanical results as expected for $I=1$. This supports the chosen approach.
  
\begin{figure}[htb]
      \centering
      \includegraphics[width=\columnwidth]{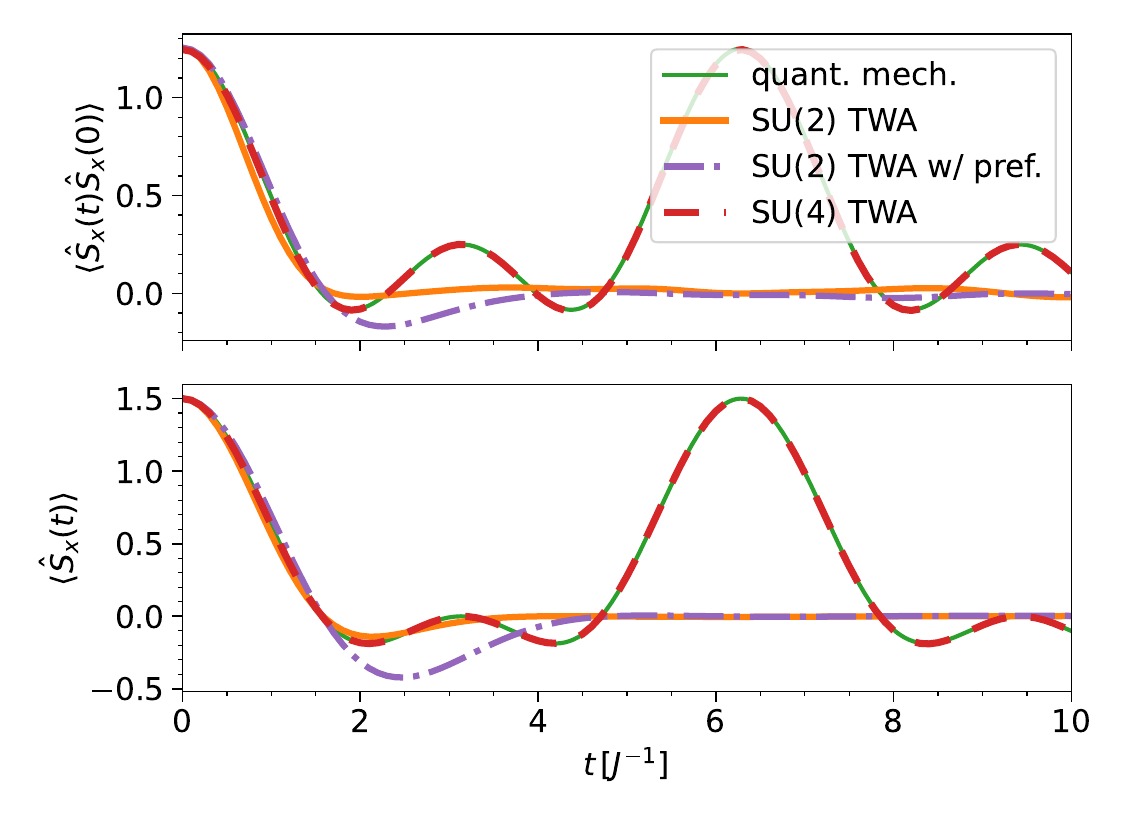}
      \caption[Comparison of the SU(2) and SU(3) TWA with the exact quantum mechanical result 
			for the quadrupolar Hamiltonian for a single spin $I=3/2$]
			{Comparison of the SU(2) TWA, the SU(3) TWA, the SU(2) TWA including the additional prefactor (denoted 
			``w/ pref.''), and the quantum mechanical result (denoted ``quant.\  mech.'') 
			for the quadrupolar Hamiltonian \eqref{H_david} with $I = 3/2$
       considering the correlation (top) and the expectation value (bottom) of the $x$-component. 
       The TWA results are averaged over $10^6$ configurations drawn from normal distributions.}
      \label{SU4_Hdavid}
   \end{figure}

 This idea can be further extended to a SU(4) TWA required for a spin $I=3/2$
	and generally to SU(2I+1) TWA for spin $I$.
	For our model, we need $I=3/2$ so that we focus on this case. 
	Figure \ref{SU4_Hdavid} depicts the comparison of various approaches for
	the simple Hamiltonian
  \begin{equation}
	\label{H_david}
      \hat{H}_{\mathrm{simpl}} = \frac{J}{2}\left(  \hat{I}_z^2 - 2\hat{I}_z \right) 
  \end{equation}
	for $I=3/2$. The lengthy equations of motion for 15 observables
	can be found in Appendix \ref{a:eomHsimpl}.
  The details of the extension to SU(4) are explained in the Supplementary Information.

  The extended TWA approaches are applied to the central spin model supplemented
	by nuclear quadrupolar interactions.
  For the simple model (A) for the potential $V$, the equations of motion for both 
	the SU(2) TWA and the SU(4) TWA are set up and simulated.
	In these simulations, we observe, see below, that there is only a factor between the
	results from SU(2) TWA and the SU(4) TWA so that one does not need to re-do
	all calculations with SU(4) TWA.
	For the stochastic model (B) for the potential $V$  the runtime is significantly 
  larger due to the increased number of dynamic variables to be followed. 
  Thus, only the SU(2) TWA is simulated in this case because these results are sufficient
	to draw conclusions up to the above mentioned factor.
	
  For the simple model (A) for $V$ given in \eqref{quadrchoiceofV}, 
	the equations of motion for the nuclear spins are given by 
	  \begin{equation}
	\label{BewglIkmitHquad}
      \frac{\mathrm{d}}{\mathrm{d}t} \pmb{I}_k 
			= \left( A_k \pmb{S} + \pmb{h}_{\mathrm{n}} + \pmb{b}_{\mathrm{Q},k} \right) 
			\times \pmb{I}_k \, , \quad k \in {1,\ldots,N} 
  \end{equation}
	in SU(2) TWA according to Eq.\  \eqref{Bewgl}. The vector $\pmb{b}_{\mathrm{Q},k}$ is given by 
  \begin{equation}\label{bvecquadV}
      \pmb{b}_{\mathrm{Q},k} = \frac{2|e|Q}{I(2I-1)} \left(\begin{array}{c} a I_{x,k} \\ b I_{y,k} \\ c I_{z,k} 
			\end{array}\right) .
  \end{equation}
  The equations of motion for the central spin do not change.
  We use the relation 
	\be
	q_k= \frac{2|e|Q}{I(2I-1)}a
	\ee
	between the quadrupolar interaction strength $q_k$ 
	and the parameter $a$. 
	
  For the SU(4) TWA, we first re-express the Hamiltonian with the extended number of observables
	$X_j$ with $j\in\{1,2,\ldots,15\}$, see Supplemental Material,
\be
      H_{\mathrm{quad}} = \sum_{k=1}^N \frac{|e|Qa}{6I(2I-1)} 
			\left[ 4 X_{8,k} + \sfrac{8}{\sqrt{5}} X_{15,k} + 5 \mathbb{1} \right] .
\ee
Using this Hamiltonian and 	Eq.\  \eqref{Bewgl2}, the equations of motion given in the 
Supplemental Material are obtained. The $q_k$ are defined  as before.

For the stochastic model (B) for $V$ for the SU(2) TWA, Eq.\  \eqref{Hquadnk} again  yields Eq.\  \eqref{BewglIkmitHquad}
with the vectors $\pmb{b}_{\mathrm{Q},k}$  given by 
	\be
  \begin{aligned}
      \pmb{b}_{\mathrm{Q},k} &= 2 q_k \left[ \left( \pmb{I}_k\cdot \pmb{n}_k^z \right) \pmb{n}_k^z 
			+ \frac{\eta_k}{3}  \left( \left( \pmb{I}_k\cdot  \pmb{n}_k^x \right) \pmb{n}_k^x \right.\right.
			 \\
			& \qquad - \left.\left. \left( \pmb{I}_k\cdot  \pmb{n}_k^y \right) \pmb{n}_k^y \right) \right] .
  \end{aligned}
	\ee
  Random numbers are drawn for the parameters $q_k$, $\eta_k$ and $\theta_k$ according to the distributions 
	computed for Ga and As in Ref.\  \cite{bulut14}. The vectors $\pmb{n}_k$ are obtained by 
    \begin{align}
      \pmb{n}_k^z = \left(\begin{array}{c} \sin(\theta_k) \cos(\varphi_k) 
			\\ \sin(\theta_k) \sin(\varphi_k) \\ \cos(\theta_k) \end{array}\right) \, , 
			\enspace & \pmb{n}_k^x = \left(\begin{array}{c} \cos(\theta_k) \cos(\varphi_k) 
			\\ \cos(\theta_k) \sin(\varphi_k) \\ -\sin(\theta_k) \end{array}\right) \, ,
			\nonumber
			\\  \pmb{n}_k^y = & \left(\begin{array}{c}  -\sin(\varphi_k) \\ \cos(\varphi_k) \\ 0 \end{array}\right)
					\end{align}
  for which the values $\varphi$ are drawn from an equal distribution in the interval $[0, 2 \pi]$.

\subsection{Application of the Pulses}
  \label{ss:pulses}
	
  The Larmor precession of the central spin, induced by the external magnetic field, dephases due to the hyperfine interaction with the bath spins  \cite{merku02}. The envelope of the Larmor precessions is given 
	in the semi-classical description by
  \begin{equation}
	\label{envelope}
      S_{\mathrm{env}}(t) \coloneqq \sqrt{ \left[\, \overline{S^z(t)S^z(0)} \, \right]^2 
			+ \left[ \, \overline{S^y(t)S^z(0)} \, \right]^2  },
  \end{equation}
  where the overbar indicates the average over the initial conditions.
  To remedy the dephasing, laser pulses are applied in periodic trains to re-align the central spins. 
 We call the way in which these pulses are treated in the present semi-classical TWA the pulse model. 
 The features of various pulse models have been investigated before in great detail
	\cite{scher18, scher20, scher21c}. 
Hence, we adapt the best suited simple pulse model in this work. It is simple in the sense
that the intermediate excitation of a trion and its subsequent decay are neglected.
We consider all pulses to be instantaneous. In order to take the quantum mechanical aspects into account, 
we interpret each pulse as a quantum mechanical measurement \cite{scher18} in which the uncertainty
needs to be accounted for. Thus, no perfect alignement can be reached. The orientation the electronic
spin directly after the pulse is given by
  \begin{equation}
      \pmb{S}_a = (X, Y, 1/2)^T \,,
  \end{equation}
  where $\pmb{S}_a$ stands for the alignment of the central spin after application of the pulse.
  The values $X$ and $Y$ are random numbers drawn from a normal distribution with $\mu=0$ and 
  $\sigma^2 = 1/4$ resulting from the quantum mechanical expectation value 
	$\langle (\hat{S}_\alpha)^2 \rangle = 1/4$. Their inclusion ensures the
	quantum mechanical uncertainty and the correct average length of the central spin of 
	$S^2=3/4$.

\subsection{Numerics}	
	\label{ss:numerics}	
	
  To solve the equations of motion the Dormand-Prince algorithm \cite{dorma80} is used.
	This is a  5th order adaptive Runge Kutta algorithm which adjusts the step
  size dynamically. The required random numbers are generated with the 
	Mersenne-Twister pseudo random number  generator \cite{matsu98b}.

  A main difficulty in the numerical simulation of the 
	TWA equations is the runtime. The required runtime increases approximately 
  quadratically with increasing magnetic field and linearly with the
  number of nuclear spins \cite{scher21c, scher18}. 
	In order to mitigate the quickly increasing runtime, we parallelize
  partial computations by MPI  (Message Passing Interface) \cite{messa15}.
  The used resources are 1600 cores on a locally available compute cluster 
	making it possible to run two simulations at the same time,
	each using 800 cores. Depending on the values selected for the parameters
  the runtime varies strongly. For one set of date, the simulations take up to several weeks.

    \section{Spin Noise}
    \label{s:spinnoise}

Prior to treating the driven case with the periodic pulse trains we
investigate spin noise, i.e., the spin-spin autocorrelation of the
electronic spin in time. No external magnetic field
and no pulses are considered. To gauge our approach with respect to 
established properties we first briefly treat the case without
quadrupolar interactions. Then, the influence of the quadrupolar
interaction on the spin-spin autorcorrelation is studied.
Due to the constraints on runtime, it is not possible to choose the 
parameters directly relevant for experimental setups.
Therefore, we vary the parameters and discuss 
what behavior can be expected upon extrapolation to the experimental regime.

\subsection{Without Quadrupolar Interaction}
\label{ss:spinnoise_woquad}

\begin{figure}[htb]
    \centering
    \includegraphics[width=\columnwidth]{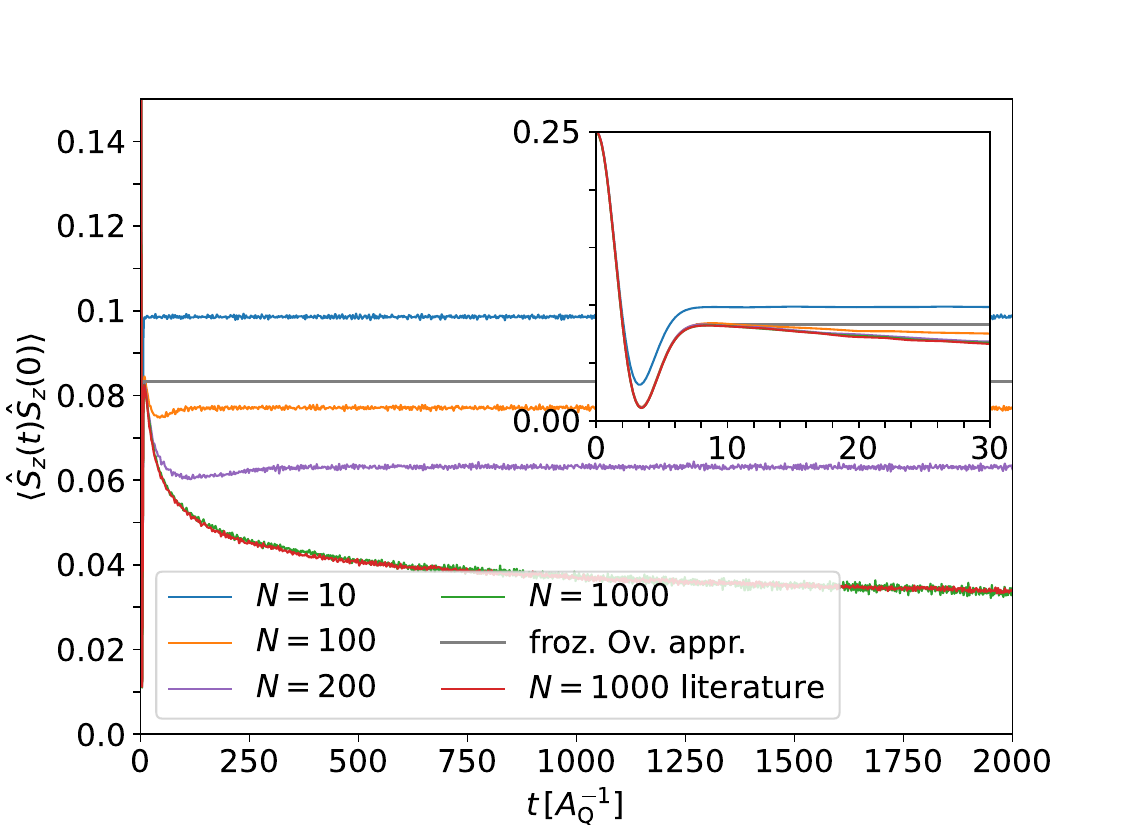}
    \caption[Long-time dynamics without quadrupolar interaction at constant $\gamma$]
    {Long-time dynamics at constant $\gamma$ of $0.01$ if $N$ and thus $r_{\mathrm{cut}}$ are
		varied.  Here, $I=1/2$ is used for the nuclear spins. 
     We use $A_{\mathrm{Q}} = 1$ and $h = 0 \,$T. The autocorrelations are
		averaged over $10^5$ initial conditions.
      The red line represents results from Ref.\  \cite{fause17a} (denoted ``literature''), 
			where  the same values and $N=1000$ are used. 
			The result of the frozen Overhauser field approximation 
			\cite{merku02} (denoted ``froz.\  Ov.\  appr.'') 
			is shown in gray for orientation.}
    \label{longtimebehaviourB0_rcovar}
\end{figure}

\begin{figure}[htb]
    \centering
    \includegraphics[width=\columnwidth]{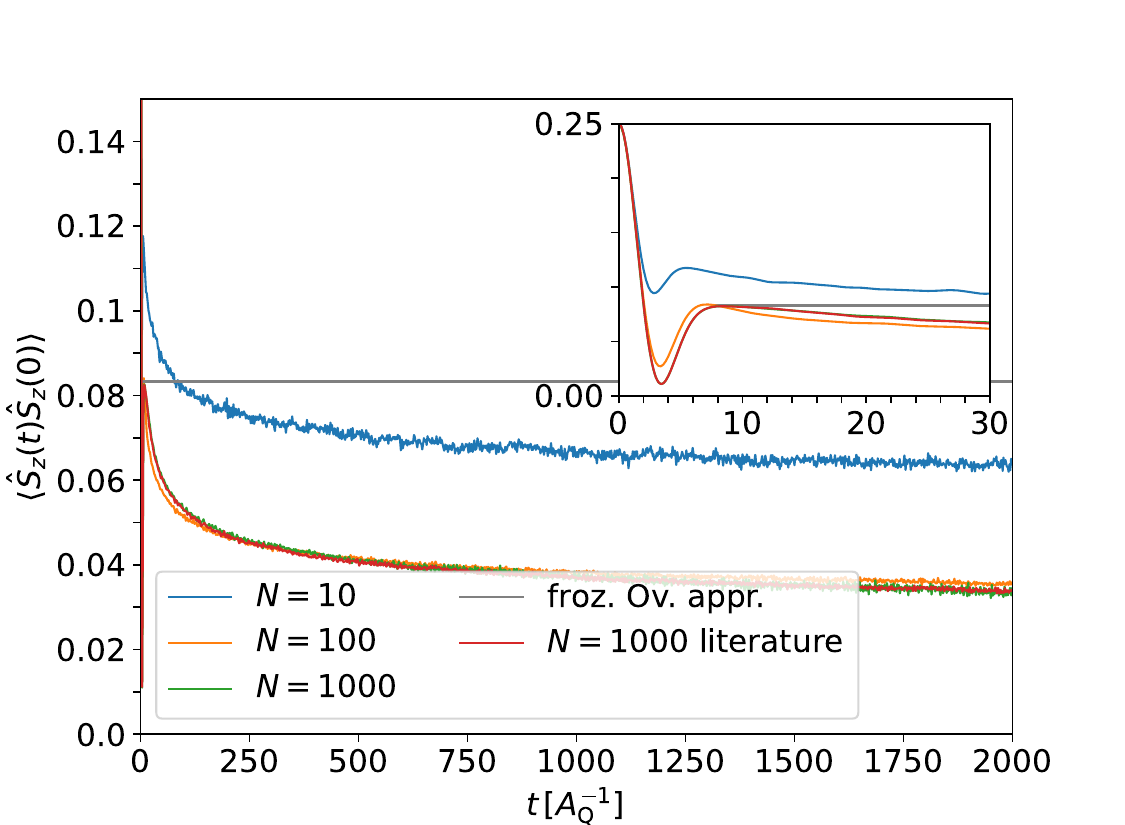}
    \caption[Long-time dynamics without quadrupolar interaction at constant cutoff radius]
    {Long-time dynamics at constant cutoff radius $r_{\mathrm{cut}} = \sqrt{10} \, l_0$
		if $N$ and thus $\gamma$ are varied. The other parameters are the same as in 
		Fig.\  \ref{longtimebehaviourB0_rcovar}.}
    \label{longtimebehaviourB0_gamvar}
\end{figure}

We investigate the spin noise excluding the quadrupolar interaction for 
varying values of $N$ keeping either $\gamma=2/N_\text{eff}$ or $r_{\mathrm{cut}}$ constant.
The resulting impact for constant $\gamma$ and varying $r_{\mathrm{cut}}$ is shown
 in Fig.\  \ref{longtimebehaviourB0_rcovar}. 
For a reduced cutoff radius, the long-time dynamics is described less reliably whereas
the minimum at short times is still rendered faithfully.
If $\gamma$ is reduced and $r_{\mathrm{cut}}$ kept constant, the minimum at the
beginning is less pronounced, but the description of the long-time dynamics is better
preserved, see Fig.\  \ref{longtimebehaviourB0_gamvar}.

In both cases, it can be seen that for a too small number of nuclear spins of $N = 10$,
there is a huge deviation in comparison to other results. In particular the red curve
has been obtained by the very efficient and reliable spectral density approach established in Ref.\
\onlinecite{fause17a} using Eq.\@ \eqref{Akindex}.
With a comparable choice of parameters, we reach a very good agreement
of our results with these reference results.
Overall, one has to keep the influence of the choice of
parameters in mind when analyzing the results of subsequent simulations. 
Especially, the number $N$ of nuclear spins needs to be chosen sufficiently
large.

\subsection{Simple Model (A) for the Potential $V$}
\label{ss:spinnoise_simple}

\begin{figure}[htb]
    \centering
    \includegraphics[width=\columnwidth]{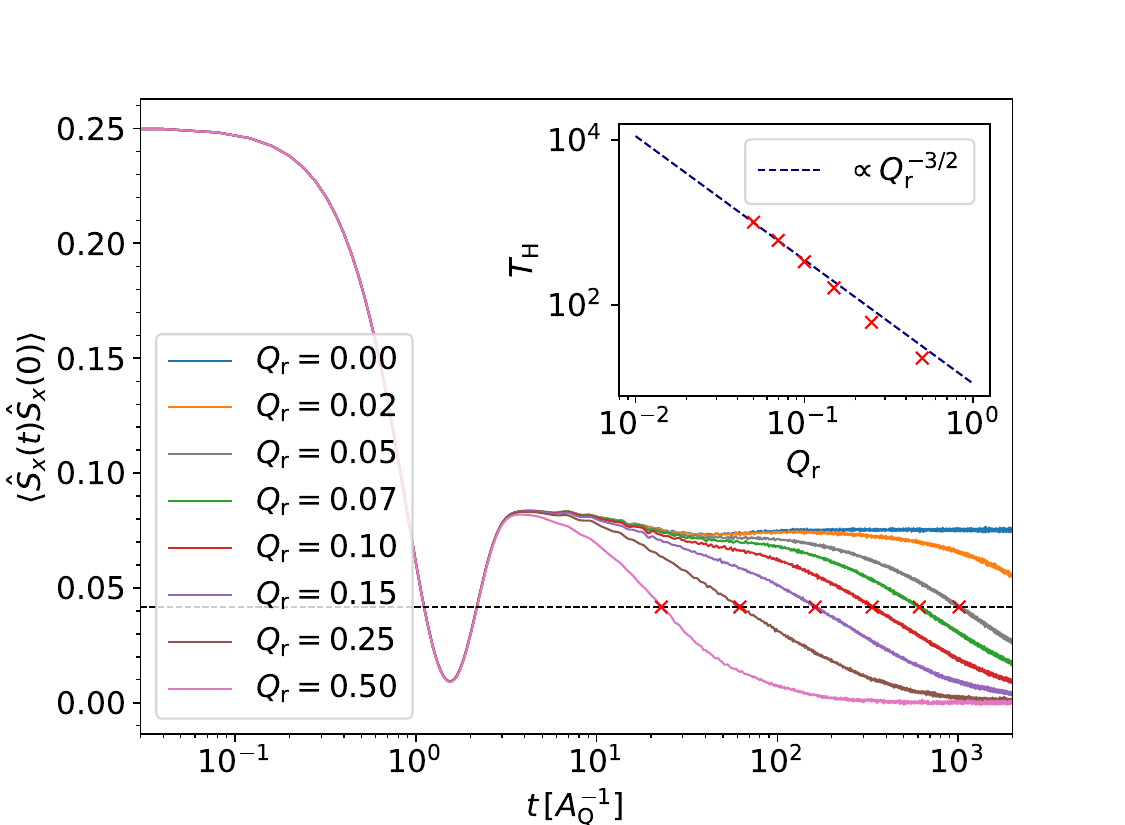}
    \caption[Long-term behavior of the autocorrelation for various $Q_{\mathrm{r}}$ in SU(2) TWA]
    {Long-term behavior of the autocorrelation without external magnetic field for various values of
		$Q_{\mathrm{r}}$ simulated with a SU(2) TWA. Note  the logarithmic time scale.
     The values $I = 3/2$, $A_{\mathrm{Q}} = 1$, $N = 100$, and $\gamma = 0.01$ are used. 
     Each curve is averaged over $10^6$ initial conditions.
     In the inset the dependence of  $T_{\mathrm{H}}$ on $Q_{\mathrm{r}}$ is shown in a
		double logarithmic plot. The time $T_{\mathrm{H}}$ is defined by 
		the instant when the black dashed line is crossed in the main panel.
		The inset clearly indicates a power law with exponent $3/2$.}
    \label{qi_vglhack15}
\end{figure}

For the spin noise, we  investigate the long-time dynamics without external magnetic field for 
the simple model (A) for $V$.
We start by considering the influence of $Q_{\mathrm{r}}$ on the results from SU(2) TWA
as depicted in Fig.\  \ref{qi_vglhack15}. 
The larger $Q_{\mathrm{r}}$ is chosen, the smaller the time $T_{\mathrm{H}}$ is
at which the autocorrelation drops below $\langle \hat{S}_x(0) \hat{S}_x(0)\rangle/6$. 
This value is indicated by the dashed black line in Fig.\  \ref{qi_vglhack15}. 
In the inset, the dependence of $T_{\mathrm{H}}$ on $Q_{\mathrm{r}}$ is
displayed in a double-logarithmic plot clearly indicating the power law
 $T_{\mathrm{H}}\propto Q_{\mathrm{r}}^{-3/2}$. This was also found previously \cite{hackm15}
in fully quantum mechanical calculations for smaller systems of $N=10$ spins with $I=3/2$.

\begin{figure}[htb]
    \centering
    \includegraphics[width=\columnwidth]{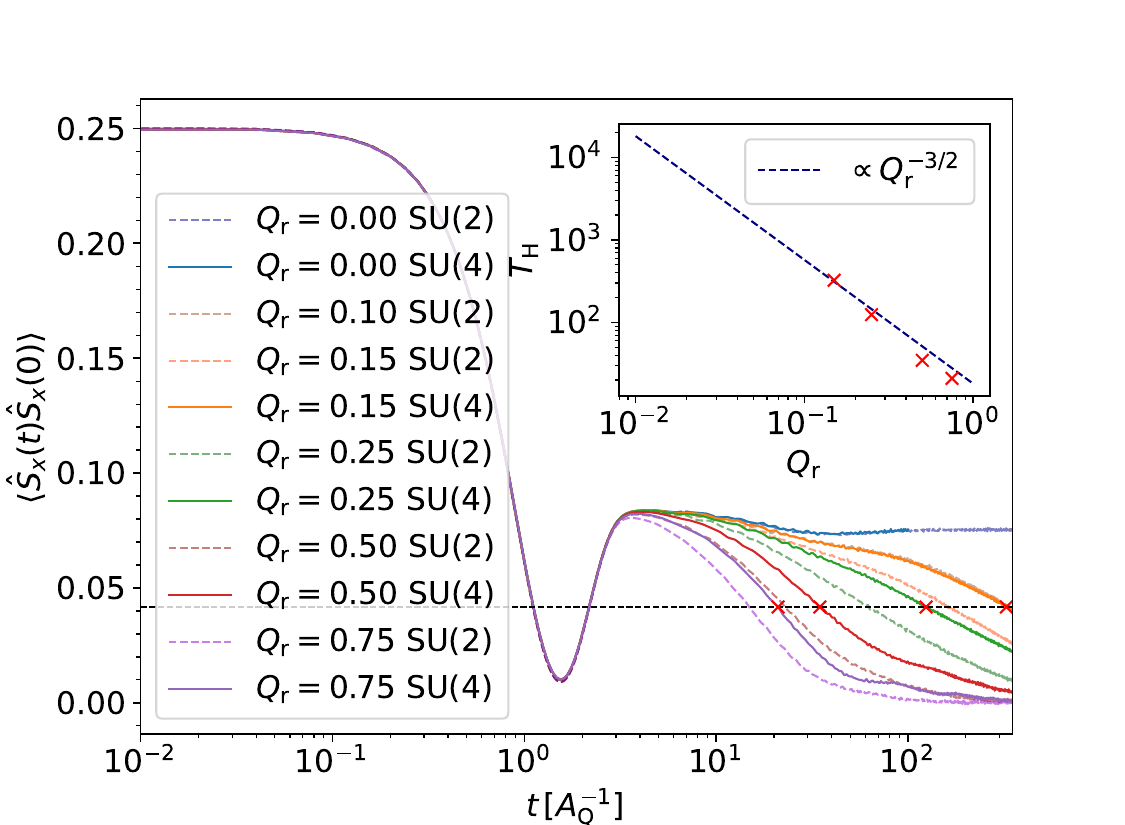}
    \caption[Comparison of the SU(2) TWA and the SU(4) TWA data for various $Q_{\mathrm{r}}$]
     {Comparison of the SU(2) TWA and the SU(4) TWA data focussing on the long-term behavior 
      of the autocorrelation without external magnetic field for various $Q_{\mathrm{r}}$. 
      The choice of parameters in the plot and in the inset is the same as in Fig.\  \ref{qi_vglhack15}.}
    \label{qi_vglsu4}
\end{figure}

Next, we compare the SU(2) data to the data computed from the more accurate SU(4) TWA representation.
Figure \ref{qi_vglsu4} shows that the behavior is very similar. Indeed, the curve
fall almost on top of each other if the value $Q_{\mathrm{r}}$ is rescaled 
by a factor of about $1.5$. For instance, the curves from SU(2) TWA at $Q_{\mathrm{r}} = 0.1$ and from 
 SU(4) TWA at $Q_{\mathrm{r}} = 0.15$ agree nicely.
Furthermore, the results for the decay times $T_\text{H}$ 
from SU(4) TWA display the same power law $Q_{\mathrm{r}}^{-3/2}$ as before
as is to be expected if a constant factor in $Q_{\mathrm{r}}$ maps their results onto each other.
There are, however, also differences. 
In the curve from SU(4) TWA at $Q_{\mathrm{r}} = 0.75$, for instance, a weak shoulder 
occurs at large times which is not present in the SU(2) TWA data. 
Slight differences in the slope of the curves can be seen as well.

\begin{figure}[htb]
    \centering
    \includegraphics[width=\columnwidth]{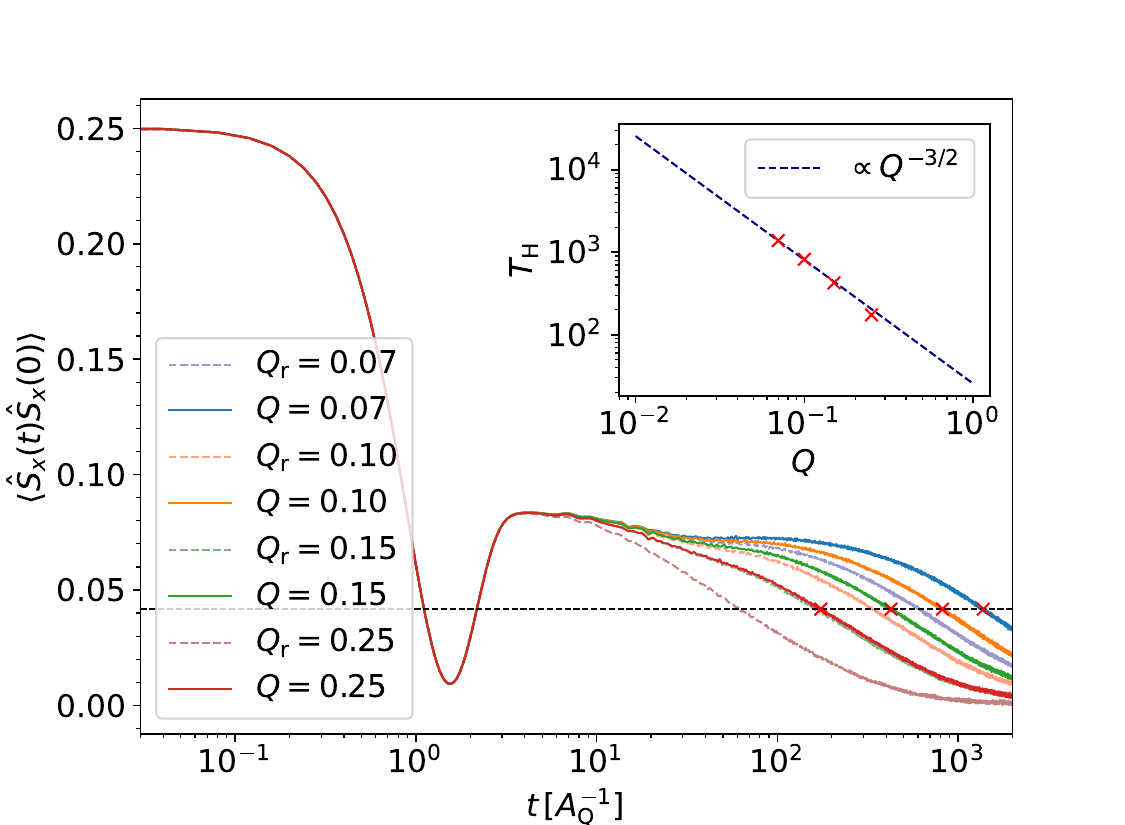}
    \caption[Comparison of data from SU(2) TWA for various $Q_{\mathrm{r}}$ and $Q$]
    {Comparison of data from SU(2) TWA for various $Q_{\mathrm{r}}$ and $Q$ without external magnetic field. 
     The choice of the parameters in the plot and in the inset is 
     the same as in Fig.\  \ref{qi_vglhack15}.}
    \label{qi_vglQ}
\end{figure}

Finally, the influence of the two measures of the strength of the quadrupolar interaction 
$Q_{\mathrm{r}}$ and $Q$  is assessed in Fig.\  \ref{qi_vglQ}. 
For simplicity, this is done for the quicker approach based on SU(2) TWA.
Qualitatively, the same curves are obtained,  but they are shifted by a factor. 
Such a shift was to be expected because the choice of the minimum value of $A_k$ 
as gauge in $Q$ implies a larger value than then one in 
$Q_{\mathrm{r}}$ for the same set of hyperfine couplings constants.

In the experimental studies of spin noise, mostly the $z$-component of the spin
is investigated \cite{becht15, glase16}.
However, in the simple model (A) for $V$, the quadrupolar interaction does not influence 
the $z$-component at all due to its rotational symmetry about the $z$-axis.
This can be seen directly on inserting Eq.\ \eqref{bvecquadV} into 
Eq.\ \eqref{BewglIkmitHquad}. The $z$-component of the cross product vanishes for 
$a=b$.

\subsection{Stochastic Model (B) for the Potential $V$}
\label{ss:spinnoise_stochastic}

So far, we assumed the simple quadratic electric potential (A) for the study of the quadrupolar interaction. 
Here, we intend to study a more realistic choice for the electric potential, namely the stochastic model
based on previous results \cite{bulut12,bulut14} on the charge distribution in quantum dots.
In the sequel, the $q_k$ are defined differently by a factor of 2 from what we did in the simple model (A). 
This allows us to stay consistent with the definitions used in previous work \cite{hackm15, fisch22}. 

\begin{figure}[htb]
    \centering
    \includegraphics[width=\columnwidth]{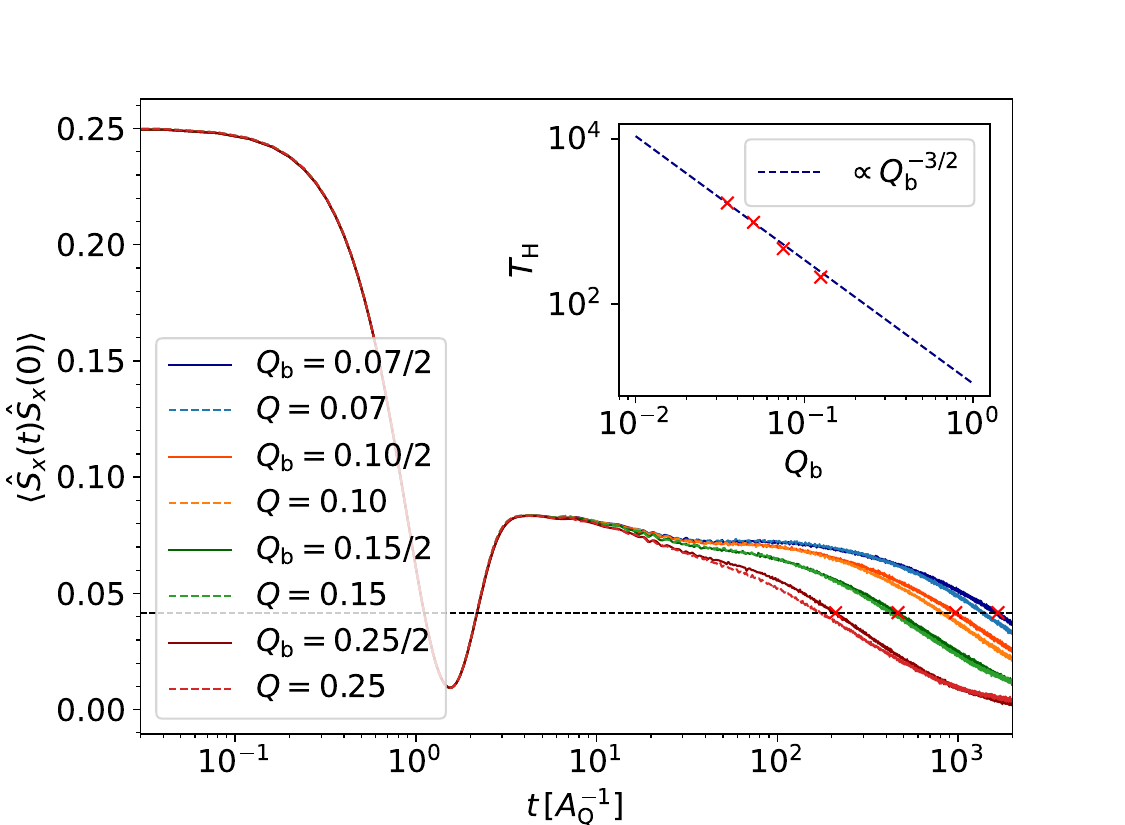}
    \caption[Comparison of the autocorrelations in the simple model (A) and in the 
		stochastic model (B) for $V$ for four values of $Q$]
    {Comparison of autocorrelation without external magnetic field using the simple 
		 model (A) and the stochastic model (B) $V$ for four values of $Q$. 
		The SU(2) TWA calculation is used for efficiency. In the inset the expected power law
		for the decay time $T_{\mathrm{H}}$ is retrieved.
     The parameters used are the same as in Fig.\ \ref{qi_vglhack15}.}
    \label{qi_bulutaylangzeit}
\end{figure}

Figure  \ref{qi_bulutaylangzeit} displays the
temporal behavior for four different $Q$ values. 
In addition, the previous SU(2) TWA results for the simple model (A) for $V$ are plotted
taking into account the factor of two between the $Q$ values. 
The $Q$ value for the stochastic model is denoted $Q_{\mathrm{b}}$ to make a clear distinction possible. 
It is related to the previous definition by $Q_{\mathrm{b}} = Q/2$ because of the different definitions of the $q_k$.
For all values, a very similar behavior of the curves resulting from the two approaches can be observed.
The power law of the lifetime $T_{\mathrm{H}} \propto Q_{\mathrm{b}}^{-3/2}$ is found again. 
Only a very small shift of the respective curves and 
a slightly steeper curve at $Q_{\mathrm{b}} = 0.25/2$ can be discerned.

    \section{Revival Amplitude}
    \label{s:revivalamplitude}

\begin{figure}[htb]
    \centering
    \includegraphics[width=\columnwidth]{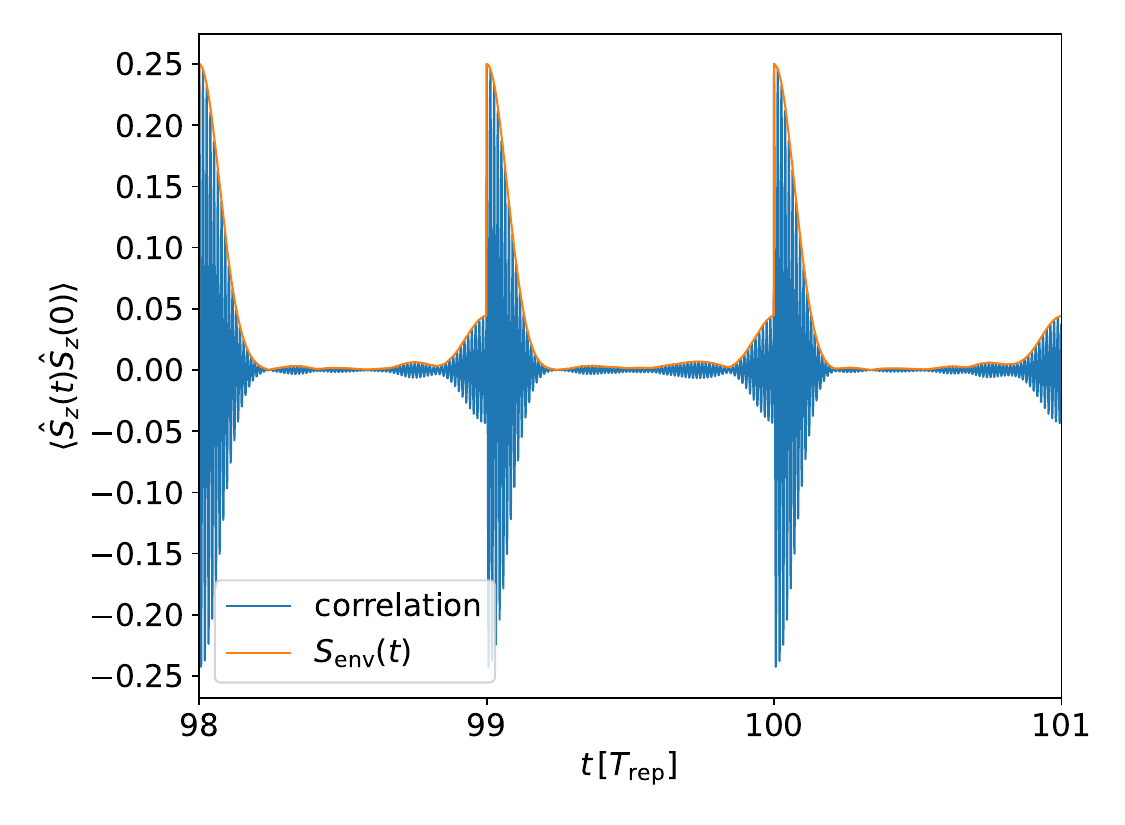}
    \caption[Central spin dynamics resulting from the application of the pulse model after 98 pulses]
    {Time dependence of the autocorrelation (blue curve) of the $z$-component of the central spin 
    and its envelope (orange curve) according to Eq.\  \eqref{envelope} 
		as a function of time in units of the repetition time $T_{\mathrm{rep}}$, i.e., the time 
		between two consecutive pulses. The signal just before each strong uprise is the revival
		amplitude.}
    \label{wiederholtesPulsen}
\end{figure}

Having established that our approach reproduces the known effects of weak quadrupolar
interactions on spin noise we pass on and investigate the quadrupolar effects on the revival 
amplitude after long trains of periodic pulses. Such periodic pulsing trains the quantum
dot so that the distribution of Overhauser fields takes values which favor 
a revival of the polarization of the electronic spin as shown in Fig.\  \ref{wiederholtesPulsen}. 
The increase is often called the revival amplitude; it is similar to a spin Hahn echo
in NMR. The revival amplitude is defined as value of the envelope function just before the next pulse
\begin{equation}
\label{defSrev}
    S_{\mathrm{rev}} (n_{\mathrm{p}} T_{\mathrm{rep}}^-) = \lim_{\delta\to 0^+} 
		S_{\mathrm{env}}(n_{\mathrm{p}} T_{\mathrm{rep}}-\delta) .
\end{equation}

The occurrence of the revival amplitude can be explained by NIFF
(nuclei induced frequency focusing) \cite{greil07a, beuge16, jasch17, scher18, klein18}
via the hyperfine interaction. The periodic realignment of the central spin 
by the pulses affects the dynamics of the bath spins. The
distribution of their precession frequencies is essentially given
by the distribution of the Overhauser field. It is being focused on  modes 
of the spin dynamics which are commensurate with the pulse repetition rate.
When this condition is fulfilled, pulsing no longer leads to a change in the 
revival amplitude. A quasi-stationary non-equilibrium steady state (NESS) is reached.

\subsection{Without Quadrupolar Interaction}
\label{ss:woquad}

First, the behavior of the occurring revival amplitude is investigated
without the quadrupolar interaction in order to retrieve known results 
and to verify the present approach.
Because of the long simulation times, it is not possible to choose the 
experimental parameter values.
In all calculations without quadrupolar interaction we use the SU(2) TWA 
because there is no need for the extension in absence of local quadratic terms.

To investigate the revival amplitude we define its relative value
\begin{equation}
\label{relativerevampl}
    S_{\mathrm{rel}}(n_{\mathrm{p}}) := \frac{S_{\mathrm{rev}}
		(n_{\mathrm{p}} T_{\mathrm{rep}}^-)}{S_{\mathrm{rev}}(n_{\mathrm{p}} T_{\mathrm{rep}}^+)} \, ,
		\quad n_{\mathrm{p}} \in \mathbb{N} ,
\end{equation}
where $n_{\mathrm{p}} T_{\mathrm{rep}}^-$ is given by Eq.\  \eqref{defSrev} 
and $n_{\mathrm{p}} T_{\mathrm{rep}}^+$ analogously, but with $+\delta$ in the argument.
The saturation value of the revival amplitude cannot be read off directly from the data
for two reasons. First, the convergence is rather slow, in particular for 
large magnetic fields because the simulation up to a fixed time requires $\propto B$ steps
and the saturation of the revival amplitude requires $\propto B^2$ pulses so that in total
a cubic scaling results \cite{scher21c}. Second, there is
substantial statistical noise in the revival amplitude. 
Therefore, a fit function is used to determine the respective values
which has proven to be adequate \cite{scher20, scher21c}
\begin{equation}
\label{Ausgleichsfunktion}
    S_{\mathrm{fit}} (n_{\mathrm{p}}) = A_{\mathrm{NIFF}} \frac{2}{\pi} \arctan \left( \frac{n_{\mathrm{p}}}{\nu} \right) + B_{\mathrm{off}} . 
\end{equation}
The value $B_{\mathrm{off}}$ describes a statistical offset and the 
finite value $A_{\mathrm{NIFF}}$ stems from the NIFF \cite{scher21c}. 
The inverse rate of NIFF is characterized by the parameter $\nu$ \cite{scher21c}. 
Finally, the saturation value is given by \cite{scher20}
\begin{equation}
\label{NESSausgleich}
    S_{\mathrm{fit, NESS}} = \mathrm{sign}(\nu) A_{\mathrm{NIFF}} + B_{\mathrm{off}} .
\end{equation}
The  error of this procedure is determined by the root-mean-square deviation of the fit 
from the last 10 percent of the data points because only the fluctuations around  the saturation value 
are of interest. This includes both the statistical fluctuations and how well the fit captures the data.

\begin{figure}[htb]
    \centering
    \includegraphics[width=\columnwidth]{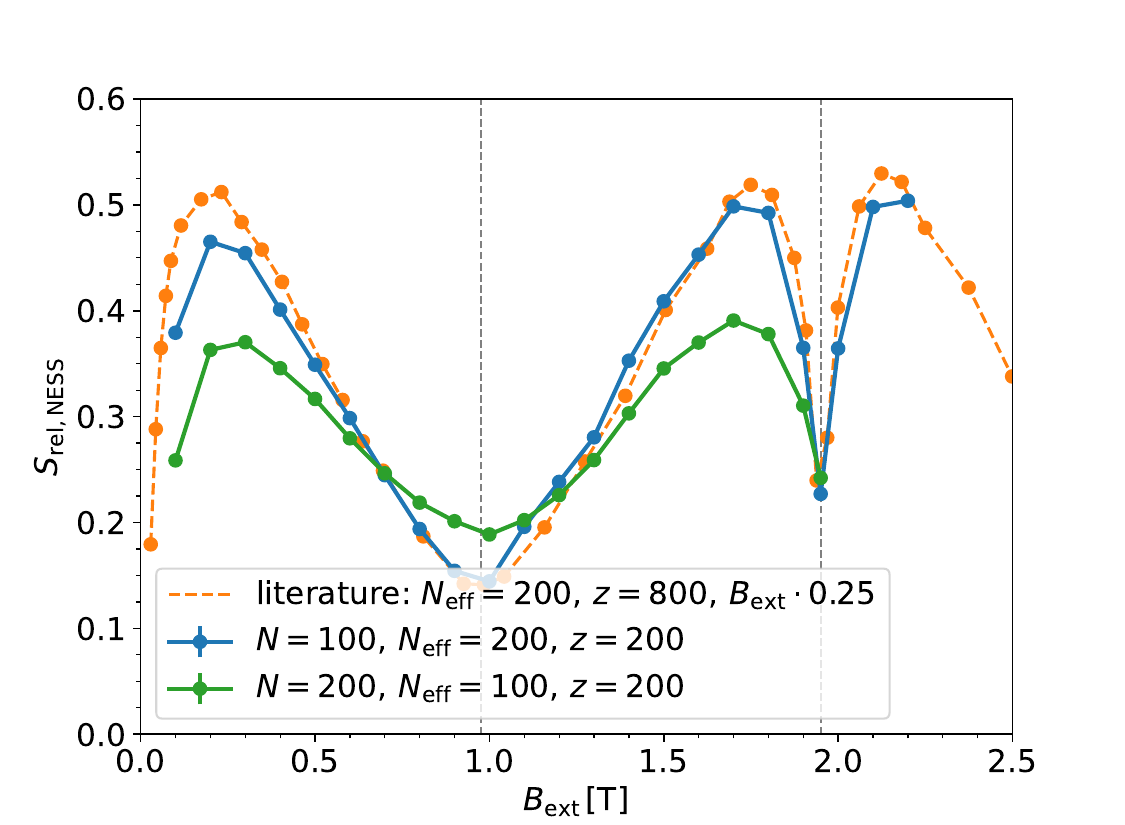}
    \caption[Relative revival amplitude in the NESS as a function of the external magnetic field without quadrupolar interaction]
		{Relative revival amplitude \eqref{relativerevampl} in the NESS as a function of the external 
		magnetic field. Data from Ref.\  \onlinecite{klein18} is included (denoted `literature').
    The values $I=3/2$, $A_{\mathrm{Q}} = 1.19$, and $T_{\mathrm{rep}} = 13.2 \,$ns 
    are used because these values are the experimental ones \cite{klein18, jasch17}. 
    The data is averaged over $10^4$ initial conditions. 
		A value of $z=200$ is used for the two simulated curves while 
		Ref.\  \onlinecite{klein18} used $z=800$. Hence, the magnetic field values of the literature data is 
		re-scaled. The lines are guides to the eye.}
    \label{Srev_vglPh}
\end{figure}

The value of the revival amplitude depends on the magnetic field. The saturation
values for varying magnetic fields is shown in Fig.\  \ref{Srev_vglPh} together with results 
from Ref.\  \cite{klein18}.
Two sets are displayed, one for $N=100$ and $N_{\mathrm{eff}} = 200$ and one 
for $N=200$ and $N_{\mathrm{eff}} = 100$. 
The former value for $N_{\mathrm{eff}}$ corresponds to the value used in Ref.\  \onlinecite{klein18}.
In contrast to the data in this reference, a value  $z=200$ is used here instead of $z=800$. 
This re-scaling has the vital advantage to reduce the runtime significantly since the 
simulation time required to approach saturation increases cubically $\propto B^3$ \cite{scher21c}.
The curve with the same $N_{\mathrm{eff}}$ as in Ref.\  \onlinecite{klein18} fits very well 
to the results of our data if the magnetic external field is re-scaled by 
a factor of four as in the $z$-factor.
The characteristic minima occur at the re-scaled smaller magnetic fields, 
but the qualitative behavior is very much the same.

When $N_{\mathrm{eff}}$ is halved and $N$ is doubled  
the revival signal does not change qualitatively, 
but there is a change in the magnitude of the maximum 
and minimum values of the curve, see Fig.\ \ref{Srev_vglPh}.
From previous results, for instance Fig.\ 11 in Ref.\ \onlinecite{klein18} 
or Fig.\ $5.18$ in Ref.\ \onlinecite{scher21c}, we know that the
quantitative changes are due to the change of $N_{\mathrm{eff}}$.
The larger $N_{\mathrm{eff}}$ the sharper the structures are, e.g., 
the higher and narrower the dips are. The precise value of $N$ does not
have a particularly high impact as long as it is not too small, i.e., 
the cutoff radius $r_\text{cut}$ should not be chosen too small.

In conclusion, the  model and the employed parameters reproduce the 
known results from previous studies well and thus provide a good starting point
for the investigation of the effects of quadrupolar interactions.
In particular, our findings so far justify to re-scale $z$ and thus the magnetic fields
down so that the simulations do not take too long.

\subsection{Revival Amplitude in the Simple Model (A)}
\label{ss:revampl_simpl}

\begin{figure}[htb]
    \centering
    \includegraphics[width=\columnwidth]{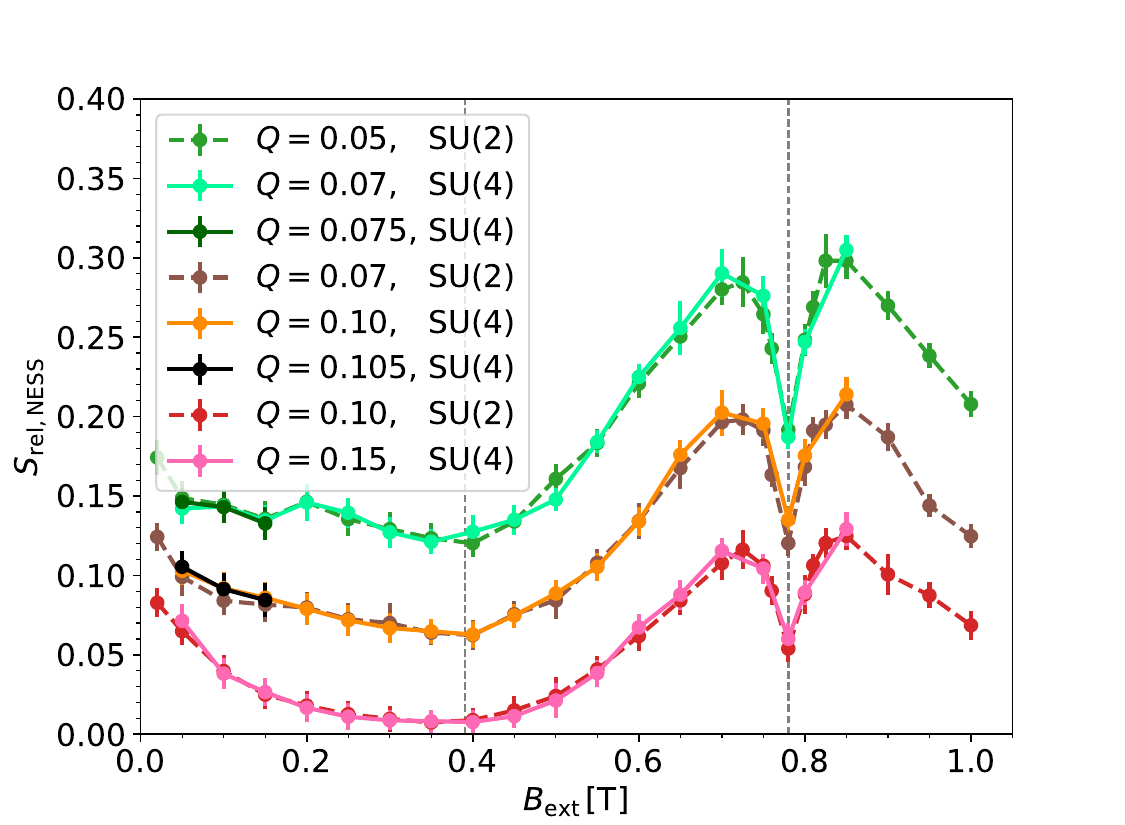}
    \caption[Comparison of the SU(2) TWA and the SU(4) TWA calculation of the magnetic field dependence of the 
		revival amplitude for three different values for $Q$]
    {Comparison of the SU(2) TWA and the SU(4) TWA calculation of the magnetic field dependence of 
		the saturated revival amplitude for three different values for $Q$.
		The values $I = 3/2$, $T_{\mathrm{rep}} = 13.2 \, \mathrm{ns}$, $A_{\mathrm{Q}} = 1.19$, $N=100$, 
		$z=80$ and $\gamma=0.01$ are used; each curve is averaged over $10^4$ initial conditions.
   The lines are guides to the eye.}
    \label{qi_komplsu4}
\end{figure}

We start with a comparison between the results from SU(2) TWA 
and from SU(4) TWA as displayed in Fig.\  \ref{qi_komplsu4}.
The upper two curves are shifted vertically by $0.05$ and $0.1$, respectively, to allow 
for better comparison.
We include the factor of 1.5 in the value of $Q$ which was found in the spin noise results
in Sect.\ \ref{ss:spinnoise_simple} to lead to agreement between the 
results of SU(2) and SU(4) calculations. It is considered here up to the second digit.
We see that even for the revival amplitudes resulting from long trains of periodic pulses
the SU(2) and SU(4) data for the rescaled $Q$ values agree very well.
Hence,  we make use of this fact and perform the following calculations by the faster SU(2) TWA. 
This greatly speeds up the simulations and thus makes it possible to investigate parameters that are closer to 
the experimental ones. However, one should stay aware of the factor of 1.5 between 
$Q$ values for the two TWA approaches.

\begin{figure}[htb]
    \centering
    \includegraphics[width=\columnwidth]{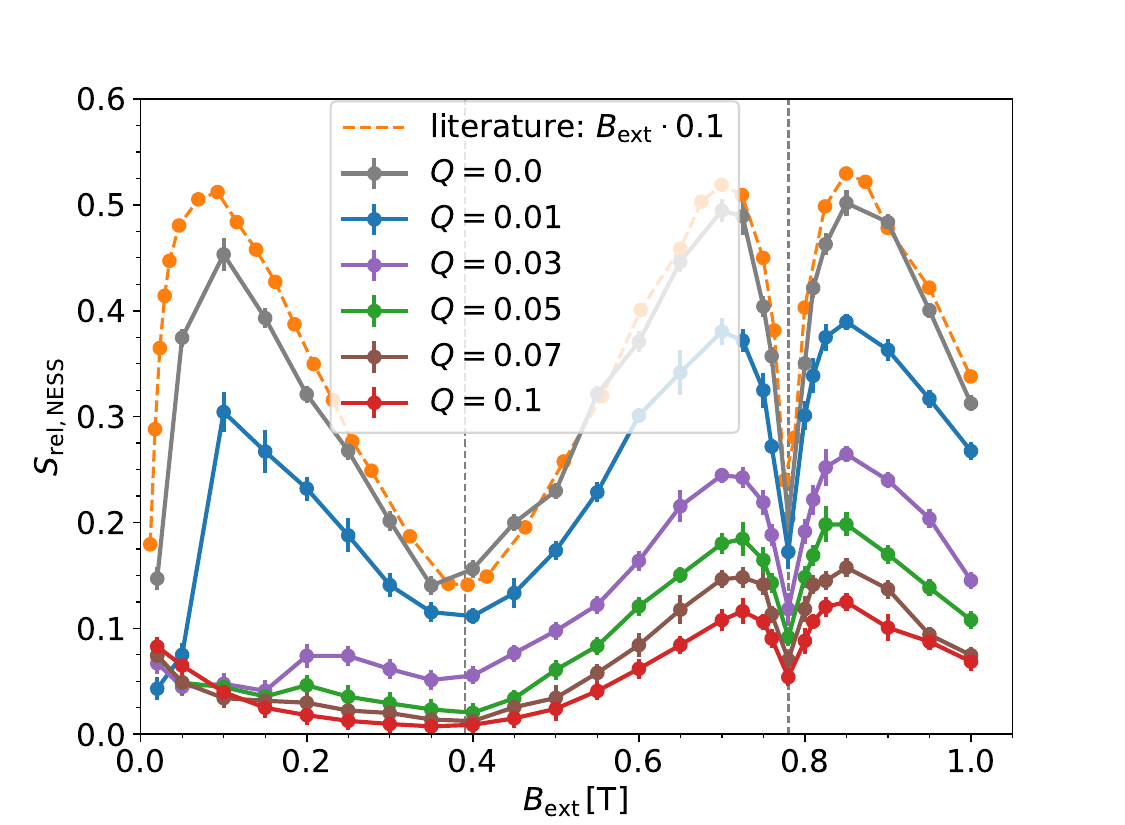}
    \caption[Magnetic field dependence of the revival amplitude in the NESS regime 
		for different values of $Q$ computed by SU(2) TWA]
    {Magnetic field dependence of the revival amplitude in the NESS regime for different values of $Q$ computed 
		by SU(2) TWA.  The parameters used are the same as in Fig.\  \ref{qi_komplsu4}. 
		The results from Ref.\  \cite{klein18} (`literature') with $z=800$ are included with 
     magnetic fields re-scaled by the factor $0.1$ because $z=80$ is used in the present simulations. 
		The lines are guides to the eye.}
    \label{qi_komplsu2}
\end{figure}

In Fig.\  \ref{qi_komplsu2}, curves for more values of $Q$ from SU(2) TWA are depicted.
In all curves the two characteristic minima can be found at about $0.39 \, \mathrm{T}$ (broad
minimum) and  $0.78 \, \mathrm{T}$ (sharp dip).

Having established the justifications for our approach, we turn to the
main point of the present study. What is the effect of the quadrupolar 
interaction on the revival amplitude, i.e., on the shape of the revival
amplitude as function of the magnetic field?

\begin{figure}[htb]
    \centering
    \includegraphics[width=\columnwidth]{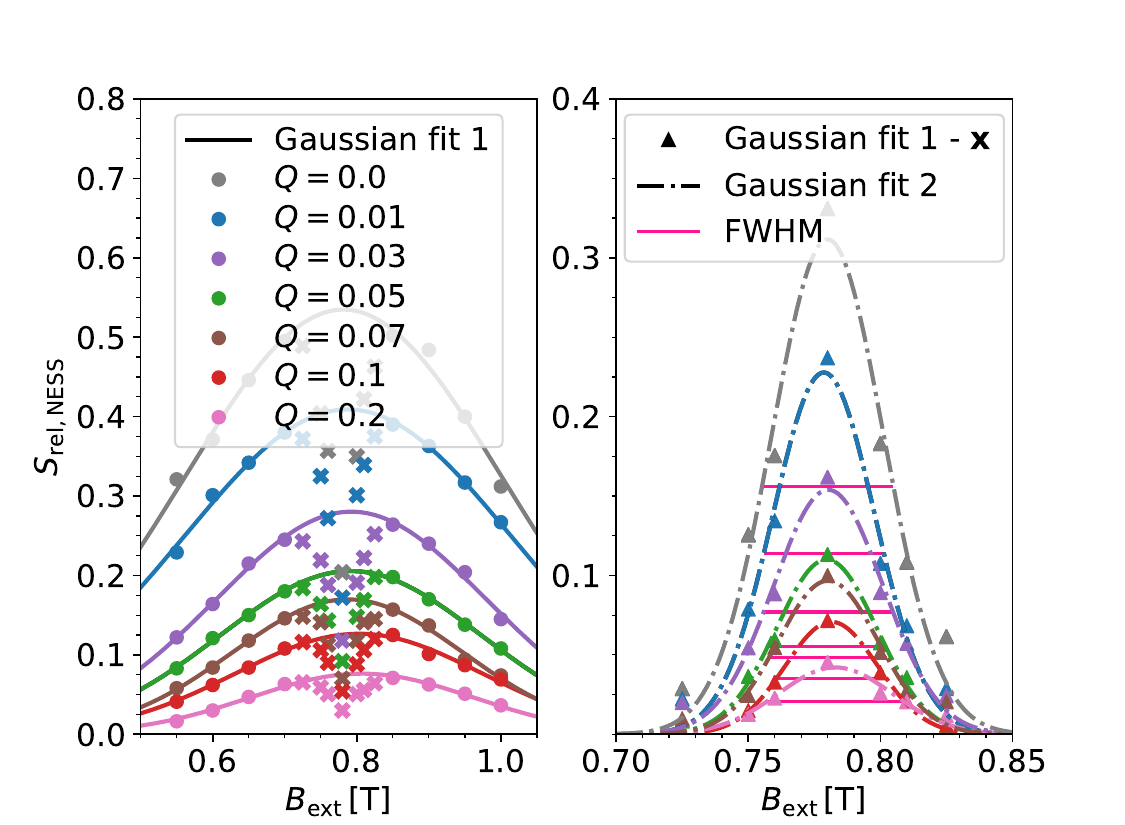}
    \caption[Gaussian fits for characterizing the dip at $0.78 \, \mathrm{T}$]
        {Gaussian fits for characterizing the dip at $0.78 \, \mathrm{T}$ based on the data from 
				Fig.\  \ref{qi_komplsu2}. In the left panel, the circles are fitted by Gaussians. 
        The cross-shaped symbols are subtracted from this fit and the resulting values 
				are plotted as triangles in the right panel. Then, these triangles are fitted  
				by Gaussians. The pink horizontal lines in the right panel indicate the ensuing FWHM.}
    \label{gaussfits}
\end{figure}

For increasing values of $Q$ the dip at $0.78 \, \mathrm{T}$ becomes smaller, but
simultaneously the curve $S_\text{rel, NESS}(B)$ becomes flatter in total.
In order to investigate this behavior in greater detail, we want to analyze the shape
of the dip quantitatively. A peak would be easier to describe, namely by
its position, its height and its width, e.g., its full width at half maximum (FWHM).
Thus, we interpret the dip as negative peak. For its analysis, we proceed in two steps.
First, we fit the large uprise between $0.55 \, \mathrm{T}$ to $0.7 \, \mathrm{T}$ and between 
 $0.85 \, \mathrm{T}$ to $1 \, \mathrm{T}$ 
by a Gaussian function in order to determine its shape if there were \emph{no} dip.
Such fits are shown in the left panel of Fig.\  \ref{gaussfits}.
Second, the data points between $0.7 \, \mathrm{T}$ and $0.85 \, \mathrm{T}$ are subtracted from the 
obtained Gaussian fits. The resulting curves are peaks by construction shown in the right panel
of Fig.\  \ref{gaussfits}. So we analyze them as peaks gain by fitting Gaussian to them
as visualized in the right panel of Fig.\  \ref{gaussfits}.

The Gaussian fits resulting from the second step provide height (more precisely depth) and  width 
of the dips. The data from Fig.\  \ref{qi_komplsu2} is used for the fits together with additional 
data at $Q=0.2$.
This procedure is repeated for other sets of parameters as well. In particular, 
other scalings of the the $z$-factor and the corresponding external magnetic field
are studied in order to approach the values relevant for experiment, see the
data listed  in Tab.\  \ref{tab:skalierungen}.

\begin{table}[htb]
    \centering
    \begin{tabular}{ c c c c c  }
        \hline
         $\sqrt{\lambda} \,$ & Titling in plot & $N_{\mathrm{eff}}$ & $z$ & $B_{\mathrm{min, scaled}}$\\ 
         \hline
         10  & f10 & 200  & 80  & $0.78  \, \mathrm{T}$ \\
         5   & f5  & 800  & 160 & $1.56  \, \mathrm{T}$ \\
         3.5 & f3  & 1600 & 226 & $2.21 \, \mathrm{T}$ \\
         \hline  
    \end{tabular}
    \vspace{0.5em}
    \caption[]{Different scalings of the real parameters including the magnetic field,
		at which the sharp dip occurs, leading
		to the data shown in Figs.\ \ref{qi_bpeaks}, \ref{qi_hpeaks}, and 
		\ref{qi_kompletthpeak}. For a smaller scaling factor $\sqrt{\lambda}$, the values of the parameters are closer to the
		experimental values.}
    \label{tab:skalierungen}
\end{table}

\begin{figure}[htb]
    \centering
    \includegraphics[width=\columnwidth]{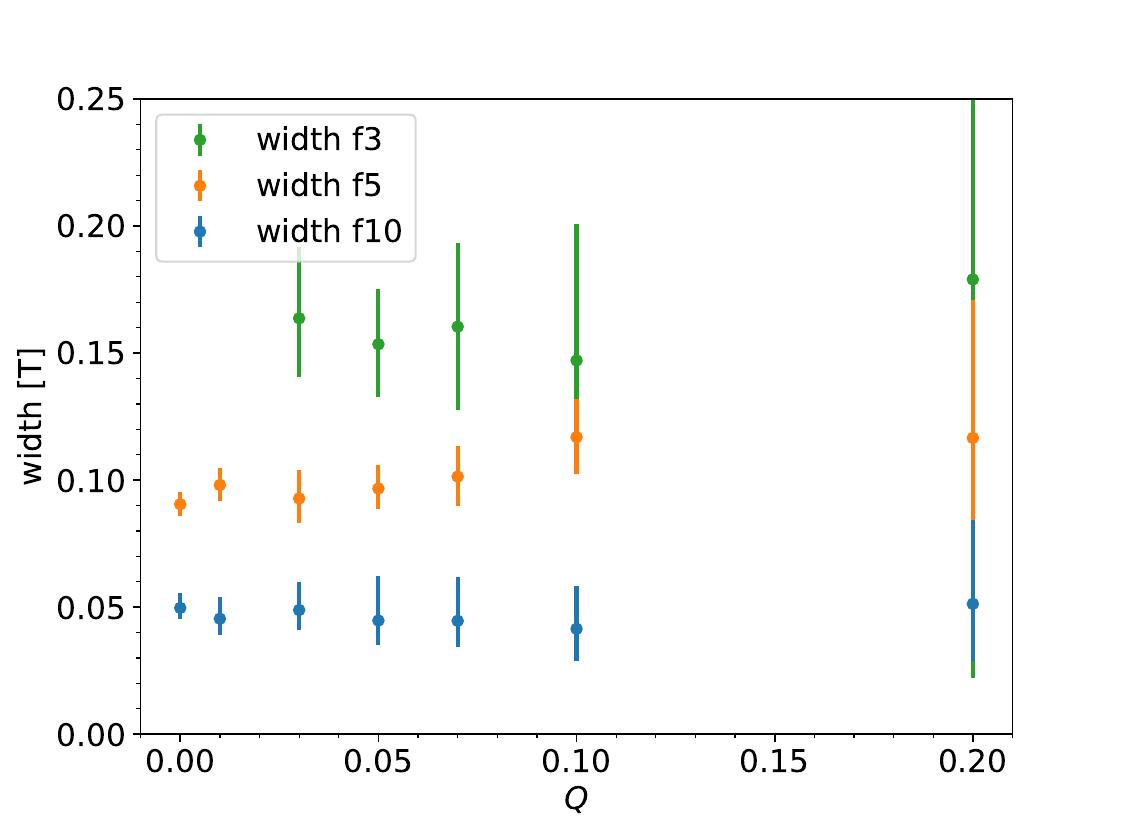}
    \caption[Width of the sharp dip as a function of $Q$ for various scalings]
		{Width of the sharp dip as a function of $Q$ for various scalings.
		The values of the parameters are listed in Tab.\  \ref{tab:skalierungen}. 
		The other parameters are those used for Fig.\  \ref{qi_komplsu4}. }
    \label{qi_bpeaks}
\end{figure}

The results for the width of the dips, determined as FWHM as in the right panel of Fig.\ \ref{gaussfits},
are displayed in Fig.\ \ref{qi_bpeaks} as function of $Q$ for various scaling factors. 
For the smallest scaling factor $3.5$ the lower values of $Q$ could not be simulated within reasonable time.
For all scaling factors we observe an almost constant width of the dip, i.e., its width
does not depend significantly on the strength of the quadrupolar interaction. 
If the dip were washed out by the quadrupolar interaction 
one would have expected an increase of the width upon increasing
$Q$. Thus, the constancy of the width indicates that the quadrupolar interaction 
does not smear out the dip. 
In addition, we observe that the width increases if the scaling factor is approaching
unity, i.e., in the limit that the $z$-factor and thus the magnetic field is
approaching the experimental values.

\begin{figure}[htb]
    \centering
    \includegraphics[width=\columnwidth]{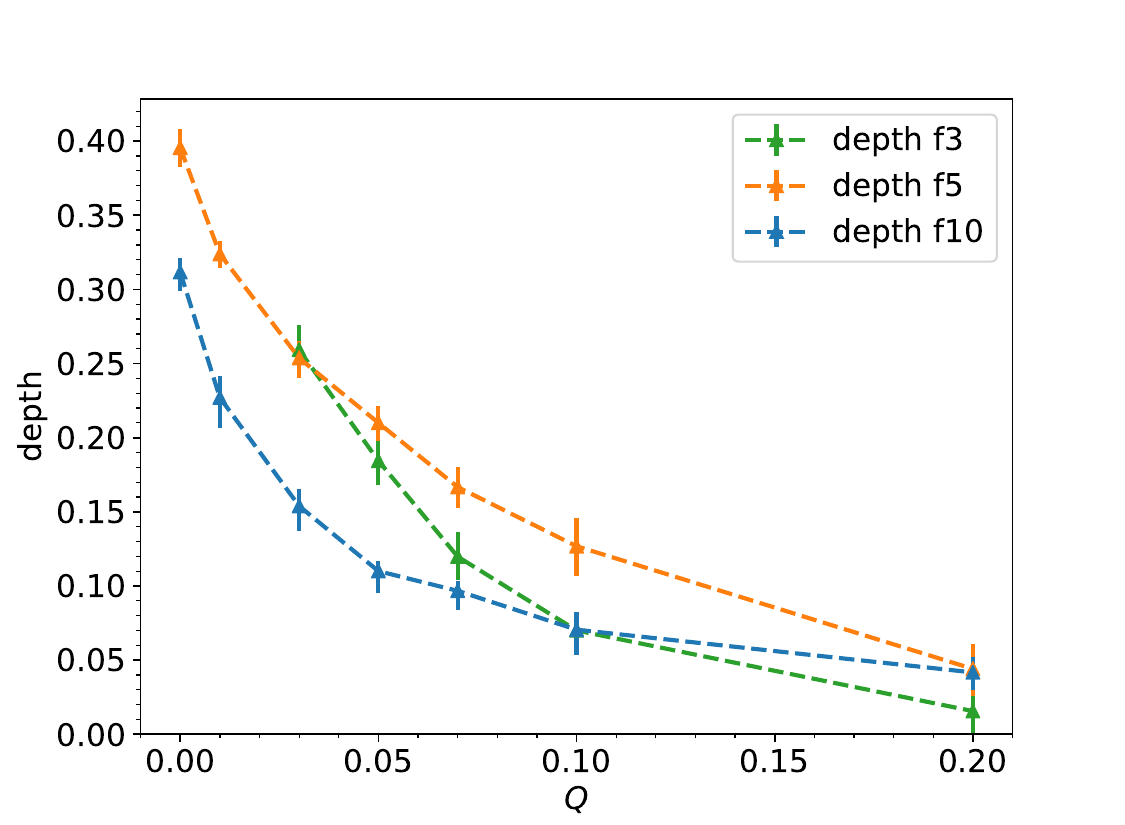}
    \caption[Depth of the sharp dip as a function of $Q$ for various scaling factors]
		{Depth of the sharp dip as a function of $Q$. The values of the parameters for the various
		scaling factors are listed in Tab.\  \ref{tab:skalierungen}. 
		The other parameters are those used for Fig.\  \ref{qi_komplsu2}. The lines are guides to the eye.}
    \label{qi_hpeaks}
\end{figure}

Next, we study the depth of the dip as determined by the procedure described above as the height of the peaks
shown in the right panel of Fig.\ \ref{gaussfits}. The results are shown in Fig.\ \ref{qi_hpeaks}.
They show a clear trend of decreasing depth of the dips as the quadrupolar interaction $Q$ is 
increased. This trend is particularly prominent for small scaling factors, i.e., it
is certainly relevant for experiment. Unfortunately, the runtime is a major difficulty here, since it 
 already took several weeks to obtain data for all values of f3.
The finding of decreasing depth agrees with the observation in the curves in Fig.\ \ref{qi_komplsu2}
that the curves of the revival signal  become
flatter. In Fig.\  \ref{qi_komplsu2}, we observe that not only the depth of the dips decreases 
but also the height of the total curve.

\begin{figure}[htb]
    \centering
    \includegraphics[width=\columnwidth]{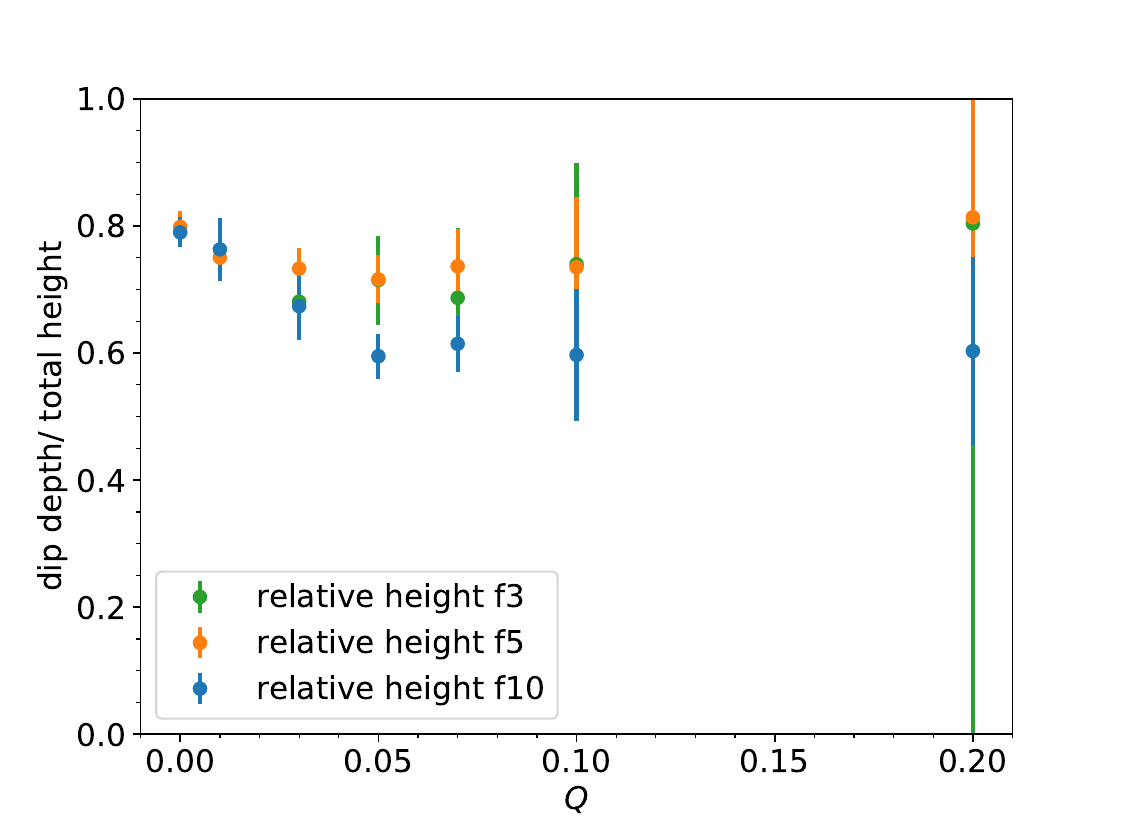}
    \caption[Relative depth of the sharp dip as function of $Q$ for various scaling factors]
		{Relative depth of the sharp dip as function of $Q$ for various scaling factors; for the corresponding
		parameters see Tab.\  \ref{tab:skalierungen}.}
    \label{qi_kompletthpeak}
\end{figure}

We point out that in experiment no absolute values of the revival amplitudes is measured in experiment.
The experimental data is provided in arbitrary units. This implies that all statements on shape
can only be made based on relative quantities. 
Therefore, we analyze the relative depth of the dips, i.e., the depth relative to the
height of the total curve. The latter is determined from the Gaussian fits
in the first step as illustrated in the left panel of Fig.\ \ref{gaussfits} and the 
smallest value of the broad minimum of the respective curve.
The resulting data is depicted in Fig.\ \ref{qi_kompletthpeak} as 
function of $Q$. 
The error bars of the values are determined using the Gaussian propagation 
of uncertainty from the errors of the individual variables. 

There are two striking features of the relative depths. First, they agree fairly well
for the shown scaling factors. This corroborates our approach to scale down the factor $z$
between the electronic and the nuclear magnetic moment in order to observe the relevant
resonance phenomena at much lower magnetic fields. Second, we observe that the relative depth
remains essentially constant as function of $Q$. Again, this agrees with the above remark
that the total curve becomes flatter with increasing $Q$ while keeping the relation between
minima and maxima constant.

One may wonder what this means for experiment.
On the one hand, the dip persists even for finite quadrupolar interaction. In arbitrary units of the revival
amplitude, it keeps its shape. On the other hand, the width of the dips stays constant in absolute terms
such that the dip may no longer appear as particularly sharp.

\subsection{Revival Amplitude in the Stochastic  Model (B)}
\label{ss:revampl_stochastic}

In a last step, we study the results for the stochastic model (B) 
for the distribution of quadrupolar interactions. In this case, the $q_k$ are defined differently 
than in the simple model (A) to be consistent with previous definitions \cite{hackm15, fisch22}.
As before in Sect.\ \ref{s:spinnoise} for the spin noise, a factor of two results in comparison to the definition 
in the simple model (A).

\begin{figure}[htb]
    \centering
    \includegraphics[width=\columnwidth]{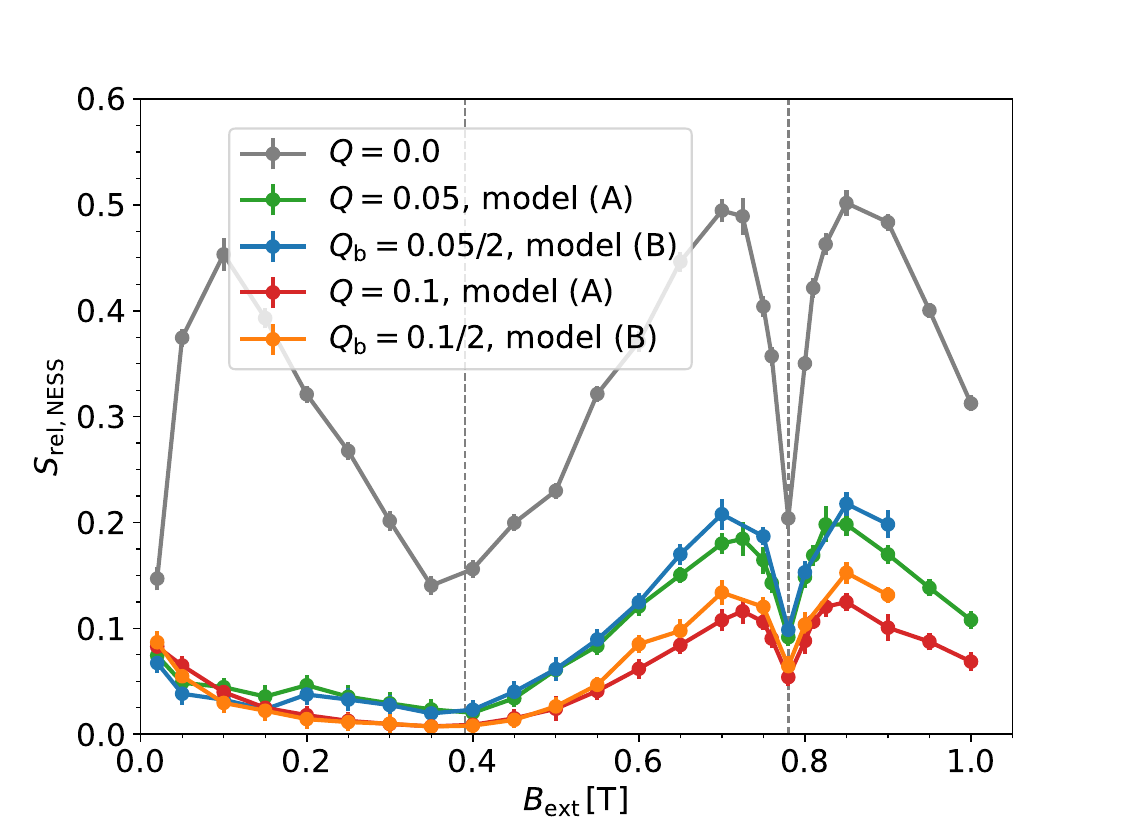}
    \caption[Comparison of the revival amplitude in model (A) and model (B) as function
		of the magnetic field for three values of $Q$.]
          {Comparison of the saturated revival amplitude in model (A) and model (B) as function
		of the magnetic field for three values of $Q$. The data is computed in SU(2) TWA. The curve for $Q=0=Q_\text{b}$ 		
    is also plotted for comparison. The parameters used correspond to those in Fig.\  \ref{qi_komplsu4}. 
		The lines are guides to the eye. }
    \label{qi_bulutaySrev}
\end{figure}

The magnetic field dependence of the saturation value is shown in Fig.\  \ref{qi_bulutaySrev} for two 
$Q$ values along with the results from the simple model using  SU(2) TWA in both cases. The scaling
factor is 10. Very similar behavior is seen. 
For a magnetic field from about $0.2 \, \mathrm{T}$ to $0.55 \, \mathrm{T}$ the data coincide very precisely. 
For smaller magnetic fields, the revival signal is slightly lower for model (B) while for larger magnetic fields, 
it is slightly higher than for model (A). The relevant behavior  around the sharp dip is very similar.
The slightly higher revival amplitudes indicate a depth of the  dip which is a bit larger, but
with essentially the same value relative to the total height of the figure at about $0.8$T.
The two model for $V$ differ greatly so that the agreement of the data is remarkable.

We conclude that the precise model for the distribution of the quadrupolar interactions does not
matter. This was to be expected in view of the analogous results for spin noise. The 
two models yielded very much the same result if the $Q$ values were properly scaled.
Thus, we are able to transfer the conclusions obtained for model (A) to model (B) which is
much closer to the experimental situation.
 
    \section{Conclusions}
    \label{s:conclusion}

		In previous studies of the electronic spin dynamics in quantum dots subjected to long trains
		of periodic pulses 	it could be established that the commensurability of the Larmor precession
		of the electronic spin and of the nuclear spins with the repetition time between two consecutive
		times plays an essential role. Nuclei induced frequency focussing leads to minima and maxima
		in the revival amplitude of the electronic spin orientation just before the next pulse arrives.
		If the number of nuclear precessions is half-integer a very broad minimum appears while an integer
		number of nuclear precession yield very sharp dips. If several nuclear isotopes are present several
		sharp dips appear which can be used in a kind of NMR spectroscopy
		to examine the sample for its isotope content \cite{scher21a}. But no such dips have so far been measured in experiment \cite{klein18}. Thus, the objective of this study was to investigate whether a certain
		additional coupling, so far neglected in the theoretical simulations 
		of large spin baths beyond exact diagonalization \cite{hackm15,glase16}, can explain the difference
		between experiment and theory. The next important interaction after the hyperfine interaction
		and the electronic and nuclear Zeeman terms is the quadrupolar interaction $Q$ of the nuclei since
		it directly influences the nuclear spin dynamics and since it breaks the spin rotation symmetry.
		
		In order to be able to simulate sizable nuclear spin baths we use a semi-classical truncated Wigner
		approximation (TWA) to simulate spin baths up to 200 nuclear spins with $I=3/2$. 
		This bath size is still fairly small compared to the experimentally relevant bath sizes. Thus
		we re-scaled the hyperfine couplings to keep the experimentally relevant 
		time scales. In addition, we re-scaled the nuclear magnetic moment in order to observe
		the relevant dips by lower magnetic fields because the simulation time increases cubically
		with the magnetic field.
		
		First, we showed that the
		local quadrupolar dynamics can be captured exactly by expressing the Hamiltonian as linear superposition
		of an extended set of operators. For $I=3/2$, an SU(4) TWA is needed. But we could show that for the issues
		of interest here we can stick to an SU(2) TWA if the size of the quadrupolar interactions are re-scaled
		by a factor $1.5$. Also, we studied two assumptions on the quadrupolar interaction.
		Model (A) assumes a very simple quadratic potential while model (B) draws random values
		for the quadrupolar interactions from the distributions computed from first principles \cite{bulut12,bulut14}.
		We find, however, that both models yield essentially the same results for spin noise and revival amplitudes
		if the proper definitions are taken into account.  (The spin noise from model (A) in $z$-direction 
		is not generic due to an exact symmetry, but the transversal spin noise is.)
		
		Based on the above observations, we  studied the influence of quadrupolar interactions
		on the sharp dips. We found that the width of the sharp dips does not depend on the value of
		$Q$ in the range of physically plausible values for the quadrupolar interaction. But the overall
		height of the feature around the sharp dips depends on $Q$. For increasing $Q$ the curves become
		flatter so that the height of the curve and the depth of the dip decrease. The relative depth, however, i.e., 
		the depth relative to the total height, does not alter upon changing $Q$.
		In this sense, the dips are not washed out and should be discernible in experiment. 
		
		The question arises why the experiment does not display the sharp dips:
		(i) One reason can be that one has not searched precisely for these dips. If the magnetic field is
		changed in large steps it is possible that the dips are missed.
		(ii) Another reason can be that the next-leading interaction is responsible. There
		is a dipole-dipole interaction between the nuclear spins which has been neglected so far because
		it is weaker by one to two orders of magnetitude \cite{merku02,schli03}. But for the saturation of
		the revival signal it can still be relevant. It can be included by spinDMFT \cite{grass21}.
		(iii) The laser pulse orients the electronic spin via the intermediate excitation of
		a trion. The laser pulse is often taken to be infinitesimally short. But this is also an approximation
		and it was shown \cite{klein18} that pulses of finite duration are less effective for
		large magnetic fields. So this can also be a factor for the absence of sharp dips in experiment.
		We stress, however, that the use of the intermediate trion as such does not lead 
		to the absence of sharp dips. This effect has already been considered when simulating
		NMR spectroscopy \cite{scher21a}.
		(iv) We cannot completely exclude that the down-scaling of the bath sizes for the
		simulations introduces some artefacts.
		(v) Any background signal added to  the revival amplitude can make it hard to 
		observe the actual revival amplitudes if these are not very pronounced.
		
		Summarizing, we showed that the quadrupolar interaction has a flattening effect on the
		magnetic field dependence of revival amplitude. But it does not influence the 
		relative depth and thus it does not provide a conclusive explanation for the absence of
		the sharp dips of NMR spectroscopy in experiment, at least not as sole effect.
		Further research in theory and in experiment are certainly called for to resolve this issue.

    \begin{acknowledgments} 
      We thank Dr.~Philipp Schering for providing program code of his work 
			and for helpful support and discussions.  
      Furthermore, we thank Dr.~Carsten Nase for his continued technical support.
    \end{acknowledgments}

%
		
    \begin{appendix} 
		 \section{Equations of motion for $\hat{H}_{\mathrm{quad}}$ with SU(2) TWA and SU(3) TWA}
      \label{a:eomHquad} 
		
      The equations of motion for the Hamiltonian 
      \begin{equation}
          \hat{H}_{\mathrm{quad}} = q \left[ \left( 3\hat{I}_z^2 - \pmb{I}^2 \right) 
					+ \eta \left( \hat{I}_x^2 - \hat{I}_y^2 \right) \right] + \pmb{h} \cdot \hat{\pmb{I}} ,
      \end{equation}
      which is investigated in Fig.\  \ref{SU3_Hquad}, are given by 
      \begin{equation}
          \frac{\mathrm{d}}{\mathrm{d}t} \pmb{I} = \left[ \, p_{\mathrm{f}} \left(\begin{array}{c} 
					2q (\eta -1) I_x \\ 2q (-\eta -1) I_y \\ 4q I_z \end{array}\right) + \frac{h}{\sqrt{3}} 
					\left(\begin{array}{c} 1 \\ 1 \\ 1 \end{array}\right) \right] \times \pmb{I} 
      \end{equation}
      using the SU(2) TWA, where $p_{\mathrm{f}} =1$ for the version without
			additional prefactor and $p_{\mathrm{f}} = 1/2$ when the 
      prefactor derived in Ref.\  \cite{fisch22} is included.

      For the SU(3) TWA, the matrices and values for $f_{\alpha \beta \gamma}$ from Ref.\  \cite{david15} are used.
      This results in the Hamiltonian
      \begin{equation}
      \begin{split}
          H_{\mathrm{W}}^{\mathrm{SU(3)}} = q \Biggl[ &(\eta -1) 
					\left( \frac{1}{3} \left( 2 \mathbb{1} + \frac{\sqrt{3}}{2} X_8 \right) + \frac{1}{2} X_4 \right) 
					\\\ -
                                      &(\eta +1) \left( \frac{1}{3} 
																			\left( 2 \mathbb{1} + \frac{\sqrt{3}}{2} X_8 \right) - \frac{1}{2} X_4 \right) 					\\\ +             &2 \left( \frac{1}{3} 
																			\left( 2 \mathbb{1} - \sqrt{3} X_8 \right) \right) \Biggr]
																			\\\ + &\frac{h}{\sqrt{3}}(X_1+X_2+X_3) .
      \end{split}
      \end{equation}
      With Eq.\  \eqref{Bewgl2}  the equations of motion follow
      \begingroup
        \flausr
        \begin{subequations}\label{bewglen_su4_david}
        \begin{equation}
          \frac{\partial}{\partial t} X_1 = \frac{h}{\sqrt{3}} (X_3- X_2) + q X_7 (\eta + 3)  , \\
        \end{equation}
        \begin{equation}
          \frac{\partial}{\partial t} X_2 = \frac{h}{\sqrt{3}} (X_1- X_3) + q X_6 (\eta - 3) , \\
        \end{equation}
        \begin{equation}  
          \frac{\partial}{\partial t} X_3 = \frac{h}{\sqrt{3}} (X_2- X_1) + q X_5 2 \eta , 
        \end{equation}
        \begin{equation}
          \frac{\partial}{\partial t} X_4 = \frac{h}{\sqrt{3}} (-2 X_5 - X_6 - X_7) , \\
        \end{equation}
        \begin{equation}  
          \frac{\partial}{\partial t} X_5 = \frac{h}{\sqrt{3}} (2 X_4 + X_6 - X_7) - q X_3 2 \eta , \\
        \end{equation}
        \begin{equation}  
          \frac{\partial}{\partial t} X_6 = \frac{h}{\sqrt{3}} (X_4 - X_5 - X_7 + \sqrt{3} X_8) - q X_2  (-\eta + 3) , \\
        \end{equation}
        \begin{equation}  
          \frac{\partial}{\partial t} X_7 = \frac{h}{\sqrt{3}} (X_4 + X_5 + X_6 - \sqrt{3} X_8)- q X_1 (-\eta - 3) , \\
        \end{equation}
        \begin{equation}  
          \frac{\partial}{\partial t} X_8 = h X_7 - h X_6  .
        \end{equation}
        \end{subequations}
    \endgroup

\section{Equations of motion for $\hat{H}_{\mathrm{simpl}}$ with SU(2) TWA and SU(4) TWA}
      \label{a:eomHsimpl} 

      The equations of motion for the Hamiltonian $\hat{H}_{\mathrm{simpl}}$ 
			\eqref{H_david} are given for SU(2) TWA by
      \begin{equation}
          \frac{\mathrm{d} \pmb{I}}{\mathrm{d}t} = \left(\begin{array}{c} 0 \\ 0 \\ (p_{\mathrm{f}} I_z - 1) J \end{array}\right) \times \pmb{I} \, .
      \end{equation}
      The SU(4) TWA is explained in the Supplemental Material where  the corresponding 
      matrices and values for $f_{\alpha \beta \gamma}$ are given as well .
      The Hamiltonian for SU(4) TWA reads
      \begin{equation}
          H_{\mathrm{W}}^{\mathrm{SU(4)}} = \left( - \frac{1}{3} X_8 - \sqrt{\sfrac{4}{45}} \, X_{15} + \frac{5}{8} \, \mathbb{1} - X_3 \right) J \, .
      \end{equation}
      Substituting this and 
      \begin{subequations}
      \begin{align}
          \frac{\partial H_{\mathrm{W}}^{\mathrm{SU(4)}}}{\partial X_3} &= - 1 J \, , \\
          \frac{\partial H_{\mathrm{W}}^{\mathrm{SU(4)}}}{\partial X_8} &= - \frac{1}{3} J \, , \\
          \frac{\partial H_{\mathrm{W}}^{\mathrm{SU(4)}}}{\partial X_{15}} &= -\sqrt{\sfrac{4}{45}} J \, .
      \end{align}
      \end{subequations}
      into Eq.\  \eqref{Bewgl2} yields the 15 equations of motion of the SU(4) TWA
            \begingroup
        \flausr
      \begin{subequations}\label{bewglen_su4_david_app}
      \begin{equation}
      \begin{split}
          \frac{\partial}{\partial t} X_{1}  =& \, \biggl[ -1 X_2 (-1) -\sqrt{\sfrac{8}{3}} \, X_7  \left(-\sfrac{1}{3}\right) \\\ & -\sqrt{\sfrac{5}{3}} \, X_{14} \left(-\sfrac{1}{3}\right)  +\sqrt{\sfrac{10}{3}} \, X_7 \left( -\sqrt{\sfrac{4}{45}} \right) \\\ &-\sqrt{\sfrac{4}{3}} \, X_{14} \left( -\sqrt{\sfrac{4}{45}} \right) \biggr] J
      \end{split}
      \end{equation}
      \begin{equation}
      \begin{split}
          \frac{\partial}{\partial t} X_{2}  =& \, \biggl[ 1 X_1 (-1) +\sqrt{\sfrac{8}{3}} \, X_6 \left(-\sfrac{1}{3}\right) \\\ & +\sqrt{\sfrac{5}{3}} \, X_{13} \left(-\sfrac{1}{3}\right) -\sqrt{\sfrac{10}{3}} \, X_6 \left( -\sqrt{\sfrac{4}{45}} \right) \\\ &+\sqrt{\sfrac{4}{3}} \, X_{13} \left( -\sqrt{\sfrac{4}{45}} \right) \biggr] J
      \end{split}
      \end{equation}
      \begin{equation}
          \frac{\partial}{\partial t} X_{3}  = 0 J
      \end{equation}
      \begin{equation}
      \begin{split}
          \frac{\partial}{\partial t} X_{4}  =& \, \biggl[ -2 X_5 (-1) +1 X_5 \left(-\sfrac{1}{3}\right) \\\ &+\sqrt{5} X_5 \left( -\sqrt{\sfrac{4}{45}} \right) \biggr] J
      \end{split}
      \end{equation}
      \begin{equation}
      \begin{split}
          \frac{\partial}{\partial t} X_{5}  =& \, \biggl[ 2 X_4 (-1) -1 X_4 \left(-\sfrac{1}{3}\right) \\\ &-\sqrt{5} X_4 \left( -\sqrt{\sfrac{4}{45}} \right) \biggr] J
      \end{split}
      \end{equation}
      \begin{equation}
      \begin{split}
          \frac{\partial}{\partial t} X_{6}  =& \, \biggl[ -1 X_7 (-1) - \sfrac{2}{3} \, X_7 \left(-\sfrac{1}{3}\right) \\\ &+\sfrac{\sqrt{10}}{3} \, X_{14} \left(-\sfrac{1}{3}\right) -\sqrt{\sfrac{8}{3}} \, X_2 \left(-\sfrac{1}{3}\right) \\\ &+\sqrt{\sfrac{10}{3}} \, X_2 \left( -\sqrt{\sfrac{4}{45}} \right) +\sfrac{\sqrt{5}}{3} \, X_7 \\\ & \left( -\sqrt{\sfrac{4}{45}} \right) +\sfrac{\sqrt{8}}{3} \, X_{14} \left( -\sqrt{\sfrac{4}{45}} \right) \biggr] J
      \end{split}
      \end{equation}
      \begin{equation}
      \begin{split}
          \frac{\partial}{\partial t} X_{7}  =& \, \biggl[ 1 X_6 (-1) +\sqrt{\sfrac{8}{3}} \, X_1 \left(-\sfrac{1}{3}\right) \\\ &+ \sfrac{2}{3} \, X_6 \left(-\sfrac{1}{3}\right) -\sfrac{\sqrt{10}}{3} \, X_{13} \left(-\sfrac{1}{3}\right) \\\ &-\sqrt{\sfrac{10}{3}} \, X_1 \left( -\sqrt{\sfrac{4}{45}} \right) -\sfrac{\sqrt{5}}{3} \, X_6 \\\ & \left( -\sqrt{\sfrac{4}{45}} \right) -\sfrac{\sqrt{8}}{3} \, X_{13} \left( -\sqrt{\sfrac{4}{45}} \right) \biggr] J
      \end{split}
      \end{equation}
      \begin{equation}
      \begin{split}
          \frac{\partial}{\partial t} X_{8}  = 0 J
      \end{split}
      \end{equation}
      \begin{equation}
      \begin{split}
          \frac{\partial}{\partial t} X_{9}  =& \, \biggl[ -3 X_{10} (-1) + \sfrac{2}{3} \, X_{10} \left(-\sfrac{1}{3}\right) \\\ &-\sfrac{\sqrt{5}}{3} \, X_{10} \left( -\sqrt{\sfrac{4}{45}} \right) \biggr] J
      \end{split}
      \end{equation}
      \begin{equation}
      \begin{split}
          \frac{\partial}{\partial t} X_{10} =& \, \biggl[ 3 X_9 (-1) - \sfrac{2}{3} \, X_9 \left(-\sfrac{1}{3}\right) \\\ &+\sfrac{\sqrt{5}}{3} \, X_9 \left( -\sqrt{\sfrac{4}{45}} \right)  \biggr] J
      \end{split}
      \end{equation}
      \begin{equation}
      \begin{split}
          \frac{\partial}{\partial t} X_{11} =& \, \biggl[ -2 X_{12} (-1) -\sfrac{7}{3} \, X_{12} \left(-\sfrac{1}{3}\right) \\\ &-\sfrac{\sqrt{5}}{3} \, X_{12} \left( -\sqrt{\sfrac{4}{45}} \right)  \biggr] J
      \end{split}
      \end{equation}
      \begin{equation}
      \begin{split}
          \frac{\partial}{\partial t} X_{12} =& \, \biggl[ 2 X_{11} (-1) +\sfrac{7}{3} \, X_{11} \left(-\sfrac{1}{3}\right) \\\ &+\sfrac{\sqrt{5}}{3} \, X_{11} \left( -\sqrt{\sfrac{4}{45}} \right)  \biggr] J
      \end{split}
      \end{equation}
      \begin{equation}
      \begin{split}
          \frac{\partial}{\partial t} X_{13} =& \, \biggl[ -1 X_{14} (-1) + \sfrac{\sqrt{10}}{3} \, X_7 \left(-\sfrac{1}{3}\right)  \\\ &-\sqrt{\sfrac{5}{3}} \, X_2 \left(-\sfrac{1}{3}\right) +\sfrac{4}{3} \, X_{14} \left(-\sfrac{1}{3}\right) \\\ &-\sqrt{\sfrac{4}{3}} \, X_2 \left( -\sqrt{\sfrac{4}{45}} \right)  +\sfrac{\sqrt{8}}{3} \, X_7 \\\ & \left( -\sqrt{\sfrac{4}{45}} \right) -\sfrac{\sqrt{20}}{3} \, X_{14} \left( -\sqrt{\sfrac{4}{45}} \right)  \biggr] J
      \end{split}
      \end{equation}
      \begin{equation}
      \begin{split}
          \frac{\partial}{\partial t} X_{14} =& \, \biggl[ 1 X_{13} (-1) +\sqrt{\sfrac{5}{3}} \, X_1 \left(-\sfrac{1}{3}\right) \\\ &-\sfrac{\sqrt{10}}{3} \, X_6 \left(-\sfrac{1}{3}\right) -\sfrac{4}{3} \, X_{13} \left(-\sfrac{1}{3}\right) \\\ &+\sqrt{\sfrac{4}{3}} \, X_1 \left( -\sqrt{\sfrac{4}{45}} \right) -\sfrac{\sqrt{8}}{3} \, X_6 \\\ & \left( -\sqrt{\sfrac{4}{45}} \right) +\sfrac{\sqrt{20}}{3} \, X_{13} \left( -\sqrt{\sfrac{4}{45}} \right)  \biggr] J
      \end{split}
      \end{equation}
      \begin{equation}
      \begin{split}
          \frac{\partial}{\partial t} X_{15} = 0 J
      \end{split}
      \end{equation}
      \end{subequations}
      \endgroup
      
      For the initial conditions, random numbers are drawn from Gaussian distributions.
      The mean and the variance of this Gaussian distribution is given by $\mu = 0$ and $\sigma^2 = 5/4$
      for the calculation of the correlation $\langle \hat{I}^x(t) \hat{I}^x(0) \rangle$.
      To calculate the expectation value $\langle \hat{I}^x(t) \rangle$, we draw random numbers from 
      a multivariate Gaussian distribution with mean values and covariance matrix determined from the equations
			\begin{subequations}
      \begin{align}
          \mu_i &= \langle s | \hat{X}_i | s \rangle \, ,\label{ew_Mittelwerte} \\ 
          \Sigma_{ij} 
					&= \frac{1}{2} \langle s | (\hat{X}_i \hat{X}_j + \hat{X}_j \hat{X}_i) 
					| s \rangle - \mu_i \mu_j \, , \label{ew_Kovarianzmatrix}
      \end{align} 
			\end{subequations}
      where $\hat{X}_i$ is one of the 15  matrices of SU(4) TWA.
\end{appendix}

    \setcounter{table}{0}
    \setcounter{section}{0}
    \setcounter{equation}{0}
    \counterwithout*{equation}{section}
    \title{Supplementary Material to the paper Influence of quadrupolar interaction on NMR spectroscopy}

    \author{Alina Joch}
    \email{alina.joch@tu-dortmund.de}
    \affiliation{Condensed Matter Theory, 
    TU Dortmund University, Otto-Hahn-Stra\ss{}e 4, 44221 Dortmund, Germany}
    \affiliation{DLR, Germany}

    \author{G\"otz S.\ Uhrig}
    \email{goetz.uhrig@tu-dortmund.de}
    \affiliation{Condensed Matter Theory, 
    TU Dortmund University, Otto-Hahn-Stra\ss{}e 4, 44221 Dortmund, Germany}

    \date{\textrm{\today}}

    \maketitle

    \section{SU(4) TWA}
    \label{s:su4twa}
		
    In order to build the Hamiltonian for the SU(4) truncated Wigner approximation (TWA), the 15 matrices of this basis must be set up analogously to the matrices in Ref.\@ \cite{david15}. 
    For this, the Gellman matrices \cite{sbaih13} are used and transformed with a unitary transformation in such a way that the first three matrices correspond to the spin matrices. 
    The unitary transformation is shown here in three partial transformations for simplicity:
    \begin{subequations}\label{unitaeretrafo}
    \begin{equation}
        \sqrt{\sfrac{5}{2}} \cdot \left( \begin{array}{c} \sqrt{\sfrac{3}{10}} \\ -\sqrt{\sfrac{1}{5}} \\ -\sqrt{\sfrac{1}{2}} \end{array}
               \begin{array}{c} \sqrt{\sfrac{2}{5}}  \\ \sqrt{\sfrac{3}{5}}  \\ 0                    \end{array}
               \begin{array}{c}\sqrt{\sfrac{3}{10}}  \\ -\sqrt{\sfrac{1}{5}} \\ \sqrt{\sfrac{1}{2}}  \end{array} \right) \cdot 
        \left( \begin{array}{c} \lambda_1 \\ \lambda_6 \\ \lambda_{13} \end{array}  \right) =
        \left( \begin{array}{c} X_1 \\ X_6 \\ X_{13} \end{array}  \right)
    \end{equation}
    and
    \begin{equation}
        \sqrt{\sfrac{5}{2}} \cdot \left( \begin{array}{c} \sqrt{\sfrac{3}{10}} \\ -\sqrt{\sfrac{1}{5}} \\ -\sqrt{\sfrac{1}{2}} \end{array}
               \begin{array}{c} \sqrt{\sfrac{2}{5}}  \\ \sqrt{\sfrac{3}{5}}  \\ 0                    \end{array}
               \begin{array}{c}\sqrt{\sfrac{3}{10}}  \\ -\sqrt{\sfrac{1}{5}} \\ \sqrt{\sfrac{1}{2}}  \end{array} \right) \cdot 
        \left( \begin{array}{c} \lambda_2 \\ \lambda_7 \\ \lambda_{14} \end{array}  \right) =
        \left( \begin{array}{c} X_2 \\ X_7 \\ X_{14} \end{array}  \right)
    \end{equation}
    and
    \begin{equation}
        \sqrt{\sfrac{5}{2}} \cdot \left( \begin{array}{c} \sqrt{\sfrac{1}{10}} \\ -\sqrt{\sfrac{9}{10}} \\ 0                    \end{array}
               \begin{array}{c} \sqrt{\sfrac{3}{10}} \\ \sqrt{\sfrac{3}{90}}  \\ -\sqrt{\sfrac{6}{9}} \end{array}
               \begin{array}{c} \sqrt{\sfrac{6}{10}} \\ \sqrt{\sfrac{6}{90}}  \\ \sqrt{\sfrac{3}{9}}  \end{array} \right) \cdot 
        \left( \begin{array}{c} \lambda_3 \\ \lambda_8 \\ \lambda_{15} \end{array}  \right) =
        \left( \begin{array}{c} X_3 \\ X_8 \\ X_{15} \end{array}  \right) \, .
    \end{equation}
    \end{subequations}
    All the remaining of the 15 matrices are multiplied by a factor $\sqrt{\sfrac{5}{2}}$ so that the trace of $X_{\alpha}^2$ is equal to five in all cases. 
    Apart from that, the remaining matrices are not changed further.
    With the matrices resulting after the transformation, see Tab.\@ \ref{tab:15matrizen}, then again analogous to Ref.\@ \cite{david15}, the terms $I_{\alpha}^2$ can be represented 
    as linear combinations of the 15 given matrices
    \begingroup
    \flausr
    \begin{subequations}\label{su4_Iquadrat}
    \begin{equation}
    \begin{split}
        I_x^2 = &\frac{1}{3} X_8 + \sqrt{\sfrac{4}{45}} \, X_{15} + \frac{5}{4}  \, \mathbb{1} + \sqrt{\sfrac{3}{10}} \, X_4 \\\ &+ \sqrt{\sfrac{3}{10}} \, X_{11} \, , 
    \end{split}
    \end{equation}
    \begin{equation}
    \begin{split}
        I_y^2 = &\frac{1}{3} X_8 + \sqrt{\sfrac{4}{45}} \, X_{15} + \frac{5}{4}  \, \mathbb{1} - \sqrt{\sfrac{3}{10}} \, X_4 \\\ &- \sqrt{\sfrac{3}{10}} \, X_{11} \, , 
    \end{split}
    \end{equation}
    \begin{equation}
    \begin{split}
        I_z^2 = &-\frac{2}{3} X_8 - \sqrt{\sfrac{16}{45}} \, X_{15} + \frac{5}{4} \, \mathbb{1} \, .
    \end{split}
    \end{equation}
    \end{subequations}
    \endgroup

    \begin{table}[htb]
        \centering
        \begin{tabular}{ c }
        \hline
        \addlinespace
        \vspace{0.5em}
            $
            \hspace{-0.5em}X_1 \hspace{-0.3em}= \hspace{-0.3em}\left(     \begin{array}{c} \hspace{-0.4em}0                    \\\hspace{-0.4em} \sqrt{\sfrac{3}{4}} \\\hspace{-0.4em} 0                  \\ \hspace{-0.4em} 0                   \hspace{-0.5em}\end{array}
                                                                        \begin{array}{c} \hspace{-0.4em}\sqrt{\sfrac{3}{4}}  \\\hspace{-0.4em} 0                   \\\hspace{-0.4em} 1                    \\ \hspace{-0.4em} 0                   \hspace{-0.5em}\end{array}
                                                                        \begin{array}{c} \hspace{-0.4em}0                    \\\hspace{-0.4em} 1                   \\\hspace{-0.4em} 0                    \\ \hspace{-0.4em} \sqrt{\sfrac{3}{4}} \hspace{-0.5em}\end{array}
                                                                        \begin{array}{c} \hspace{-0.4em}0                    \\\hspace{-0.4em} 0                   \\\hspace{-0.4em} \sqrt{\sfrac{3}{4}}  \\ \hspace{-0.4em} 0                   \hspace{-0.5em}\end{array} \right)  
            $
            , \, $ 
            \hspace{-0.5em}X_2 \hspace{-0.3em}= \hspace{-0.3em}\left(    \begin{array}{c} \hspace{-0.4em} 0                       \\\hspace{-0.4em} \sqrt{\sfrac{3}{4}}\, i \\\hspace{-0.4em}  0                      \\\hspace{-0.4em} 0                     \hspace{-0.5em} \end{array}
                    \begin{array}{c} \hspace{-0.4em}-\sqrt{\sfrac{3}{4}}\, i   \\\hspace{-0.4em}  0                    \\\hspace{-0.4em} 1 i                     \\\hspace{-0.4em} 0                     \hspace{-0.5em} \end{array}
                    \begin{array}{c} \hspace{-0.4em} 0                       \\\hspace{-0.4em} -1 i                   \\\hspace{-0.4em}  0                      \\\hspace{-0.4em} \sqrt{\sfrac{3}{4}}\, i \hspace{-0.5em} \end{array}
                    \begin{array}{c} \hspace{-0.4em} 0                       \\\hspace{-0.4em} 0                     \\\hspace{-0.4em} -\sqrt{\sfrac{3}{4}}\, i  \\ \hspace{-0.4em} 0                    \hspace{-0.5em} \end{array} \right) 
            $\\
            \vspace{0.5em}$
            \hspace{-0.5em}X_3 \hspace{-0.3em}= \hspace{-0.3em}\left(     \begin{array}{c} \hspace{-0.2em}\sfrac{3}{2}\\ 0            \\ 0            \\ 0             \hspace{-0.5em} \end{array}
                    \begin{array}{c} \hspace{-0.2em}0           \\ \sfrac{1}{2} \\ 0            \\ 0             \hspace{-0.5em} \end{array}
                    \begin{array}{c} \hspace{-0.2em}0           \\ 0            \\ -\sfrac{1}{2}\\ 0             \hspace{-0.5em} \end{array}
                    \begin{array}{c} \hspace{-0.2em}0           \\ 0            \\ 0            \\ -\sfrac{3}{2} \hspace{-0.5em} \end{array} \right)  
            $
            , \, $ 
            \hspace{-0.5em}X_4 \hspace{-0.3em}= \hspace{-0.3em}\left(    \begin{array}{c} \hspace{-0.4em} 0                   \\ 0        \\  \sqrt{\sfrac{5}{2}} \\  0        \hspace{-0.5em} \end{array}
                    \begin{array}{c} \hspace{-0.4em} 0                   \\ 0        \\  0                   \\  0        \hspace{-0.5em} \end{array}
                    \begin{array}{c} \hspace{-0.4em} \sqrt{\sfrac{5}{2}} \\ 0        \\  0                   \\  0        \hspace{-0.5em} \end{array}
                    \begin{array}{c} \hspace{-0.4em} 0                   \\ 0        \\  0                   \\  0        \hspace{-0.5em} \end{array} \right)   
            $\\
            \vspace{0.5em}$ 
            \hspace{-0.5em}X_5 \hspace{-0.3em}= \hspace{-0.3em}\left(     \begin{array}{c} \hspace{-0.4em}0                        \\ 0        \\ \sqrt{\sfrac{5}{2}}\, i  \\ 0  \hspace{-0.5em} \end{array}
                    \begin{array}{c} \hspace{-0.4em}0                        \\ 0        \\ 0                      \\ 0  \hspace{-0.5em} \end{array}
                    \begin{array}{c} \hspace{-0.4em}-\sqrt{\sfrac{5}{2}}\, i   \\ 0        \\ 0    \\ 0  \hspace{-0.5em} \end{array}
                    \begin{array}{c} \hspace{-0.4em}0                        \\ 0        \\ 0                      \\ 0  \hspace{-0.5em} \end{array} \right)   
            $
            , \, \vspace{0.5em}$
            \hspace{-0.5em}X_6 \hspace{-0.3em}= \hspace{-0.3em}\left(    \begin{array}{c} \hspace{-0.5em} 0                     \\ \hspace{-0.4em} -\sqrt{\sfrac{1}{2}}  \\ \hspace{-0.4em}  0                      \\ \hspace{-0.4em}  0                     \hspace{-0.5em} \end{array}
                    \begin{array}{c} \hspace{-0.5em} -\sqrt{\sfrac{1}{2}}  \\ \hspace{-0.4em} 0                     \\ \hspace{-0.4em}  \sqrt{\sfrac{3}{2}}    \\ \hspace{-0.4em}  0                     \hspace{-0.5em} \end{array}
                    \begin{array}{c} \hspace{-0.5em} 0                     \\ \hspace{-0.4em} \sqrt{\sfrac{3}{2}}   \\ \hspace{-0.4em}  0                      \\ \hspace{-0.4em}  -\sqrt{\sfrac{1}{2}}  \hspace{-0.5em} \end{array}
                    \begin{array}{c} \hspace{-0.5em} 0                     \\ \hspace{-0.4em} 0                     \\ \hspace{-0.4em}  -\sqrt{\sfrac{1}{2}}   \\ \hspace{-0.4em}  0                     \hspace{-0.5em} \end{array} \right)   
            $\\
            \vspace{0.5em}$
            \hspace{-0.5em}X_7 \hspace{-0.3em}= \hspace{-0.3em}\left(     \begin{array}{c} \hspace{-0.6em} 0                         \\\hspace{-0.5em} -\sqrt{\sfrac{1}{2}}\, i\\\hspace{-0.5em}  0                        \\\hspace{-0.5em}  0                       \hspace{-0.6em} \end{array}
                    \begin{array}{c} \hspace{-0.6em} \sqrt{\sfrac{1}{2}}\, i   \\\hspace{-0.5em} 0                       \\\hspace{-0.5em}  \sqrt{\sfrac{3}{2}}\, i  \\\hspace{-0.5em}  0                       \hspace{-0.6em} \end{array}
                    \begin{array}{c} \hspace{-0.6em} 0                         \\\hspace{-0.5em} -\sqrt{\sfrac{3}{2}}\, i\\\hspace{-0.5em}  0                        \\\hspace{-0.5em} -\sqrt{\sfrac{1}{2}}\, i \hspace{-0.6em} \end{array}
                    \begin{array}{c} \hspace{-0.6em} 0                         \\\hspace{-0.5em} 0                       \\\hspace{-0.5em}  \sqrt{\sfrac{1}{2}}\, i  \\\hspace{-0.5em}  0                       \hspace{-0.6em} \end{array} \right)  \,\,\,\,
            $
            \hspace{-0.7em}, \, \vspace{0.5em}$ 
            \hspace{-0.5em}X_8 \hspace{-0.3em}= \hspace{-0.3em}\left(    \begin{array}{c} \hspace{-0.4em}-\sfrac{7}{6}     \\ 0               \\ 0                \\ 0             \hspace{-0.5em} \end{array}
                    \begin{array}{c} \hspace{-0.4em}0                 \\ \sfrac{11}{6}   \\ 0                \\ 0             \hspace{-0.5em} \end{array}
                    \begin{array}{c} \hspace{-0.4em}0                 \\ 0               \\ -\sfrac{1}{6}    \\ 0             \hspace{-0.5em} \end{array}
                    \begin{array}{c} \hspace{-0.4em}0                 \\ 0               \\ 0                \\ -\sfrac{1}{2} \hspace{-0.5em} \end{array} \right)   
            $\\
            \vspace{0.5em}$ 
            \hspace{-0.5em}X_9 \hspace{-0.3em}= \hspace{-0.3em}\left(     \begin{array}{c} \hspace{-0.4em}0                  \\ 0        \\ 0          \\ \sqrt{\sfrac{5}{2}}  \hspace{-0.5em} \end{array}
                    \begin{array}{c} \hspace{-0.4em}0                  \\ 0        \\ 0          \\ 0                    \hspace{-0.5em} \end{array}
                    \begin{array}{c} \hspace{-0.4em}0                  \\ 0        \\ 0          \\ 0                    \hspace{-0.5em} \end{array}
                    \begin{array}{c} \hspace{-0.4em}\sqrt{\sfrac{5}{2}}\\ 0        \\ 0          \\ 0                    \hspace{-0.5em} \end{array} \right)   
            $
            , \, \vspace{0.5em}$ 
            \hspace{-0.5em}X_{10} \hspace{-0.3em}= \hspace{-0.3em}\left( \begin{array}{c} \hspace{-0.4em}0                        \\ 0        \\ 0          \\ \sqrt{\sfrac{5}{2}}\, i   \hspace{-0.5em} \end{array}
                    \begin{array}{c} \hspace{-0.4em}0                        \\ 0        \\ 0          \\ 0                       \hspace{-0.5em} \end{array}
                    \begin{array}{c} \hspace{-0.4em}0                        \\ 0        \\ 0          \\ 0                       \hspace{-0.5em} \end{array}
                    \begin{array}{c} \hspace{-0.4em}-\sqrt{\sfrac{5}{2}}\, i   \\ 0        \\ 0          \\ 0                       \hspace{-0.5em} \end{array} \right)   
            $\\
            \vspace{0.5em}$
            \hspace{-0.5em}X_{11} \hspace{-0.3em}= \hspace{-0.3em}\left(  \begin{array}{c} \hspace{-0.4em}0          \\ 0                     \\ 0          \\ 0                      \hspace{-0.5em} \end{array}
                    \begin{array}{c} \hspace{-0.4em}0          \\ 0                     \\ 0          \\ \sqrt{\sfrac{5}{2}}    \hspace{-0.5em} \end{array}
                    \begin{array}{c} \hspace{-0.4em}0          \\ 0   \\ 0          \\ 0                      \hspace{-0.5em} \end{array}
                    \begin{array}{c} \hspace{-0.4em}0          \\ \sqrt{\sfrac{5}{2}}                     \\ 0          \\ 0                      \hspace{-0.5em} \end{array} \right)  
            $
            , \, \vspace{0.5em}$
            \hspace{-0.5em}X_{12} \hspace{-0.3em}= \hspace{-0.3em}\left( \begin{array}{c} \hspace{-0.4em}0           \\ 0                     \\ 0          \\ 0                       \hspace{-0.5em} \end{array}
                    \begin{array}{c} \hspace{-0.4em}0           \\ 0                     \\ 0          \\ \sqrt{\sfrac{5}{2}}\, i   \hspace{-0.5em} \end{array}
                    \begin{array}{c} \hspace{-0.4em}0           \\ 0 \\ 0          \\ 0                       \hspace{-0.5em} \end{array}
                    \begin{array}{c} \hspace{-0.4em}0           \\ -\sqrt{\sfrac{5}{2}}\, i         \\ 0          \\ 0                       \hspace{-0.5em} \end{array} \right)   
            $\\
            \vspace{0.5em}$
            \hspace{-0.5em}X_{13} \hspace{-0.3em}= \hspace{-0.3em}\left(  \begin{array}{c} \hspace{-0.7em}0                        \\\hspace{-0.6em} -\sqrt{\sfrac{5}{4}}  \\\hspace{-0.6em} 0                    \\\hspace{-0.6em} 0                      \hspace{-1.1em} \end{array}
                    \begin{array}{c} \hspace{-0.7em}-\sqrt{\sfrac{5}{4}}     \\\hspace{-0.6em} 0                     \\\hspace{-0.6em} 0                    \\\hspace{-0.6em} 0                      \hspace{-1.1em} \end{array}
                    \begin{array}{c} \hspace{-0.7em}0                        \\\hspace{-0.6em} 0                     \\\hspace{-0.6em} 0                    \\\hspace{-0.6em} \sqrt{\sfrac{5}{4}}    \hspace{-1.1em} \end{array}
                    \begin{array}{c} \hspace{-0.7em}0                        \\\hspace{-0.6em} 0                     \\\hspace{-0.6em} \sqrt{\sfrac{5}{4}}  \\\hspace{-0.6em} 0                      \hspace{-1.1em} \end{array} \right)   
            $
            , \, $ 
            \hspace{-0.5em}X_{14} \hspace{-0.3em}= \hspace{-0.3em}\left( \begin{array}{c} \hspace{-0.8em}0                     \\ \hspace{-0.6em} -\sqrt{\sfrac{5}{4}}\, i \\ \hspace{-0.6em} 0                       \\ \hspace{-0.6em} 0                        \hspace{-1.2em} \end{array}
                    \begin{array}{c} \hspace{-0.8em}\sqrt{\sfrac{5}{4}}\, i \\\hspace{-0.6em} 0                     \\ \hspace{-0.6em} 0                       \\ \hspace{-0.6em} 0                        \hspace{-1.2em} \end{array}
                    \begin{array}{c} \hspace{-0.8em}0                     \\\hspace{-0.6em} 0                     \\ \hspace{-0.6em} 0                         \\ \hspace{-0.6em} \sqrt{\sfrac{5}{4}}\, i    \hspace{-1.2em} \end{array}
                    \begin{array}{c} \hspace{-0.8em}0                     \\\hspace{-0.6em} 0                     \\ \hspace{-0.6em}-\sqrt{\sfrac{5}{4}}\, i   \\ \hspace{-0.6em} 0                        \hspace{-1.2em} \end{array} \right)   
            $\\
            \vspace{0.5em}$
            \hspace{-0.5em}X_{15} \hspace{-0.3em}= \hspace{-0.3em}\left(  \begin{array}{c} \hspace{-0.4em}-\sfrac{\sqrt{5}}{6}     \\ 0        \\ 0          \\ 0         \hspace{-0.5em} \end{array}
                    \begin{array}{c} \hspace{-0.4em}0           \\-\sfrac{\sqrt{5}}{6}   \\ 0          \\ 0         \hspace{-0.5em} \end{array}
                    \begin{array}{c} \hspace{-0.4em}0           \\ 0        \\ \sfrac{5 \sqrt{5}}{6}  \\ 0         \hspace{-0.5em} \end{array}
                    \begin{array}{c} \hspace{-0.4em}0           \\ 0        \\ 0          \\-\sqrt{\sfrac{5}{4}}    \hspace{-0.5em} \end{array} \right)  
            $
             \\
            \hline
    \end{tabular}
    \vspace{0.5em}
    \caption[]{15 matrices for the SU(4) TWA obtained by unitary transformation \eqref{unitaeretrafo} of the Gellman matrices.}
    \label{tab:15matrizen}
    \end{table}

    In order to set up the equations of motion, the values of $f_{\alpha \beta \gamma}$ are needed. These can be determined by
    \begin{equation}\label{fabcbestimmen}
        f_{\alpha \beta \gamma} = - \frac{1}{5} i \, \mathrm{Tr} \left[ X_{\alpha} [X_{\beta}, X_{\gamma}] \right] \, 
    \end{equation}
    and are listed in Tab.\@ \ref{tab:fabc}.

    \begin{table}[htb]
    \centering
    \begin{tabular}{ c c c c | c c c c | c c c c  }
        \hline
        $\alpha$ & $\beta$ & $\gamma$ & $f_{\alpha \beta \gamma}$ & $\alpha$ & $\beta$ & $\gamma$ & $f_{\alpha \beta \gamma}$ & $\alpha$ & $\beta$ & $\gamma$ & $f_{\alpha \beta \gamma}$ \\
        \hline
        1   &  2    &  3    & $1                    $ &      2   &  7    &  12   & $\sfrac{\sqrt{5}}{2} $  &      6   &  7    &  8    & $\sfrac{2}{3}         $ \\             
        1   &  4    &  7    & $\sfrac{\sqrt{5}}{2}  $ &      2   &  8    &  13   & $\sqrt{\sfrac{5}{3}} $  &      6   &  7    &  15   & $-\sfrac{\sqrt{5}}{3} $ \\       
        1   &  4    &  10   & $-\sfrac{\sqrt{3}}{2} $ &      2   &  9    &  11   & $\sfrac{\sqrt{3}}{2} $  &      6   &  8    &  14   & $\sfrac{\sqrt{10}}{3} $ \\      
        1   &  4    &  14   & $\sfrac{1}{\sqrt{2}}  $ &      2   &  10   &  12   & $\sfrac{\sqrt{3}}{2} $  &      6   &  9    &  12   & $-\sfrac{1}{\sqrt{2}} $ \\       
        1   &  5    &  6    & $-\sfrac{\sqrt{5}}{2} $ &      2   &  11   &  13   & $\sfrac{1}{\sqrt{2}} $  &      6   &  10   &  11   & $\sfrac{1}{\sqrt{2}}  $ \\       
        1   &  5    &  9    & $\sfrac{\sqrt{3}}{2}  $ &      2   &  12   &  14   & $\sfrac{1}{\sqrt{2}} $  &      6   &  11   &  14   & $\sfrac{\sqrt{3}}{2}  $ \\       
        1   &  5    &  13   & $-\sfrac{1}{\sqrt{2}} $ &      2   &  13   &  15   & $-\sqrt{\sfrac{4}{3}}$  &      6   &  12   &  13   & $-\sfrac{\sqrt{3}}{2} $ \\    
        1   &  6    &  12   & $-\sfrac{\sqrt{5}}{2} $ &      3   &  4    &  5    & $2                   $  &      6   &  14   &  15   & $-\sfrac{\sqrt{8}}{3} $ \\        
        1   &  7    &  8    & $\sqrt{\sfrac{8}{3}}  $ &      3   &  6    &  7    & $1                   $  &      7   &  8    &  13   & $-\sfrac{\sqrt{10}}{3}$ \\        
        1   &  7    &  11   & $\sfrac{\sqrt{5}}{2}  $ &      3   &  9    &  10   & $3                   $  &      7   &  9    &  11   & $-\sfrac{1}{\sqrt{2}} $ \\        
        1   &  7    &  15   & $-\sqrt{\sfrac{10}{3}}$ &      3   &  11   &  12   & $2                   $  &      7   &  10   &  12   & $-\sfrac{1}{\sqrt{2}} $ \\            
        1   &  8    &  14   & $-\sqrt{\sfrac{5}{3}} $ &      3   &  13   &  14   & $1                   $  &      7   &  11   &  13   & $\sfrac{\sqrt{3}}{2}  $ \\            
        1   &  9    &  12   & $\sfrac{\sqrt{3}}{2}  $ &      4   &  5    &  8    & $-1                  $  &      7   &  12   &  14   & $\sfrac{\sqrt{3}}{2}  $ \\            
        1   &  10   &  11   & $-\sfrac{\sqrt{3}}{2} $ &      4   &  5    &  15   & $-\sqrt{5}           $  &      7   &  13   &  15   & $\sfrac{\sqrt{8}}{3}  $ \\          
        1   &  11   &  14   & $\sfrac{1}{\sqrt{2}}  $ &      4   &  6    &  10   & $-\sfrac{1}{\sqrt{2}}$  &      8   &  9    &  10   & $-\sfrac{2}{3}        $ \\     
        1   &  12   &  13   & $-\sfrac{1}{\sqrt{2}} $ &      4   &  6    &  14   & $-\sfrac{\sqrt{3}}{2}$  &      8   &  11   &  12   & $\sfrac{7}{3}         $ \\ 
        1   &  14   &  15   & $\sqrt{\sfrac{4}{3}}  $ &      4   &  7    &  9    & $\sfrac{1}{\sqrt{2}} $  &      8   &  13   &  14   & $-\sfrac{4}{3}        $ \\ 
        2   &  4    &  6    & $\sfrac{\sqrt{5}}{2}  $ &      4   &  7    &  13   & $-\sfrac{\sqrt{3}}{2}$  &      9   &  10   &  15   & $\sfrac{\sqrt{5}}{3}  $ \\ 
        2   &  4    &  9    & $\sfrac{\sqrt{3}}{2}  $ &      4   &  9    &  14   & $\sfrac{\sqrt{5}}{2} $  &      9   &  11   &  14   & $-\sfrac{\sqrt{5}}{2} $ \\ 
        2   &  4    &  13   & $\sfrac{1}{\sqrt{2}}  $ &      4   &  10   &  13   & $-\sfrac{\sqrt{5}}{2}$  &      9   &  12   &  13   & $-\sfrac{\sqrt{5}}{2} $ \\ 
        2   &  5    &  7    & $\sfrac{\sqrt{5}}{2}  $ &      5   &  6    &  9    & $\sfrac{1}{\sqrt{2}} $  &      10  &  11   &  13   & $\sfrac{\sqrt{5}}{2}  $ \\ 
        2   &  5    &  10   & $\sfrac{\sqrt{3}}{2}  $ &      5   &  6    &  13   & $\sfrac{\sqrt{3}}{2} $  &      10  &  12   &  14   & $-\sfrac{\sqrt{5}}{2} $ \\ 
        2   &  5    &  14   & $\sfrac{1}{\sqrt{2}}  $ &      5   &  7    &  10   & $\sfrac{1}{\sqrt{2}} $  &      11  &  12   &  15   & $\sfrac{\sqrt{5}}{3}  $ \\ 
        2   &  6    &  8    & $-\sqrt{\sfrac{8}{3}} $ &      5   &  7    &  14   & $-\sfrac{\sqrt{3}}{2}$  &      13  &  14   &  15   & $\sfrac{\sqrt{20}}{3} $ \\ 
        2   &  6    &  11   & $\sfrac{\sqrt{5}}{2}  $ &      5   &  9    &  13   & $\sfrac{\sqrt{5}}{2} $  &          &       &       & $                     $ \\
        2   &  6    &  15   & $\sqrt{\sfrac{10}{3}} $ &      5   &  10   &  14   & $\sfrac{\sqrt{5}}{2} $  &          &       &       & $                     $ \\
        \hline  
    \end{tabular}
    \vspace{0.5em}
    \caption[]{Structure constants for the SU(4) TWA according to Eq.\@ \eqref{fabcbestimmen} with the matrices from Tab.\@ \ref{tab:15matrizen} rounded to 4 decimal digits.}
    \label{tab:fabc}
    \end{table}
    
    \section{Equations of motion with the SU(4) TWA for the CSM including the quadrupolar interaction using the quadratic choice of $V$}
    \label{s:eom}

    The Hamiltonian for the SU(4) TWA for the central spin model (CSM) including the quadrupolar interaction 
    using the simple model for $V$ is given by 
    \begin{equation}
    \begin{split}
        H_{\mathrm{W}}^{\mathrm{SU(4)}} = &\left( \sum_{k=1}^N  A_k \left(\begin{array}{c} X_{1,k} \\ X_{2,k} \\ X_{3,k} \end{array}\right) + \pmb{h} \right) \cdot
                                                             \left(\begin{array}{c} X_{1} \\ X_{2} \\ X_{3} \end{array}\right) 
                                                \\\ &+ \sum_{k=1}^N \pmb{h}_{{\mathrm{n}}} \cdot \left(\begin{array}{c} X_{1,k} \\ X_{2,k} \\ X_{3,k} \end{array}\right) 
                                                \\\ &+ \sum_{k=1}^N q_k 
                                                \left[ \frac{1}{3} X_{8,k} + \sqrt{\sfrac{4}{45}} \, X_{15,k} + \frac{5}{12} \mathbb{1} \right] \, .
    \end{split}
    \end{equation}
    according to Eq. \eqref{su4_Iquadrat} and correspondingly the derivatives by
    \begin{subequations}
    \begin{align}
        \frac{\partial H_{\mathrm{W}}^{\mathrm{SU(4)}}}{\partial X_{1}}  &= \sum_{k=1}^N A_k X_{1,k} + h_{x} = B_{\mathrm{ov},x} + h \, , \\
        \frac{\partial H_{\mathrm{W}}^{\mathrm{SU(4)}}}{\partial X_{2}}  &= \sum_{k=1}^N A_k X_{2,k} + h_{y} = B_{\mathrm{ov},y} + 0 \, , \\
        \frac{\partial H_{\mathrm{W}}^{\mathrm{SU(4)}}}{\partial X_{3}}  &= \sum_{k=1}^N A_k X_{3,k} + h_{z} = B_{\mathrm{ov},z} + 0 \, , \\
        \frac{\partial H_{\mathrm{W}}^{\mathrm{SU(4)}}}{\partial X_{1, k}}  &= A_k X_1 + h_{\mathrm{n},x} = A_k X_1 + h_{\mathrm{n}} \, , \\
        \frac{\partial H_{\mathrm{W}}^{\mathrm{SU(4)}}}{\partial X_{2, k}}  &= A_k X_2 + h_{\mathrm{n},y} = A_k X_2 + 0 \, , \\
        \frac{\partial H_{\mathrm{W}}^{\mathrm{SU(4)}}}{\partial X_{3, k}}  &= A_k X_3 + h_{\mathrm{n},z} = A_k X_3 + 0 \, , \\
        \frac{\partial H_{\mathrm{W}}^{\mathrm{SU(4)}}}{\partial X_{8, k}}  &= \frac{1}{3}  q_k  \, , \\
        \frac{\partial H_{\mathrm{W}}^{\mathrm{SU(4)}}}{\partial X_{15, k}} &= \sqrt{\sfrac{4}{45}} \, q_k \, .
    \end{align}
    \end{subequations}
    It holds: $S_x = X_{1}$, $S_y = X_{2}$ and $S_z = X_{3}$ and $I_{x,k} = X_{1,k}$, $I_{y,k} = X_{2,k}$ and $I_{z,k} = X_{3,k}$.
    With the structure constants \ref{tab:fabc}, the following is obtained for the central spin:

    \begingroup
      \flausr
    \begin{subequations}\label{bewglen_su4_cs}
    \begin{equation}
    \begin{split}
        X_1 = (B_{\mathrm{ov},y})  X_3 - (B_{\mathrm{ov},z})  X_2
    \end{split}
    \end{equation}
    \begin{equation}
    \begin{split}
        X_2 = (B_{\mathrm{ov},z})  X_1 - (B_{\mathrm{ov},x}+h)  X_3
    \end{split}
    \end{equation}
    \begin{equation}
    \begin{split}
        X_3 = (B_{\mathrm{ov},x}+h)  X_2 - (B_{\mathrm{ov},y})  X_1
    \end{split}
    \end{equation}
    \begin{equation}
    \begin{split}
        X_4 = &\, (B_{\mathrm{ov},x}+h) \, (-\sfrac{\sqrt{5}}{2} \, X_7+\sfrac{\sqrt{3}}{2} \, X_{10} -\sfrac{1}{\sqrt{2}} \, X_{14} ) \\\ &+ (B_{\mathrm{ov},y}) \, (-\sfrac{\sqrt{5}}{2} \, X_6 -\sfrac{\sqrt{3}}{2} \, X_9-\sfrac{1}{\sqrt{2}} \, X_{13}) \\\ &- (B_{\mathrm{ov},z}) 2  X_5
    \end{split}
    \end{equation}
    \begin{equation}
    \begin{split}
        X_5 = &\, (B_{\mathrm{ov},x}+h) \, (\sfrac{\sqrt{5}}{2} \, X_6-\sfrac{\sqrt{3}}{2} \, X_9 +\sfrac{1}{\sqrt{2}} \, X_{13}) \\\ &+ (B_{\mathrm{ov},y}) \, (-\sfrac{\sqrt{5}}{2} \, X_7 -\sfrac{\sqrt{3}}{2} \, X_{10}-\sfrac{1}{\sqrt{2}} \, X_{14}) \\\ &+ (B_{\mathrm{ov},z}) 2 X_4
    \end{split}
    \end{equation}
    \begin{equation}
    \begin{split}
        X_6 = &\, (B_{\mathrm{ov},x}+h) \, (\sfrac{\sqrt{5}}{2} \, X_{12}-\sfrac{\sqrt{5}}{2} \, X_5) \\\ &+ (B_{\mathrm{ov},y}) \, (\sfrac{\sqrt{5}}{2} \, X_4+\sqrt{\sfrac{8}{3}} \, X_8 \\\ &-\sfrac{\sqrt{5}}{2} \, X_{11}-\sqrt{\sfrac{10}{3}} \, X_{15}) - (B_{\mathrm{ov},z}) 1 X_7
    \end{split}
    \end{equation}
    \begin{equation}
    \begin{split}
        X_7 = &\, (B_{\mathrm{ov},x}+h) \, (\sfrac{\sqrt{5}}{2} \, X_4-\sqrt{\sfrac{8}{3}} \, X_8 -\sfrac{\sqrt{5}}{2} \, X_{11}\\\ &+\sqrt{\sfrac{10}{3}} \, X_{15}) + (B_{\mathrm{ov},y}) \, (\sfrac{\sqrt{5}}{2} \, X_5-\sfrac{\sqrt{5}}{2} \, X_{12}) \\\ &+ (B_{\mathrm{ov},z}) 1 X_6
    \end{split}
    \end{equation}
    \begin{equation}
    \begin{split}
        X_8 = &\, (B_{\mathrm{ov},x}+h) \, (\sqrt{\sfrac{8}{3}} \, X_7+\sqrt{\sfrac{5}{3}} \, X_{14}) \\\ &+ (B_{\mathrm{ov},y}) \, (-\sqrt{\sfrac{8}{3}} \, X_6-\sqrt{\sfrac{5}{3}} \, X_{13})
    \end{split}
    \end{equation}
    \begin{equation}
    \begin{split}
        X_9 = &\, (B_{\mathrm{ov},x}+h) \, (\sfrac{\sqrt{3}}{2} \, X_5-\sfrac{\sqrt{3}}{2} \, X_{12}) \\\ &+ (B_{\mathrm{ov},y}) \, (\sfrac{\sqrt{3}}{2} \, X_4-\sfrac{\sqrt{3}}{2} \, X_{11}) - (B_{\mathrm{ov},z}) 3  X_{10}
    \end{split}
    \end{equation}
    \begin{equation}
    \begin{split}
        X_{10} = &\, (B_{\mathrm{ov},x}+h) \, (-\sfrac{\sqrt{3}}{2} \, X_4+\sfrac{\sqrt{3}}{2} \, X_{11}) \\\ &+ (B_{\mathrm{ov},y}) \, (\sfrac{\sqrt{3}}{2} \, X_5-\sfrac{\sqrt{3}}{2} \, X_{12}) + (B_{\mathrm{ov},z}) 3 X_9
    \end{split}
    \end{equation}
    \begin{equation}
    \begin{split}
        X_{11} = &\, (B_{\mathrm{ov},x}+h) \, (\sfrac{\sqrt{5}}{2} \, X_7-\sfrac{\sqrt{3}}{2} \, X_{10} -\sfrac{1}{\sqrt{2}} \, X_{14}) \\\ &+ (B_{\mathrm{ov},y}) \, (\sfrac{\sqrt{5}}{2} \, X_6 +\sfrac{\sqrt{3}}{2} \, X_9-\sfrac{1}{\sqrt{2}} \, X_{13}) \\\ &- (B_{\mathrm{ov},z}) 2 X_{12}
    \end{split}
    \end{equation}
    \begin{equation}
    \begin{split}
        X_{12} = &\, (B_{\mathrm{ov},x}+h) \, (-\sfrac{\sqrt{5}}{2} \, X_6+\sfrac{\sqrt{3}}{2} \, X_9 +\sfrac{1}{\sqrt{2}} \, X_{13}) \\\ &+ (B_{\mathrm{ov},y}) \, (\sfrac{\sqrt{5}}{2} \, X_7 +\sfrac{\sqrt{3}}{2} \, X_{10}-\sfrac{1}{\sqrt{2}} \, X_{14}) \\\ &+ (B_{\mathrm{ov},z}) 2 X_{11}
    \end{split}
    \end{equation}
    \begin{equation}
    \begin{split}
        X_{13} = &\, (B_{\mathrm{ov},x}+h) \, (-\sfrac{1}{\sqrt{2}} \, X_5-\sfrac{1}{\sqrt{2}} \, X_{12}) \\\ &+ (B_{\mathrm{ov},y}) \, (\sfrac{1}{\sqrt{2}} \, X_4+\sqrt{\sfrac{5}{3}} \, X_8 +\sfrac{1}{\sqrt{2}} \, X_{11} \\\ &+\sqrt{\sfrac{4}{3}} \, X_{15}) - (B_{\mathrm{ov},z}) 1 X_{14}
    \end{split}
    \end{equation}
    \begin{equation}
    \begin{split}
        X_{14} = &\, (B_{\mathrm{ov},x}+h) \, (\sfrac{1}{\sqrt{2}} \, X_4-\sqrt{\sfrac{5}{3}} \, X_8 +\sfrac{1}{\sqrt{2}} \, X_{11} \\\ &-\sqrt{\sfrac{4}{3}} \, X_{15}) + (B_{\mathrm{ov},y}) \, (\sfrac{1}{\sqrt{2}} \, X_5+\sfrac{1}{\sqrt{2}} \, X_{12}) \\\ &+ (B_{\mathrm{ov},z}) 1 X_{13}
    \end{split}
    \end{equation}
    \begin{equation}
    \begin{split}
        X_{15} = &\, (B_{\mathrm{ov},x}+h) \, (\sqrt{\sfrac{4}{3}} \, X_{14}-\sqrt{\sfrac{10}{3}} \, X_7) \\\ &+ (B_{\mathrm{ov},y}) \, (\sqrt{\sfrac{10}{3}} \, X_6-\sqrt{\sfrac{4}{3}} \, X_{13})
    \end{split}
    \end{equation}
    \end{subequations}
    \endgroup

    For the nuclear spins we obtain:

    \begingroup
      \flausr
    \begin{subequations}\label{bewglen_su4_ns}
    \begin{equation}
    \begin{split}
        X_{1,k} = &\,  (A_k X_2)  X_{3,k} - (A_k  X_3) X_{2,k}   \\\ &+1/3 \, q_k (-\sqrt{\sfrac{8}{3}} \, X_{7,k} -\sqrt{\sfrac{5}{3}} \, X_{14,k}) \\\ &+ \sqrt{\sfrac{4}{45}} \, q_k (\sqrt{\sfrac{10}{3}} \, X_{7,k} -\sqrt{\sfrac{4}{3}} \, X_{14,k})
    \end{split}
    \end{equation}
    \begin{equation}
    \begin{split}
        X_{2,k} = &\, (A_k X_3) X_{1,k} - (A_k  X_1+h_{\mathrm{n}})  X_{3,k} \\\ &+1/3 \, q_k (\sqrt{\sfrac{8}{3}} \, X_{6,k} +\sqrt{\sfrac{5}{3}} \, X_{13,k}) \\\ &+\sqrt{\sfrac{4}{45}} \, q_k (-\sqrt{\sfrac{10}{3}} \, X_{6,k} +\sqrt{\sfrac{4}{3}} \, X_{13,k})
    \end{split}
    \end{equation}
    \begin{equation}
    \begin{split}
        X_{3,k} = (A_k X_1+h_{\mathrm{n}})  X_{2,k} - (A_k X_2)  X_{1,k}    
    \end{split}
    \end{equation}
    \begin{equation}
    \begin{split}
        X_{4,k} = &\, (A_k X_1+h_{\mathrm{n}}) \, (-\sfrac{\sqrt{5}}{2} \, X_{7,k} +\sfrac{\sqrt{3}}{2} \, X_{10,k} \\\ &-\sfrac{1}{\sqrt{2}} \, X_{14,k}) +(A_k X_2) \, (-\sfrac{\sqrt{5}}{2} \, X_{6,k} \\\ &-\sfrac{\sqrt{3}}{2} \, X_{9,k} -\sfrac{1}{\sqrt{2}} \, X_{13,k}) + X_{5,k} (-2 (A_k X_3) \\\ &+1/3 \, q_k +\sqrt{5} \, \sqrt{\sfrac{4}{45}} \, q_k)
    \end{split}
    \end{equation}
    \begin{equation}
    \begin{split}
        X_{5,k} = &\, (A_k X_1+h_{\mathrm{n}}) \, (\sfrac{\sqrt{5}}{2} \, X_{6,k} -\sfrac{\sqrt{3}}{2} \, X_{9,k} \\\ &+\sfrac{1}{\sqrt{2}} \, X_{13,k}) +(A_k X_2) \, (-\sfrac{\sqrt{5}}{2} \, X_{7,k} \\\ &-\sfrac{\sqrt{3}}{2} \, X_{10,k} -\sfrac{1}{\sqrt{2}} \, X_{14,k}) +X_{4,k} (2 (A_k X_3) \\\ & -1/3 \, q_k-\sqrt{5} \, \sqrt{\sfrac{4}{45}} \, q_k)
    \end{split}
    \end{equation}
    \begin{equation}
    \begin{split}
        X_{6,k} = &\, (A_k X_1+h_{\mathrm{n}}) \, (\sfrac{\sqrt{5}}{2} \, X_{12,k} -\sfrac{\sqrt{5}}{2} \, X_{5,k}) \\\ &+ (A_k X_2) \, (\sfrac{\sqrt{5}}{2} \, X_{4,k} +\sqrt{\sfrac{8}{3}} \, X_{8,k} \\\ &-\sfrac{\sqrt{5}}{2} \, X_{11,k} -\sqrt{\sfrac{10}{3}} \, X_{15,k}) -1 (A_k X_3) X_{7,k}  \\\ &+ 1/3 \, q_k (-\sfrac{2}{3} \, X_{7,k}+\sfrac{\sqrt{10}}{3} \, X_{14,k} \\\ & -\sqrt{\sfrac{8}{3}} \, X_{2,k}) + \sqrt{\sfrac{4}{45}} \, q_k (\sqrt{\sfrac{10}{3}} \, X_{2,k} \\\ &+\sfrac{\sqrt{5}}{3} \, X_{7,k}+\sfrac{\sqrt{8}}{3} \, X_{14,k})
    \end{split}
    \end{equation}
    \begin{equation}
    \begin{split}
        X_{7,k} = &\, (A_k X_1+h_{\mathrm{n}}) \, (\sfrac{\sqrt{5}}{2} \, X_{4,k}  -\sqrt{\sfrac{8}{3}} \, X_{8,k} \\\ &-\sfrac{\sqrt{5}}{2} \, X_{11,k} +\sqrt{\sfrac{10}{3}} \, X_{15,k}) \\\ &+ (A_k X_2) \, (\sfrac{\sqrt{5}}{2} \, X_{5,k} -\sfrac{\sqrt{5}}{2} \, X_{12,k})  \\\ &+ 1 (A_k X_3) X_{6,k}  + 1/3 \, q_k (\sqrt{\sfrac{8}{3}} \, X_{1,k} \\\ &+\sfrac{2}{3} \, X_{6,k} -\sfrac{\sqrt{10}}{3} \, X_{13,k}) \\\ &+ \sqrt{\sfrac{4}{45}} \, q_k (-\sqrt{\sfrac{10}{3}} \, X_{1,k} -\sfrac{\sqrt{5}}{3} \, X_{6,k} \\\ &-\sfrac{\sqrt{8}}{3} \, X_{13,k})
    \end{split}
    \end{equation}
    \begin{equation}
    \begin{split}
        X_{8,k} = &\, (A_k X_1+h_{\mathrm{n}}) \, (\sqrt{\sfrac{8}{3}} \, X_{7,k}+\sqrt{\sfrac{5}{3}} \, X_{14,k}) \\\ &+ (A_k X_2) \, (-\sqrt{\sfrac{8}{3}} \, X_{6,k}-\sqrt{\sfrac{5}{3}} \, X_{13,k})
    \end{split}
    \end{equation}
    \begin{equation}
    \begin{split}
        X_{9,k} = &\, (A_k X_1+h_{\mathrm{n}}) \, (\sfrac{\sqrt{3}}{2} \, X_{5,k}-\sfrac{\sqrt{3}}{2} \, X_{12,k}) \\\ &+ (A_k X_2) \, (\sfrac{\sqrt{3}}{2} \, X_{4,k}-\sfrac{\sqrt{3}}{2} \, X_{11,k}) \\\ &+ X_{10,k} (-3 (A_k X_3) +\sfrac{2}{9} \, q_k -\sfrac{\sqrt{5}}{3} \, \sqrt{\sfrac{4}{45}} \, q_k)
    \end{split}
    \end{equation}
    \begin{equation}
    \begin{split}
        X_{10,k} = &\, (A_k X_1+h_{\mathrm{n}}) \, (-\sfrac{\sqrt{3}}{2} \, X_{4,k} +\sfrac{\sqrt{3}}{2} \, X_{11,k}) \\\ &+ (A_k X_2) \, (\sfrac{\sqrt{3}}{2} \, X_{5,k} -\sfrac{\sqrt{3}}{2} \, X_{12,k}) \\\ &+ X_{9,k} (3 (A_k X_3) -\sfrac{2}{9} \, q_k +\sfrac{\sqrt{5}}{3} \, \sqrt{\sfrac{4}{45}} \, q_k)
    \end{split}
    \end{equation}
    \begin{equation}
    \begin{split}
        X_{11,k} = &\, (A_k X_1+h_{\mathrm{n}}) \, (\sfrac{\sqrt{5}}{2} \, X_{7,k} -\sfrac{\sqrt{3}}{2} \, X_{10,k} \\\ &-\sfrac{1}{\sqrt{2}} \, X_{14,k}) + (A_k X_2) \, (\sfrac{\sqrt{5}}{2} \, X_{6,k} \\\ &+\sfrac{\sqrt{3}}{2} \, X_{9,k} -\sfrac{1}{\sqrt{2}} \, X_{13,k}) + X_{12,k} \\\ & (-2 (A_k X_3) -\sfrac{7}{9} \, q_k-\sfrac{\sqrt{5}}{3} \, \sqrt{\sfrac{4}{45}} \, q_k)
    \end{split}
    \end{equation}
    \begin{equation}
    \begin{split}
        X_{12,k} = &\, (A_k X_1+h_{\mathrm{n}}) \, (-\sfrac{\sqrt{5}}{2} \, X_{6,k} +\sfrac{\sqrt{3}}{2} \, X_{9,k} \\\ &+\sfrac{1}{\sqrt{2}} \, X_{13,k}) + (A_k X_2) \, (\sfrac{\sqrt{5}}{2} \, X_{7,k} \\\ &+\sfrac{\sqrt{3}}{2} \, X_{10,k} -\sfrac{1}{\sqrt{2}} \, X_{14,k}) + X_{11,k} \\\ & (2 (A_k X_3) +\sfrac{7}{9} \, q_k+\sfrac{\sqrt{5}}{3} \, \sqrt{\sfrac{4}{45}} \, q_k)
    \end{split}
    \end{equation}
    \begin{equation}
    \begin{split}
        X_{13,k} = &\, (A_k X_1+h_{\mathrm{n}}) \, (-\sfrac{1}{\sqrt{2}} \, X_{5,k} -\sfrac{1}{\sqrt{2}} \, X_{12,k}) \\\ &+ (A_k X_2) \, (\sfrac{1}{\sqrt{2}} \, X_{4,k} +\sqrt{\sfrac{5}{3}} \, X_{8,k} \\\ &+\sfrac{1}{\sqrt{2}} \, X_{11,k} +\sqrt{\sfrac{4}{3}} \, X_{15,k}) \\\ &- 1 (A_k X_3) X_{14,k}  + 1/3 \, q_k (\sfrac{\sqrt{10}}{3} \, X_{7,k} \\\ &+\sfrac{4}{3} \, X_{14,k} -\sqrt{\sfrac{5}{3}} \, X_{2,k}) + \sqrt{\sfrac{4}{45}} \, q_k \\\ &(-\sqrt{\sfrac{4}{3}} \, X_{2,k} +\sfrac{\sqrt{8}}{3} \, X_{7,k}-\sfrac{\sqrt{20}}{3} \, X_{14,k})
    \end{split}
    \end{equation}
    \begin{equation}
    \begin{split}
        X_{14,k} = &\, (A_k X_1+h_{\mathrm{n}}) \, (\sfrac{1}{\sqrt{2}} \, X_{4,k}-\sqrt{\sfrac{5}{3}} \, X_{8,k} \\\ &+\sfrac{1}{\sqrt{2}} \, X_{11,k}-\sqrt{\sfrac{4}{3}} \, X_{15,k}) \\\ &+ (A_k X_2) \, (\sfrac{1}{\sqrt{2}} \, X_{5,k}+\sfrac{1}{\sqrt{2}} \, X_{12,k}) \\\ &+ 1 (A_k X_3) X_{13,k}  + 1/3 \, q_k (\sqrt{\sfrac{5}{3}} \, X_{1,k} \\\ &-\sfrac{\sqrt{10}}{3} \, X_{6,k} -\sfrac{4}{3} \, X_{13,k}) + \sqrt{\sfrac{4}{45}} \, q_k \\\ & (\sqrt{\sfrac{4}{3}} \, X_{1,k} -\sfrac{\sqrt{8}}{3} \, X_{6,k}+\sfrac{\sqrt{20}}{3} \, X_{13,k})
    \end{split}
    \end{equation}
    \begin{equation}
    \begin{split}
        X_{15,k} = &\, (A_k X_1+h_{\mathrm{n}}) \, (\sqrt{\sfrac{4}{3}} \, X_{14,k}  -\sqrt{\sfrac{10}{3}} \, X_{7,k}) \\\ &+ (A_k  X_2) \, (\sqrt{\sfrac{10}{3}} \, X_{6,k} -\sqrt{\sfrac{4}{3}} \, X_{13,k})
    \end{split}
    \end{equation}
    \end{subequations}
    \endgroup
\vspace{7em}

\newpage


\begin{thebibliography}{53}%
  \makeatletter
  \providecommand \@ifxundefined [1]{%
   \@ifx{#1\undefined}
  }%
  \providecommand \@ifnum [1]{%
   \ifnum #1\expandafter \@firstoftwo
   \else \expandafter \@secondoftwo
   \fi
  }%
  \providecommand \@ifx [1]{%
   \ifx #1\expandafter \@firstoftwo
   \else \expandafter \@secondoftwo
   \fi
  }%
  \providecommand \natexlab [1]{#1}%
  \providecommand \enquote  [1]{``#1''}%
  \providecommand \bibnamefont  [1]{#1}%
  \providecommand \bibfnamefont [1]{#1}%
  \providecommand \citenamefont [1]{#1}%
  \providecommand \href@noop [0]{\@secondoftwo}%
  \providecommand \href [0]{\begingroup \@sanitize@url \@href}%
  \providecommand \@href[1]{\@@startlink{#1}\@@href}%
  \providecommand \@@href[1]{\endgroup#1\@@endlink}%
  \providecommand \@sanitize@url [0]{\catcode `\\12\catcode `\$12\catcode
    `\&12\catcode `\#12\catcode `\^12\catcode `\_12\catcode `\%12\relax}%
  \providecommand \@@startlink[1]{}%
  \providecommand \@@endlink[0]{}%
  \providecommand \url  [0]{\begingroup\@sanitize@url \@url }%
  \providecommand \@url [1]{\endgroup\@href {#1}{\urlprefix }}%
  \providecommand \urlprefix  [0]{URL }%
  \providecommand \Eprint [0]{\href }%
  \providecommand \doibase [0]{https://doi.org/}%
  \providecommand \selectlanguage [0]{\@gobble}%
  \providecommand \bibinfo  [0]{\@secondoftwo}%
  \providecommand \bibfield  [0]{\@secondoftwo}%
  \providecommand \translation [1]{[#1]}%
  \providecommand \BibitemOpen [0]{}%
  \providecommand \bibitemStop [0]{}%
  \providecommand \bibitemNoStop [0]{.\EOS\space}%
  \providecommand \EOS [0]{\spacefactor3000\relax}%
  \providecommand \BibitemShut  [1]{\csname bibitem#1\endcsname}%
  \let\auto@bib@innerbib\@empty
  \bibitem [{\citenamefont {Loss}\ and\ \citenamefont
    {DiVincenzo}(1998)}]{loss98}%
    \BibitemOpen
    \bibfield  {author} {\bibinfo {author} {\bibfnamefont {D.}~\bibnamefont
    {Loss}}\ and\ \bibinfo {author} {\bibfnamefont {D.~P.}\ \bibnamefont
    {DiVincenzo}},\ }\bibfield  {title} {\bibinfo {title} {Quantum computation
    with quantum dots},\ }\href {https://doi.org/10.1103/PhysRevA.57.120}
    {\bibfield  {journal} {\bibinfo  {journal} {Phys. Rev. A}\ }\textbf {\bibinfo
    {volume} {57}},\ \bibinfo {pages} {120} (\bibinfo {year} {1998})}\BibitemShut
    {NoStop}%
  \bibitem [{\citenamefont {Nielsen}\ and\ \citenamefont
    {Chuang}(2000)}]{niels00}%
    \BibitemOpen
    \bibfield  {author} {\bibinfo {author} {\bibfnamefont {M.~A.}\ \bibnamefont
    {Nielsen}}\ and\ \bibinfo {author} {\bibfnamefont {I.~L.}\ \bibnamefont
    {Chuang}},\ }\href@noop {} {\emph {\bibinfo {title} {Quantum Computation and
    Quantum Information}}}\ (\bibinfo  {publisher} {Cambridge University Press},\
    \bibinfo {address} {Cambridge},\ \bibinfo {year} {2000})\BibitemShut
    {NoStop}%
  \bibitem [{\citenamefont {Ladd}\ \emph {et~al.}(2010)\citenamefont {Ladd},
    \citenamefont {Jelezko}, \citenamefont {Laflamme}, \citenamefont {Nakamura},
    \citenamefont {Monroe},\ and\ \citenamefont {O'Brien}}]{ladd10b}%
    \BibitemOpen
    \bibfield  {author} {\bibinfo {author} {\bibfnamefont {T.~D.}\ \bibnamefont
    {Ladd}}, \bibinfo {author} {\bibfnamefont {F.}~\bibnamefont {Jelezko}},
    \bibinfo {author} {\bibfnamefont {R.}~\bibnamefont {Laflamme}}, \bibinfo
    {author} {\bibfnamefont {Y.}~\bibnamefont {Nakamura}}, \bibinfo {author}
    {\bibfnamefont {C.}~\bibnamefont {Monroe}},\ and\ \bibinfo {author}
    {\bibfnamefont {J.~L.}\ \bibnamefont {O'Brien}},\ }\bibfield  {title}
    {\bibinfo {title} {{Quantum computers}},\ }\href
    {https://doi.org/10.1038/nature08812} {\bibfield  {journal} {\bibinfo
    {journal} {Nature}\ }\textbf {\bibinfo {volume} {464}},\ \bibinfo {pages}
    {45} (\bibinfo {year} {2010})}\BibitemShut {NoStop}%
  \bibitem [{\citenamefont {Urbaszek}\ \emph {et~al.}(2013)\citenamefont
    {Urbaszek}, \citenamefont {Xavier}, \citenamefont {Amand}, \citenamefont
    {Krebs}, \citenamefont {Voisin}, \citenamefont {Maletinsky}, \citenamefont
    {H\"ogele},\ and\ \citenamefont {Imamoglu}}]{urbas13}%
    \BibitemOpen
    \bibfield  {author} {\bibinfo {author} {\bibfnamefont {B.}~\bibnamefont
    {Urbaszek}}, \bibinfo {author} {\bibfnamefont {M.}~\bibnamefont {Xavier}},
    \bibinfo {author} {\bibfnamefont {T.}~\bibnamefont {Amand}}, \bibinfo
    {author} {\bibfnamefont {O.}~\bibnamefont {Krebs}}, \bibinfo {author}
    {\bibfnamefont {P.}~\bibnamefont {Voisin}}, \bibinfo {author} {\bibfnamefont
    {P.}~\bibnamefont {Maletinsky}}, \bibinfo {author} {\bibfnamefont
    {A.}~\bibnamefont {H\"ogele}},\ and\ \bibinfo {author} {\bibfnamefont
    {A.}~\bibnamefont {Imamoglu}},\ }\bibfield  {title} {\bibinfo {title}
    {Nuclear spin physics in quantum dots: An optical investigation},\ }\href
    {https://doi.org/10.1103/RevModPhys.85.79} {\bibfield  {journal} {\bibinfo
    {journal} {Rev. Mod. Phys.}\ }\textbf {\bibinfo {volume} {85}},\ \bibinfo
    {pages} {79} (\bibinfo {year} {2013})}\BibitemShut {NoStop}%
  \bibitem [{\citenamefont {Gangloff}\ \emph {et~al.}(2019)\citenamefont
    {Gangloff}, \citenamefont {\'Ethier-Majcher}, \citenamefont {Lang},
    \citenamefont {Denning}, \citenamefont {Bodey}, \citenamefont {Jackson},
    \citenamefont {Clarke}, \citenamefont {Hugues}, \citenamefont {Gall},\ and\
    \citenamefont {Atat\"ure}}]{gangl19}%
    \BibitemOpen
    \bibfield  {author} {\bibinfo {author} {\bibfnamefont {D.~A.}\ \bibnamefont
    {Gangloff}}, \bibinfo {author} {\bibfnamefont {G.}~\bibnamefont
    {\'Ethier-Majcher}}, \bibinfo {author} {\bibfnamefont {C.}~\bibnamefont
    {Lang}}, \bibinfo {author} {\bibfnamefont {E.~V.}\ \bibnamefont {Denning}},
    \bibinfo {author} {\bibfnamefont {J.~H.}\ \bibnamefont {Bodey}}, \bibinfo
    {author} {\bibfnamefont {D.~M.}\ \bibnamefont {Jackson}}, \bibinfo {author}
    {\bibfnamefont {E.}~\bibnamefont {Clarke}}, \bibinfo {author} {\bibfnamefont
    {M.}~\bibnamefont {Hugues}}, \bibinfo {author} {\bibfnamefont {C.~L.}\
    \bibnamefont {Gall}},\ and\ \bibinfo {author} {\bibfnamefont
    {M.}~\bibnamefont {Atat\"ure}},\ }\bibfield  {title} {\bibinfo {title}
    {Quantum interface of an electron and a nuclear ensemble},\ }\href
    {https://doi.org/10.1126/science.aaw2906} {\bibfield  {journal} {\bibinfo
    {journal} {Science}\ }\textbf {\bibinfo {volume} {364}},\ \bibinfo {pages}
    {62} (\bibinfo {year} {2019})}\BibitemShut {NoStop}%
  \bibitem [{\citenamefont {Chekhovich}\ \emph {et~al.}(2020)\citenamefont
    {Chekhovich}, \citenamefont {da~Silva},\ and\ \citenamefont
    {Rastelli}}]{chekh20}%
    \BibitemOpen
    \bibfield  {author} {\bibinfo {author} {\bibfnamefont {E.~A.}\ \bibnamefont
    {Chekhovich}}, \bibinfo {author} {\bibfnamefont {S.~F.~C.}\ \bibnamefont
    {da~Silva}},\ and\ \bibinfo {author} {\bibfnamefont {A.}~\bibnamefont
    {Rastelli}},\ }\bibfield  {title} {\bibinfo {title} {Nuclear spin quantum
    register in an optically active semiconductor quantum dot},\ }\href
    {https://doi.org/10.1038/s41565-020-0769-3} {\bibfield  {journal} {\bibinfo
    {journal} {Nat. Nanotech.}\ }\textbf {\bibinfo {volume} {15}},\ \bibinfo
    {pages} {999} (\bibinfo {year} {2020})}\BibitemShut {NoStop}%
  \bibitem [{\citenamefont {Awschalom}\ \emph {et~al.}(2002)\citenamefont
    {Awschalom}, \citenamefont {Loss},\ and\ \citenamefont {Samarth}}]{awsch02}%
    \BibitemOpen
    \bibfield  {author} {\bibinfo {author} {\bibfnamefont {D.~D.}\ \bibnamefont
    {Awschalom}}, \bibinfo {author} {\bibfnamefont {D.}~\bibnamefont {Loss}},\
    and\ \bibinfo {author} {\bibfnamefont {N.}~\bibnamefont {Samarth}},\ }\href
    {https://doi.org/10.1007/978-3-662-05003-3} {\emph {\bibinfo {title}
    {Semiconductor Spintronics and Quantum Computation}}},\ NanoScience and
    Technology\ (\bibinfo  {publisher} {Springer},\ \bibinfo {address} {Berlin},\
    \bibinfo {year} {2002})\BibitemShut {NoStop}%
  \bibitem [{\citenamefont {Dyakonov}(2017)}]{dyako17}%
    \BibitemOpen
    \bibinfo {editor} {\bibfnamefont {M.~I.}\ \bibnamefont {Dyakonov}},\ ed.,\
    \href {https://doi.org/10.1007/978-3-319-65436-2} {\emph {\bibinfo {title}
    {Spin Physics in Semiconductors}}},\ \bibinfo {series} {Springer Series in
    Solid-State Sciences}, Vol.\ \bibinfo {volume} {157}\ (\bibinfo  {publisher}
    {Springer International Publishing 2017},\ \bibinfo {year}
    {2017})\BibitemShut {NoStop}%
  \bibitem [{\citenamefont {Slavcheva}\ and\ \citenamefont
    {Roussignol}(2010)}]{slavc10}%
    \BibitemOpen
    \bibfield  {author} {\bibinfo {author} {\bibfnamefont {G.}~\bibnamefont
    {Slavcheva}}\ and\ \bibinfo {author} {\bibfnamefont {P.}~\bibnamefont
    {Roussignol}},\ }\href {https://doi.org/10.1007/978-3-642-12491-4} {\emph
    {\bibinfo {title} {Optical Generation and Control of Quantum Coherence in
    Semiconductor Nanostructures}}},\ NanoScience and Technology\ (\bibinfo
    {publisher} {Springer},\ \bibinfo {address} {Berlin Heidelberg},\ \bibinfo
    {year} {2010})\BibitemShut {NoStop}%
  \bibitem [{\citenamefont {Greilich}\ \emph {et~al.}(2006)\citenamefont
    {Greilich}, \citenamefont {Yakovlev}, \citenamefont {Shabaev}, \citenamefont
    {Efros}, \citenamefont {Yugova}, \citenamefont {Oulton}, \citenamefont
    {Stavarache}, \citenamefont {Reuter}, \citenamefont {Wieck},\ and\
    \citenamefont {Bayer}}]{greil06b}%
    \BibitemOpen
    \bibfield  {author} {\bibinfo {author} {\bibfnamefont {A.}~\bibnamefont
    {Greilich}}, \bibinfo {author} {\bibfnamefont {D.~R.}\ \bibnamefont
    {Yakovlev}}, \bibinfo {author} {\bibfnamefont {A.}~\bibnamefont {Shabaev}},
    \bibinfo {author} {\bibfnamefont {A.~L.}\ \bibnamefont {Efros}}, \bibinfo
    {author} {\bibfnamefont {I.~A.}\ \bibnamefont {Yugova}}, \bibinfo {author}
    {\bibfnamefont {R.}~\bibnamefont {Oulton}}, \bibinfo {author} {\bibfnamefont
    {V.}~\bibnamefont {Stavarache}}, \bibinfo {author} {\bibfnamefont
    {D.}~\bibnamefont {Reuter}}, \bibinfo {author} {\bibfnamefont
    {A.}~\bibnamefont {Wieck}},\ and\ \bibinfo {author} {\bibfnamefont
    {M.}~\bibnamefont {Bayer}},\ }\bibfield  {title} {\bibinfo {title} {Mode
    locking of electron spin coherences in quantum dots},\ }\href
    {https://doi.org/10.1126/science.1128215} {\bibfield  {journal} {\bibinfo
    {journal} {Science}\ }\textbf {\bibinfo {volume} {313}},\ \bibinfo {pages}
    {341} (\bibinfo {year} {2006})}\BibitemShut {NoStop}%
  \bibitem [{\citenamefont {Greilich}\ \emph
    {et~al.}(2009{\natexlab{a}})\citenamefont {Greilich}, \citenamefont
    {Economou}, \citenamefont {Spatzek}, \citenamefont {Yakovlev}, \citenamefont
    {Reuter}, \citenamefont {D.Wieck}, \citenamefont {Reineck},\ and\
    \citenamefont {M.Bayer}}]{greil09a}%
    \BibitemOpen
    \bibfield  {author} {\bibinfo {author} {\bibfnamefont {A.}~\bibnamefont
    {Greilich}}, \bibinfo {author} {\bibfnamefont {S.~E.}\ \bibnamefont
    {Economou}}, \bibinfo {author} {\bibfnamefont {S.}~\bibnamefont {Spatzek}},
    \bibinfo {author} {\bibfnamefont {D.~R.}\ \bibnamefont {Yakovlev}}, \bibinfo
    {author} {\bibfnamefont {D.}~\bibnamefont {Reuter}}, \bibinfo {author}
    {\bibfnamefont {A.}~\bibnamefont {D.Wieck}}, \bibinfo {author} {\bibfnamefont
    {T.~L.}\ \bibnamefont {Reineck}},\ and\ \bibinfo {author} {\bibnamefont
    {M.Bayer}},\ }\bibfield  {title} {\bibinfo {title} {Ultrafast optical
    rotations of electron spins in quantum dots},\ }\href@noop {} {\bibfield
    {journal} {\bibinfo  {journal} {Nucl. Phys.}\ }\textbf {\bibinfo {volume}
    {5}},\ \bibinfo {pages} {262} (\bibinfo {year}
    {2009}{\natexlab{a}})}\BibitemShut {NoStop}%
  \bibitem [{\citenamefont {Greilich}\ \emph
    {et~al.}(2009{\natexlab{b}})\citenamefont {Greilich}, \citenamefont
    {Spatzek}, \citenamefont {Yugova}, \citenamefont {Akimov}, \citenamefont
    {Yakovlev}, \citenamefont {Efros}, \citenamefont {Reuter}, \citenamefont
    {Wieck},\ and\ \citenamefont {Bayer}}]{greil09b}%
    \BibitemOpen
    \bibfield  {author} {\bibinfo {author} {\bibfnamefont {A.}~\bibnamefont
    {Greilich}}, \bibinfo {author} {\bibfnamefont {S.}~\bibnamefont {Spatzek}},
    \bibinfo {author} {\bibfnamefont {I.~A.}\ \bibnamefont {Yugova}}, \bibinfo
    {author} {\bibfnamefont {I.~A.}\ \bibnamefont {Akimov}}, \bibinfo {author}
    {\bibfnamefont {D.~R.}\ \bibnamefont {Yakovlev}}, \bibinfo {author}
    {\bibfnamefont {A.~L.}\ \bibnamefont {Efros}}, \bibinfo {author}
    {\bibfnamefont {D.}~\bibnamefont {Reuter}}, \bibinfo {author} {\bibfnamefont
    {A.~D.}\ \bibnamefont {Wieck}},\ and\ \bibinfo {author} {\bibfnamefont
    {M.}~\bibnamefont {Bayer}},\ }\bibfield  {title} {\bibinfo {title}
    {Collective single-mode precession of electron spins in an ensemble of singly
    charged (in,ga)as/gaas quantum dots},\ }\href@noop {} {\bibfield  {journal}
    {\bibinfo  {journal} {Physical Review B}\ }\textbf {\bibinfo {volume} {79}},\
    \bibinfo {pages} {201305(R)} (\bibinfo {year}
    {2009}{\natexlab{b}})}\BibitemShut {NoStop}%
  \bibitem [{\citenamefont {Belykh}\ \emph {et~al.}(2016)\citenamefont {Belykh},
    \citenamefont {Evers}, \citenamefont {Yakovlev}, \citenamefont {Fobbe},
    \citenamefont {Greilich},\ and\ \citenamefont {Bayer}}]{belyk16}%
    \BibitemOpen
    \bibfield  {author} {\bibinfo {author} {\bibfnamefont {V.~V.}\ \bibnamefont
    {Belykh}}, \bibinfo {author} {\bibfnamefont {E.}~\bibnamefont {Evers}},
    \bibinfo {author} {\bibfnamefont {D.~R.}\ \bibnamefont {Yakovlev}}, \bibinfo
    {author} {\bibfnamefont {F.}~\bibnamefont {Fobbe}}, \bibinfo {author}
    {\bibfnamefont {A.}~\bibnamefont {Greilich}},\ and\ \bibinfo {author}
    {\bibfnamefont {M.}~\bibnamefont {Bayer}},\ }\bibfield  {title} {\bibinfo
    {title} {Extended {pump-probe Faraday rotation spectroscopy of the
    submicrosecond electron spin dynamics in $n$-type GaAs}},\ }\href
    {https://doi.org/10.1103/PhysRevB.94.241202} {\bibfield  {journal} {\bibinfo
    {journal} {Physical Review B}\ }\textbf {\bibinfo {volume} {94}},\ \bibinfo
    {pages} {241202(R)} (\bibinfo {year} {2016})}\BibitemShut {NoStop}%
  \bibitem [{\citenamefont {Evers}\ \emph {et~al.}(2021)\citenamefont {Evers},
    \citenamefont {Kopteva}, \citenamefont {Yugova}, \citenamefont {Yakovlev},
    \citenamefont {Reuter}, \citenamefont {Wieck}, \citenamefont {Bayer},\ and\
    \citenamefont {Greilich}}]{evers21}%
    \BibitemOpen
    \bibfield  {author} {\bibinfo {author} {\bibfnamefont {E.}~\bibnamefont
    {Evers}}, \bibinfo {author} {\bibfnamefont {N.~E.}\ \bibnamefont {Kopteva}},
    \bibinfo {author} {\bibfnamefont {I.~A.}\ \bibnamefont {Yugova}}, \bibinfo
    {author} {\bibfnamefont {D.~R.}\ \bibnamefont {Yakovlev}}, \bibinfo {author}
    {\bibfnamefont {D.}~\bibnamefont {Reuter}}, \bibinfo {author} {\bibfnamefont
    {A.~D.}\ \bibnamefont {Wieck}}, \bibinfo {author} {\bibfnamefont
    {M.}~\bibnamefont {Bayer}},\ and\ \bibinfo {author} {\bibfnamefont
    {A.}~\bibnamefont {Greilich}},\ }\bibfield  {title} {\bibinfo {title}
    {Suppression of nuclear spin fluctuations in an {InGaAs quantum dot ensemble
    by GHz}-pulsed optical excitation},\ }\href
    {https://doi.org/10.1038/s41534-021-00395-1} {\bibfield  {journal} {\bibinfo
    {journal} {npj Quant. Inf.}\ }\textbf {\bibinfo {volume} {7}},\ \bibinfo
    {pages} {60} (\bibinfo {year} {2021})}\BibitemShut {NoStop}%
  \bibitem [{\citenamefont {Petrov}\ and\ \citenamefont
    {Yakovlev}(2012)}]{petro12}%
    \BibitemOpen
    \bibfield  {author} {\bibinfo {author} {\bibfnamefont {M.~Y.}\ \bibnamefont
    {Petrov}}\ and\ \bibinfo {author} {\bibfnamefont {S.~V.}\ \bibnamefont
    {Yakovlev}},\ }\bibfield  {title} {\bibinfo {title} {Comparison of
    quantum-mechanical and semiclassical approaches for an analysis of spin
    dynamics in quantum dots},\ }\href
    {https://doi.org/10.1134/S1063776112060131} {\bibfield  {journal} {\bibinfo
    {journal} {Sov. Phys. JETP}\ }\textbf {\bibinfo {volume} {115}},\ \bibinfo
    {pages} {326} (\bibinfo {year} {2012})}\BibitemShut {NoStop}%
  \bibitem [{\citenamefont {Beugeling}\ \emph {et~al.}(2016)\citenamefont
    {Beugeling}, \citenamefont {Uhrig},\ and\ \citenamefont {Anders}}]{beuge16}%
    \BibitemOpen
    \bibfield  {author} {\bibinfo {author} {\bibfnamefont {W.}~\bibnamefont
    {Beugeling}}, \bibinfo {author} {\bibfnamefont {G.~S.}\ \bibnamefont
    {Uhrig}},\ and\ \bibinfo {author} {\bibfnamefont {F.~B.}\ \bibnamefont
    {Anders}},\ }\bibfield  {title} {\bibinfo {title} {Quantum model for mode
    locking in pulsed semiconductor quantum dots},\ }\href
    {https://doi.org/10.1103/PhysRevB.94.245308} {\bibfield  {journal} {\bibinfo
    {journal} {Physical Review B}\ }\textbf {\bibinfo {volume} {94}},\ \bibinfo
    {pages} {245308} (\bibinfo {year} {2016})}\BibitemShut {NoStop}%
  \bibitem [{\citenamefont {J\"aschke}\ \emph {et~al.}(2017)\citenamefont
    {J\"aschke}, \citenamefont {Fischer}, \citenamefont {Evers}, \citenamefont
    {Belykh}, \citenamefont {Greilich}, \citenamefont {Bayer},\ and\
    \citenamefont {Anders}}]{jasch17}%
    \BibitemOpen
    \bibfield  {author} {\bibinfo {author} {\bibfnamefont {N.}~\bibnamefont
    {J\"aschke}}, \bibinfo {author} {\bibfnamefont {A.}~\bibnamefont {Fischer}},
    \bibinfo {author} {\bibfnamefont {E.}~\bibnamefont {Evers}}, \bibinfo
    {author} {\bibfnamefont {V.~V.}\ \bibnamefont {Belykh}}, \bibinfo {author}
    {\bibfnamefont {A.}~\bibnamefont {Greilich}}, \bibinfo {author}
    {\bibfnamefont {M.}~\bibnamefont {Bayer}},\ and\ \bibinfo {author}
    {\bibfnamefont {F.~B.}\ \bibnamefont {Anders}},\ }\bibfield  {title}
    {\bibinfo {title} {Non-equilibrium nuclear spin distribution function in
    quantum dots subject to periodic pulses},\ }\href
    {https://doi.org/10.1103/PhysRevB.96.205419} {\bibfield  {journal} {\bibinfo
    {journal} {Physical Review B}\ }\textbf {\bibinfo {volume} {96}},\ \bibinfo
    {pages} {205419} (\bibinfo {year} {2017})}\BibitemShut {NoStop}%
  \bibitem [{\citenamefont {Kleinjohann}\ \emph {et~al.}(2018)\citenamefont
    {Kleinjohann}, \citenamefont {Evers}, \citenamefont {Schering}, \citenamefont
    {Greilich}, \citenamefont {Uhrig}, \citenamefont {Bayer},\ and\ \citenamefont
    {Anders}}]{klein18}%
    \BibitemOpen
    \bibfield  {author} {\bibinfo {author} {\bibfnamefont {I.}~\bibnamefont
    {Kleinjohann}}, \bibinfo {author} {\bibfnamefont {E.}~\bibnamefont {Evers}},
    \bibinfo {author} {\bibfnamefont {P.}~\bibnamefont {Schering}}, \bibinfo
    {author} {\bibfnamefont {A.}~\bibnamefont {Greilich}}, \bibinfo {author}
    {\bibfnamefont {G.~S.}\ \bibnamefont {Uhrig}}, \bibinfo {author}
    {\bibfnamefont {M.}~\bibnamefont {Bayer}},\ and\ \bibinfo {author}
    {\bibfnamefont {F.~B.}\ \bibnamefont {Anders}},\ }\bibfield  {title}
    {\bibinfo {title} {Magnetic field dependency of the electron spin revival
    amplitude in periodically pulsed quantum dots},\ }\href
    {https://doi.org/10.1103/PhysRevB.98.155318} {\bibfield  {journal} {\bibinfo
    {journal} {Physical Review B}\ }\textbf {\bibinfo {volume} {98}},\ \bibinfo
    {pages} {155318} (\bibinfo {year} {2018})}\BibitemShut {NoStop}%
  \bibitem [{\citenamefont {Schering}\ \emph {et~al.}(2018)\citenamefont
    {Schering}, \citenamefont {H\"u{}depohl}, \citenamefont {Uhrig},\ and\
    \citenamefont {Fauseweh}}]{scher18}%
    \BibitemOpen
    \bibfield  {author} {\bibinfo {author} {\bibfnamefont {P.}~\bibnamefont
    {Schering}}, \bibinfo {author} {\bibfnamefont {J.}~\bibnamefont
    {H\"u{}depohl}}, \bibinfo {author} {\bibfnamefont {G.~S.}\ \bibnamefont
    {Uhrig}},\ and\ \bibinfo {author} {\bibfnamefont {B.}~\bibnamefont
    {Fauseweh}},\ }\bibfield  {title} {\bibinfo {title} {Nuclear frequency
    focusing in periodically pulsed semiconductor quantum dots described by
    infinite classical central spin models},\ }\href
    {https://doi.org/10.1103/PhysRevB.98.024305} {\bibfield  {journal} {\bibinfo
    {journal} {Physical Review B}\ }\textbf {\bibinfo {volume} {98}},\ \bibinfo
    {pages} {024305} (\bibinfo {year} {2018})}\BibitemShut {NoStop}%
  \bibitem [{\citenamefont {Schering}\ \emph {et~al.}(2020)\citenamefont
    {Schering}, \citenamefont {Scherer},\ and\ \citenamefont {Uhrig}}]{scher20}%
    \BibitemOpen
    \bibfield  {author} {\bibinfo {author} {\bibfnamefont {P.}~\bibnamefont
    {Schering}}, \bibinfo {author} {\bibfnamefont {P.~W.}\ \bibnamefont
    {Scherer}},\ and\ \bibinfo {author} {\bibfnamefont {G.~S.}\ \bibnamefont
    {Uhrig}},\ }\bibfield  {title} {\bibinfo {title} {{Interplay of spin mode
    locking and nuclei-induced frequency focusing in quantum dots}},\ }\href
    {https://doi.org/10.1103/PhysRevB.102.115301} {\bibfield  {journal} {\bibinfo
     {journal} {Physical Review B}\ }\textbf {\bibinfo {volume} {102}},\ \bibinfo
    {pages} {115301} (\bibinfo {year} {2020})}\BibitemShut {NoStop}%
  \bibitem [{\citenamefont {Uhrig}(2020)}]{uhrig20}%
    \BibitemOpen
    \bibfield  {author} {\bibinfo {author} {\bibfnamefont {G.~S.}\ \bibnamefont
    {Uhrig}},\ }\bibfield  {title} {\bibinfo {title} {Quantum coherence from
    commensurate driving with laser pulses and decay},\ }\href
    {https://doi.org/10.21468/SciPostPhys.8.3.040} {\bibfield  {journal}
    {\bibinfo  {journal} {SciPost Phys.}\ }\textbf {\bibinfo {volume} {8}},\
    \bibinfo {pages} {040} (\bibinfo {year} {2020})}\BibitemShut {NoStop}%
  \bibitem [{\citenamefont {Davidson}\ and\ \citenamefont
    {Polkovnikov}(2015)}]{david15}%
    \BibitemOpen
    \bibfield  {author} {\bibinfo {author} {\bibfnamefont {S.~M.}\ \bibnamefont
    {Davidson}}\ and\ \bibinfo {author} {\bibfnamefont {A.}~\bibnamefont
    {Polkovnikov}},\ }\bibfield  {title} {\bibinfo {title} {$su(3)$ semiclassical
    representation of quantum dynamics of interacting spins},\ }\href
    {https://doi.org/10.1103/PhysRevLett.114.045701} {\bibfield  {journal}
    {\bibinfo  {journal} {Physical Review Letters}\ }\textbf {\bibinfo {volume}
    {114}},\ \bibinfo {pages} {045701} (\bibinfo {year} {2015})}\BibitemShut
    {NoStop}%
  \bibitem [{\citenamefont {Kleinjohann}(2022)}]{klein22}%
    \BibitemOpen
    \bibfield  {author} {\bibinfo {author} {\bibfnamefont {I.}~\bibnamefont
    {Kleinjohann}},\ }\href {https://doi.org/10.17877/DE290R-23025} {\emph
    {\bibinfo {title} {Nonequilibrium nuclear spin states in singly charged
    semiconductor quantum dots}}}\ (\bibinfo  {publisher} {{PhD thesis, TU
    Dortmund University}},\ \bibinfo {address} {Dortmund},\ \bibinfo {year}
    {2022})\BibitemShut {NoStop}%
  \bibitem [{\citenamefont {Beugeling}\ \emph {et~al.}(2017)\citenamefont
    {Beugeling}, \citenamefont {Uhrig},\ and\ \citenamefont {Anders}}]{beuge17}%
    \BibitemOpen
    \bibfield  {author} {\bibinfo {author} {\bibfnamefont {W.}~\bibnamefont
    {Beugeling}}, \bibinfo {author} {\bibfnamefont {G.~S.}\ \bibnamefont
    {Uhrig}},\ and\ \bibinfo {author} {\bibfnamefont {F.~B.}\ \bibnamefont
    {Anders}},\ }\bibfield  {title} {\bibinfo {title} {Influence of the nuclear
    {Zeeman} effect on mode locking in pulsed semiconductor quantum dots},\
    }\href {https://doi.org/10.1103/PhysRevB.96.115303} {\bibfield  {journal}
    {\bibinfo  {journal} {Physical Review B}\ }\textbf {\bibinfo {volume} {96}},\
    \bibinfo {pages} {115303} (\bibinfo {year} {2017})}\BibitemShut {NoStop}%
  \bibitem [{\citenamefont {Schering}\ and\ \citenamefont
    {Uhrig}(2021)}]{scher21a}%
    \BibitemOpen
    \bibfield  {author} {\bibinfo {author} {\bibfnamefont {P.}~\bibnamefont
    {Schering}}\ and\ \bibinfo {author} {\bibfnamefont {G.~S.}\ \bibnamefont
    {Uhrig}},\ }\bibfield  {title} {\bibinfo {title} {Nuclear magnetic resonance
    spectroscopy of nonequilibrium steady states in quantum dots},\ }\href
    {https://doi.org/10.1209/0295-5075/133/57003} {\bibfield  {journal} {\bibinfo
     {journal} {Europhys. Lett.}\ }\textbf {\bibinfo {volume} {133}},\ \bibinfo
    {pages} {57003} (\bibinfo {year} {2021})}\BibitemShut {NoStop}%
  \bibitem [{\citenamefont {Bulutay}(2012)}]{bulut12}%
    \BibitemOpen
    \bibfield  {author} {\bibinfo {author} {\bibfnamefont {C.}~\bibnamefont
    {Bulutay}},\ }\bibfield  {title} {\bibinfo {title} {Quadrupolar spectra of
    nuclear spins in strained {In$_x$Ga$_{1-x}$As} quantum dots},\ }\href
    {https://doi.org/10.1103/PhysRevB.85.115313} {\bibfield  {journal} {\bibinfo
    {journal} {Physical Review B}\ }\textbf {\bibinfo {volume} {85}},\ \bibinfo
    {pages} {115313} (\bibinfo {year} {2012})}\BibitemShut {NoStop}%
  \bibitem [{\citenamefont {Bulutay}\ \emph {et~al.}(2014)\citenamefont
    {Bulutay}, \citenamefont {Chekhovich},\ and\ \citenamefont
    {Tartakovskii}}]{bulut14}%
    \BibitemOpen
    \bibfield  {author} {\bibinfo {author} {\bibfnamefont {C.}~\bibnamefont
    {Bulutay}}, \bibinfo {author} {\bibfnamefont {E.~A.}\ \bibnamefont
    {Chekhovich}},\ and\ \bibinfo {author} {\bibfnamefont {A.~I.}\ \bibnamefont
    {Tartakovskii}},\ }\bibfield  {title} {\bibinfo {title} {Nuclear magnetic
    resonance inverse spectra of ingaas quantum dots: Atomistic level structural
    information},\ }\href {https://doi.org/10.1103/PhysRevB.90.205425} {\bibfield
     {journal} {\bibinfo  {journal} {Physical Review B}\ }\textbf {\bibinfo
    {volume} {90}},\ \bibinfo {pages} {205425} (\bibinfo {year}
    {2014})}\BibitemShut {NoStop}%
  \bibitem [{\citenamefont {Hackmann}\ \emph {et~al.}(2015)\citenamefont
    {Hackmann}, \citenamefont {Glasenapp}, \citenamefont {Greilich},
    \citenamefont {Bayer},\ and\ \citenamefont {Anders}}]{hackm15}%
    \BibitemOpen
    \bibfield  {author} {\bibinfo {author} {\bibfnamefont {J.}~\bibnamefont
    {Hackmann}}, \bibinfo {author} {\bibfnamefont {P.}~\bibnamefont {Glasenapp}},
    \bibinfo {author} {\bibfnamefont {A.}~\bibnamefont {Greilich}}, \bibinfo
    {author} {\bibfnamefont {M.}~\bibnamefont {Bayer}},\ and\ \bibinfo {author}
    {\bibfnamefont {F.~B.}\ \bibnamefont {Anders}},\ }\bibfield  {title}
    {\bibinfo {title} {Influence of the nuclear electric quadrupolar interaction
    on the coherence time of hole and electron spins confined in semiconductor
    quantum dots},\ }\href@noop {} {\bibfield  {journal} {\bibinfo  {journal}
    {Physical Review Letters}\ }\textbf {\bibinfo {volume} {115}},\ \bibinfo
    {pages} {207401} (\bibinfo {year} {2015})}\BibitemShut {NoStop}%
  \bibitem [{\citenamefont {Glasenapp}\ \emph {et~al.}(2016)\citenamefont
    {Glasenapp}, \citenamefont {Smirnov}, \citenamefont {Greilich}, \citenamefont
    {Hackmann}, \citenamefont {Glazov}, \citenamefont {Anders},\ and\
    \citenamefont {Bayer}}]{glase16}%
    \BibitemOpen
    \bibfield  {author} {\bibinfo {author} {\bibfnamefont {P.}~\bibnamefont
    {Glasenapp}}, \bibinfo {author} {\bibfnamefont {D.~S.}\ \bibnamefont
    {Smirnov}}, \bibinfo {author} {\bibfnamefont {A.}~\bibnamefont {Greilich}},
    \bibinfo {author} {\bibfnamefont {J.}~\bibnamefont {Hackmann}}, \bibinfo
    {author} {\bibfnamefont {M.~M.}\ \bibnamefont {Glazov}}, \bibinfo {author}
    {\bibfnamefont {F.~B.}\ \bibnamefont {Anders}},\ and\ \bibinfo {author}
    {\bibfnamefont {M.}~\bibnamefont {Bayer}},\ }\bibfield  {title} {\bibinfo
    {title} {Spin noise of electrons and holes in (in,ga)as quantum dots:
    Experiment and theory},\ }\href@noop {} {\bibfield  {journal} {\bibinfo
    {journal} {Physical Review B}\ }\textbf {\bibinfo {volume} {93}},\ \bibinfo
    {pages} {205429} (\bibinfo {year} {2016})}\BibitemShut {NoStop}%
  \bibitem [{\citenamefont {Glazov}(2018)}]{glazo18}%
    \BibitemOpen
    \bibfield  {author} {\bibinfo {author} {\bibfnamefont {M.~M.}\ \bibnamefont
    {Glazov}},\ }\href@noop {} {\emph {\bibinfo {title} {{Electron {\&} Nuclear
    Spin Dynamics in Semiconductor Nanostructures}}}}\ (\bibinfo  {publisher}
    {Oxford University Press},\ \bibinfo {year} {2018})\BibitemShut {NoStop}%
  \bibitem [{\citenamefont {Merkulov}\ \emph {et~al.}(2002)\citenamefont
    {Merkulov}, \citenamefont {Efros},\ and\ \citenamefont {Rosen}}]{merku02}%
    \BibitemOpen
    \bibfield  {author} {\bibinfo {author} {\bibfnamefont {I.~A.}\ \bibnamefont
    {Merkulov}}, \bibinfo {author} {\bibfnamefont {A.~L.}\ \bibnamefont
    {Efros}},\ and\ \bibinfo {author} {\bibfnamefont {M.}~\bibnamefont {Rosen}},\
    }\bibfield  {title} {\bibinfo {title} {Electron spin relaxation by nuclei in
    semiconductor quantum dots},\ }\href
    {https://doi.org/10.1103/PhysRevB.65.205309} {\bibfield  {journal} {\bibinfo
    {journal} {Physical Review B}\ }\textbf {\bibinfo {volume} {65}},\ \bibinfo
    {pages} {205309} (\bibinfo {year} {2002})}\BibitemShut {NoStop}%
  \bibitem [{\citenamefont {Schliemann}\ \emph {et~al.}(2003)\citenamefont
    {Schliemann}, \citenamefont {Khaetskii},\ and\ \citenamefont
    {Loss}}]{schli03}%
    \BibitemOpen
    \bibfield  {author} {\bibinfo {author} {\bibfnamefont {J.}~\bibnamefont
    {Schliemann}}, \bibinfo {author} {\bibfnamefont {A.}~\bibnamefont
    {Khaetskii}},\ and\ \bibinfo {author} {\bibfnamefont {D.}~\bibnamefont
    {Loss}},\ }\bibfield  {title} {\bibinfo {title} {Electron spin dynamics in
    quantum dots and related nanostructures due to hyperfine interaction with
    nuclei},\ }\href {https://doi.org/10.1088/0953-8984/15/50/R01} {\bibfield
    {journal} {\bibinfo  {journal} {J. Phys.: Condens. Matter}\ }\textbf
    {\bibinfo {volume} {15}},\ \bibinfo {pages} {R1809} (\bibinfo {year}
    {2003})}\BibitemShut {NoStop}%
  \bibitem [{\citenamefont {Lee}\ \emph {et~al.}(2005)\citenamefont {Lee},
    \citenamefont {von Allmen}, \citenamefont {Oyafuso}, \citenamefont
    {Klimeck},\ and\ \citenamefont {Whaley}}]{lee05}%
    \BibitemOpen
    \bibfield  {author} {\bibinfo {author} {\bibfnamefont {S.}~\bibnamefont
    {Lee}}, \bibinfo {author} {\bibfnamefont {P.}~\bibnamefont {von Allmen}},
    \bibinfo {author} {\bibfnamefont {F.}~\bibnamefont {Oyafuso}}, \bibinfo
    {author} {\bibfnamefont {G.}~\bibnamefont {Klimeck}},\ and\ \bibinfo {author}
    {\bibfnamefont {K.~B.}\ \bibnamefont {Whaley}},\ }\bibfield  {title}
    {\bibinfo {title} {Effect of electron-nuclear spin interactions for
    electron-spin qubits localized in ingaas self-assembled quantum dots},\
    }\href@noop {} {\bibfield  {journal} {\bibinfo  {journal} {J. Appl. Phys.}\
    }\textbf {\bibinfo {volume} {97}},\ \bibinfo {pages} {043706} (\bibinfo
    {year} {2005})}\BibitemShut {NoStop}%
  \bibitem [{\citenamefont {Gaudin}(1976)}]{gaudi76}%
    \BibitemOpen
    \bibfield  {author} {\bibinfo {author} {\bibfnamefont {M.}~\bibnamefont
    {Gaudin}},\ }\bibfield  {title} {\bibinfo {title} {Diagonalisation d'une
    classe d'hamiltoniens de spin},\ }\href@noop {} {\bibfield  {journal}
    {\bibinfo  {journal} {J. Phys. France}\ }\textbf {\bibinfo {volume} {37}},\
    \bibinfo {pages} {1087} (\bibinfo {year} {1976})}\BibitemShut {NoStop}%
  \bibitem [{\citenamefont {Gaudin}(1983)}]{gaudi83}%
    \BibitemOpen
    \bibfield  {author} {\bibinfo {author} {\bibfnamefont {M.}~\bibnamefont
    {Gaudin}},\ }\href@noop {} {\emph {\bibinfo {title} {La Fonction d'Onde de
    Bethe}}}\ (\bibinfo  {publisher} {Masson, Paris},\ \bibinfo {year}
    {1983})\BibitemShut {NoStop}%
  \bibitem [{\citenamefont {Stone}(2014)}]{stone14}%
    \BibitemOpen
    \bibfield  {author} {\bibinfo {author} {\bibfnamefont {N.~J.}\ \bibnamefont
    {Stone}},\ }\href@noop {} {\bibinfo {title} {Table of nuclear magnetic dipole
    and electric quadrupole moments,}},\ \bibinfo {howpublished} {Tech.\ Rep.\
    INDC(NDS)-0658 (IAEA, Vienna)} (\bibinfo {year} {2014})\BibitemShut {NoStop}%
  \bibitem [{\citenamefont {Stanek}\ \emph {et~al.}(2013)\citenamefont {Stanek},
    \citenamefont {Raas},\ and\ \citenamefont {Uhrig}}]{stane13}%
    \BibitemOpen
    \bibfield  {author} {\bibinfo {author} {\bibfnamefont {D.}~\bibnamefont
    {Stanek}}, \bibinfo {author} {\bibfnamefont {C.}~\bibnamefont {Raas}},\ and\
    \bibinfo {author} {\bibfnamefont {G.~S.}\ \bibnamefont {Uhrig}},\ }\bibfield
    {title} {\bibinfo {title} {Dynamics and decoherence in the central spin model
    in the low-field limit},\ }\href@noop {} {\bibfield  {journal} {\bibinfo
    {journal} {Physical Review B}\ }\textbf {\bibinfo {volume} {88}},\ \bibinfo
    {pages} {155305} (\bibinfo {year} {2013})}\BibitemShut {NoStop}%
  \bibitem [{\citenamefont {Stanek}\ \emph {et~al.}(2014)\citenamefont {Stanek},
    \citenamefont {Raas},\ and\ \citenamefont {Uhrig}}]{stane14b}%
    \BibitemOpen
    \bibfield  {author} {\bibinfo {author} {\bibfnamefont {D.}~\bibnamefont
    {Stanek}}, \bibinfo {author} {\bibfnamefont {C.}~\bibnamefont {Raas}},\ and\
    \bibinfo {author} {\bibfnamefont {G.~S.}\ \bibnamefont {Uhrig}},\ }\bibfield
    {title} {\bibinfo {title} {From quantum-mechanical to classical dynamics in
    the central-spin model},\ }\href {https://doi.org/10.1103/PhysRevB.90.064301}
    {\bibfield  {journal} {\bibinfo  {journal} {Physical Review B}\ }\textbf
    {\bibinfo {volume} {90}},\ \bibinfo {pages} {064301} (\bibinfo {year}
    {2014})}\BibitemShut {NoStop}%
  \bibitem [{\citenamefont {Gravert}\ \emph {et~al.}(2016)\citenamefont
    {Gravert}, \citenamefont {Lorenz}, \citenamefont {Nase}, \citenamefont
    {Stolze},\ and\ \citenamefont {Uhrig}}]{grave16}%
    \BibitemOpen
    \bibfield  {author} {\bibinfo {author} {\bibfnamefont {L.~B.}\ \bibnamefont
    {Gravert}}, \bibinfo {author} {\bibfnamefont {P.}~\bibnamefont {Lorenz}},
    \bibinfo {author} {\bibfnamefont {C.}~\bibnamefont {Nase}}, \bibinfo {author}
    {\bibfnamefont {J.}~\bibnamefont {Stolze}},\ and\ \bibinfo {author}
    {\bibfnamefont {G.~S.}\ \bibnamefont {Uhrig}},\ }\bibfield  {title} {\bibinfo
    {title} {Increased coherence time in narrowed bath states in quantum dots},\
    }\href@noop {} {\bibfield  {journal} {\bibinfo  {journal} {Physical Review
    B}\ }\textbf {\bibinfo {volume} {94}},\ \bibinfo {pages} {094416} (\bibinfo
    {year} {2016})}\BibitemShut {NoStop}%
  \bibitem [{\citenamefont {Faribault}\ and\ \citenamefont
    {Schuricht}(2013{\natexlab{a}})}]{farib13a}%
    \BibitemOpen
    \bibfield  {author} {\bibinfo {author} {\bibfnamefont {A.}~\bibnamefont
    {Faribault}}\ and\ \bibinfo {author} {\bibfnamefont {D.}~\bibnamefont
    {Schuricht}},\ }\bibfield  {title} {\bibinfo {title} {Integrability-based
    analysis of the hyperfine-interaction-induced decoherence in quantum dots},\
    }\href {https://doi.org/10.1103/PhysRevLett.110.040405} {\bibfield  {journal}
    {\bibinfo  {journal} {Physical Review Letters}\ }\textbf {\bibinfo {volume}
    {110}},\ \bibinfo {pages} {040405} (\bibinfo {year}
    {2013}{\natexlab{a}})}\BibitemShut {NoStop}%
  \bibitem [{\citenamefont {Faribault}\ and\ \citenamefont
    {Schuricht}(2013{\natexlab{b}})}]{farib13b}%
    \BibitemOpen
    \bibfield  {author} {\bibinfo {author} {\bibfnamefont {A.}~\bibnamefont
    {Faribault}}\ and\ \bibinfo {author} {\bibfnamefont {D.}~\bibnamefont
    {Schuricht}},\ }\bibfield  {title} {\bibinfo {title} {Spin decoherence due to
    a randomly fluctuating spin bath},\ }\href
    {https://doi.org/10.1103/PhysRevB.88.085323} {\bibfield  {journal} {\bibinfo
    {journal} {Physical Review B}\ }\textbf {\bibinfo {volume} {88}},\ \bibinfo
    {pages} {085323} (\bibinfo {year} {2013}{\natexlab{b}})}\BibitemShut
    {NoStop}%
  \bibitem [{\citenamefont {Fauseweh}\ \emph {et~al.}(2017)\citenamefont
    {Fauseweh}, \citenamefont {Schering}, \citenamefont {H\"udepohl},\ and\
    \citenamefont {Uhrig}}]{fause17a}%
    \BibitemOpen
    \bibfield  {author} {\bibinfo {author} {\bibfnamefont {B.}~\bibnamefont
    {Fauseweh}}, \bibinfo {author} {\bibfnamefont {P.}~\bibnamefont {Schering}},
    \bibinfo {author} {\bibfnamefont {J.}~\bibnamefont {H\"udepohl}},\ and\
    \bibinfo {author} {\bibfnamefont {G.~S.}\ \bibnamefont {Uhrig}},\ }\bibfield
    {title} {\bibinfo {title} {Efficient algorithms for the dynamics of large and
    infinite classical central spin models},\ }\href
    {https://doi.org/10.1103/PhysRevB.96.054415} {\bibfield  {journal} {\bibinfo
    {journal} {Physical Review B}\ }\textbf {\bibinfo {volume} {96}},\ \bibinfo
    {pages} {054415} (\bibinfo {year} {2017})}\BibitemShut {NoStop}%
  \bibitem [{\citenamefont {Schering}(2021)}]{scher21c}%
    \BibitemOpen
    \bibfield  {author} {\bibinfo {author} {\bibfnamefont {P.}~\bibnamefont
    {Schering}},\ }\href {https://doi.org/10.17877/DE290R-22402} {\emph {\bibinfo
    {title} {Nonequilibrium spin phenomena in quantum dots induced by periodic
    optical excitation}}}\ (\bibinfo  {publisher} {{PhD thesis, TU Dortmund
    University}},\ \bibinfo {address} {Dortmund},\ \bibinfo {year}
    {2021})\BibitemShut {NoStop}%
  \bibitem [{\citenamefont {Greilich}\ \emph {et~al.}(2007)\citenamefont
    {Greilich}, \citenamefont {Shabaev}, \citenamefont {Yakovlev}, \citenamefont
    {Efros}, \citenamefont {Yugova}, \citenamefont {Reuter}, \citenamefont
    {Wieck},\ and\ \citenamefont {Bayer}}]{greil07a}%
    \BibitemOpen
    \bibfield  {author} {\bibinfo {author} {\bibfnamefont {A.}~\bibnamefont
    {Greilich}}, \bibinfo {author} {\bibfnamefont {A.}~\bibnamefont {Shabaev}},
    \bibinfo {author} {\bibfnamefont {D.~R.}\ \bibnamefont {Yakovlev}}, \bibinfo
    {author} {\bibfnamefont {A.~L.}\ \bibnamefont {Efros}}, \bibinfo {author}
    {\bibfnamefont {I.~A.}\ \bibnamefont {Yugova}}, \bibinfo {author}
    {\bibfnamefont {D.}~\bibnamefont {Reuter}}, \bibinfo {author} {\bibfnamefont
    {A.~D.}\ \bibnamefont {Wieck}},\ and\ \bibinfo {author} {\bibfnamefont
    {M.}~\bibnamefont {Bayer}},\ }\bibfield  {title} {\bibinfo {title}
    {Nuclei-induced frequency focusing of electron spin coherence},\ }\href
    {https://doi.org/10.1126/science.1146850} {\bibfield  {journal} {\bibinfo
    {journal} {Science}\ }\textbf {\bibinfo {volume} {317}},\ \bibinfo {pages}
    {1896} (\bibinfo {year} {2007})}\BibitemShut {NoStop}%
  \bibitem [{\citenamefont {Abragam}(1978)}]{abrag78}%
    \BibitemOpen
    \bibfield  {author} {\bibinfo {author} {\bibfnamefont {A.}~\bibnamefont
    {Abragam}},\ }\href@noop {} {\emph {\bibinfo {title} {The Principles of
    Nuclear Magnetism}}},\ The International Series of Monographs on Physics\
    (\bibinfo  {publisher} {Clarendon Press},\ \bibinfo {address} {Oxford},\
    \bibinfo {year} {1978})\BibitemShut {NoStop}%
  \bibitem [{\citenamefont {Fischer}\ \emph {et~al.}(2021)\citenamefont
    {Fischer}, \citenamefont {Kleinjohann}, \citenamefont {Sinitsyn},\ and\
    \citenamefont {Anders}}]{fisch22}%
    \BibitemOpen
    \bibfield  {author} {\bibinfo {author} {\bibfnamefont {A.}~\bibnamefont
    {Fischer}}, \bibinfo {author} {\bibfnamefont {I.}~\bibnamefont
    {Kleinjohann}}, \bibinfo {author} {\bibfnamefont {N.~A.}\ \bibnamefont
    {Sinitsyn}},\ and\ \bibinfo {author} {\bibfnamefont {F.~B.}\ \bibnamefont
    {Anders}},\ }\bibfield  {title} {\bibinfo {title} {{Cross-correlation spectra
    in interacting quantum dot systems}},\ }\href
    {https://doi.org/10.1103/PhysRevB.105.035303} {\bibfield  {journal} {\bibinfo
     {journal} {Physical Review B}\ }\textbf {\bibinfo {volume} {105}},\ \bibinfo
    {pages} {035303} (\bibinfo {year} {2021})}\BibitemShut {NoStop}%
  \bibitem [{\citenamefont {Warburton}(2013)}]{warbu13}%
    \BibitemOpen
    \bibfield  {author} {\bibinfo {author} {\bibfnamefont {R.~J.}\ \bibnamefont
    {Warburton}},\ }\bibfield  {title} {\bibinfo {title} {Single spins in
    self-assembled quantum dots},\ }\href@noop {} {\bibfield  {journal} {\bibinfo
     {journal} {Nat. Mat.}\ }\textbf {\bibinfo {volume} {12}},\ \bibinfo {pages}
    {483} (\bibinfo {year} {2013})}\BibitemShut {NoStop}%
  \bibitem [{\citenamefont {Polkovnikov}(2010)}]{polko10}%
    \BibitemOpen
    \bibfield  {author} {\bibinfo {author} {\bibfnamefont {A.}~\bibnamefont
    {Polkovnikov}},\ }\bibfield  {title} {\bibinfo {title} {Phase space
    representation of quantum dynamics},\ }\href
    {https://doi.org/10.1016/j.aop.2010.02.006} {\bibfield  {journal} {\bibinfo
    {journal} {Ann. of Phys.}\ }\textbf {\bibinfo {volume} {325}},\ \bibinfo
    {pages} {1790} (\bibinfo {year} {2010})}\BibitemShut {NoStop}%
  \bibitem [{\citenamefont {Gr\"a\ss{}er}\ \emph {et~al.}(2021)\citenamefont
    {Gr\"a\ss{}er}, \citenamefont {Bleicker}, \citenamefont {Hering},
    \citenamefont {Yarmohammadi},\ and\ \citenamefont {Uhrig}}]{grass21}%
    \BibitemOpen
    \bibfield  {author} {\bibinfo {author} {\bibfnamefont {T.}~\bibnamefont
    {Gr\"a\ss{}er}}, \bibinfo {author} {\bibfnamefont {P.}~\bibnamefont
    {Bleicker}}, \bibinfo {author} {\bibfnamefont {D.-B.}\ \bibnamefont
    {Hering}}, \bibinfo {author} {\bibfnamefont {M.}~\bibnamefont
    {Yarmohammadi}},\ and\ \bibinfo {author} {\bibfnamefont {G.~S.}\ \bibnamefont
    {Uhrig}},\ }\bibfield  {title} {\bibinfo {title} {Dynamic mean-field theory
    for dense spin systems at infinite temperature},\ }\href
    {https://doi.org/10.1103/PhysRevResearch.3.043168} {\bibfield  {journal}
    {\bibinfo  {journal} {Physical Review Research}\ }\textbf {\bibinfo {volume}
    {3}},\ \bibinfo {pages} {043168} (\bibinfo {year} {2021})}\BibitemShut
    {NoStop}%
  \bibitem [{\citenamefont {Dormand}\ and\ \citenamefont
    {Prince}(1980)}]{dorma80}%
    \BibitemOpen
    \bibfield  {author} {\bibinfo {author} {\bibfnamefont {J.~R.}\ \bibnamefont
    {Dormand}}\ and\ \bibinfo {author} {\bibfnamefont {P.~J.}\ \bibnamefont
    {Prince}},\ }\bibfield  {title} {\bibinfo {title} {A family of embedded
    runge-kutta formulae},\ }\href {https://doi.org/10.1016/0771-050X(80)90013-3}
    {\bibfield  {journal} {\bibinfo  {journal} {Journal of Computational and
    Applied Mathematics}\ }\textbf {\bibinfo {volume} {6}},\ \bibinfo {pages}
    {19} (\bibinfo {year} {1980})}\BibitemShut {NoStop}%
  \bibitem [{\citenamefont {Matsumoto}\ and\ \citenamefont
    {Nishimura}(1998)}]{matsu98b}%
    \BibitemOpen
    \bibfield  {author} {\bibinfo {author} {\bibfnamefont {M.}~\bibnamefont
    {Matsumoto}}\ and\ \bibinfo {author} {\bibfnamefont {T.}~\bibnamefont
    {Nishimura}},\ }\bibfield  {title} {\bibinfo {title} {Mersenne twister: A
    623-dimensionally equidistributed uniform pseudo-random number generator},\
    }\href@noop {} {\bibfield  {journal} {\bibinfo  {journal} {ACM Trans. Model.
    Comput. Simul.}\ }\textbf {\bibinfo {volume} {8}},\ \bibinfo {pages} {3}
    (\bibinfo {year} {1998})}\BibitemShut {NoStop}%
  \bibitem [{\citenamefont {{Message Passing Interface Forum}}(2015)}]{messa15}%
    \BibitemOpen
    \bibfield  {author} {\bibinfo {author} {\bibnamefont {{Message Passing
    Interface Forum}}},\ }\href@noop {} {\emph {\bibinfo {title} {{MPI: A
    Message-Passing Interface Standard, Version 3.1}}}}\ (\bibinfo  {publisher}
    {University of Tennessee, available at \url{www.mpi-forum.org}},\ \bibinfo
    {address} {Knoxville, USA},\ \bibinfo {year} {2015})\BibitemShut {NoStop}%
  \bibitem [{\citenamefont {Bechtold}\ \emph {et~al.}(2015)\citenamefont
    {Bechtold}, \citenamefont {Rauch}, \citenamefont {Li}, \citenamefont
    {Simmet}, \citenamefont {Ardelt}, \citenamefont {Regler}, \citenamefont
    {M\"uller}, \citenamefont {Sinitsyn},\ and\ \citenamefont
    {Finley}}]{becht15}%
    \BibitemOpen
    \bibfield  {author} {\bibinfo {author} {\bibfnamefont {A.}~\bibnamefont
    {Bechtold}}, \bibinfo {author} {\bibfnamefont {D.}~\bibnamefont {Rauch}},
    \bibinfo {author} {\bibfnamefont {F.}~\bibnamefont {Li}}, \bibinfo {author}
    {\bibfnamefont {T.}~\bibnamefont {Simmet}}, \bibinfo {author} {\bibfnamefont
    {P.-L.}\ \bibnamefont {Ardelt}}, \bibinfo {author} {\bibfnamefont
    {A.}~\bibnamefont {Regler}}, \bibinfo {author} {\bibfnamefont
    {K.}~\bibnamefont {M\"uller}}, \bibinfo {author} {\bibfnamefont {N.~A.}\
    \bibnamefont {Sinitsyn}},\ and\ \bibinfo {author} {\bibfnamefont {J.~J.}\
    \bibnamefont {Finley}},\ }\bibfield  {title} {\bibinfo {title} {Three-stage
    decoherence dynamics of an electron spin qubit in an optically active quantum
    dot},\ }\href@noop {} {\bibfield  {journal} {\bibinfo  {journal} {Nature
    Physics}\ }\textbf {\bibinfo {volume} {11}},\ \bibinfo {pages} {1005}
    (\bibinfo {year} {2015})}\BibitemShut {NoStop}%
\end{thebibliography}

\begin{thebibliography}{2}%
        \makeatletter
        \providecommand \@ifxundefined [1]{%
         \@ifx{#1\undefined}
        }%
        \providecommand \@ifnum [1]{%
         \ifnum #1\expandafter \@firstoftwo
         \else \expandafter \@secondoftwo
         \fi
        }%
        \providecommand \@ifx [1]{%
         \ifx #1\expandafter \@firstoftwo
         \else \expandafter \@secondoftwo
         \fi
        }%
        \providecommand \natexlab [1]{#1}%
        \providecommand \enquote  [1]{``#1''}%
        \providecommand \bibnamefont  [1]{#1}%
        \providecommand \bibfnamefont [1]{#1}%
        \providecommand \citenamefont [1]{#1}%
        \providecommand \href@noop [0]{\@secondoftwo}%
        \providecommand \href [0]{\begingroup \@sanitize@url \@href}%
        \providecommand \@href[1]{\@@startlink{#1}\@@href}%
        \providecommand \@@href[1]{\endgroup#1\@@endlink}%
        \providecommand \@sanitize@url [0]{\catcode `\\12\catcode `\$12\catcode
          `\&12\catcode `\#12\catcode `\^12\catcode `\_12\catcode `\%12\relax}%
        \providecommand \@@startlink[1]{}%
        \providecommand \@@endlink[0]{}%
        \providecommand \url  [0]{\begingroup\@sanitize@url \@url }%
        \providecommand \@url [1]{\endgroup\@href {#1}{\urlprefix }}%
        \providecommand \urlprefix  [0]{URL }%
        \providecommand \Eprint [0]{\href }%
        \providecommand \doibase [0]{https://doi.org/}%
        \providecommand \selectlanguage [0]{\@gobble}%
        \providecommand \bibinfo  [0]{\@secondoftwo}%
        \providecommand \bibfield  [0]{\@secondoftwo}%
        \providecommand \translation [1]{[#1]}%
        \providecommand \BibitemOpen [0]{}%
        \providecommand \bibitemStop [0]{}%
        \providecommand \bibitemNoStop [0]{.\EOS\space}%
        \providecommand \EOS [0]{\spacefactor3000\relax}%
        \providecommand \BibitemShut  [1]{\csname bibitem#1\endcsname}%
        \let\auto@bib@innerbib\@empty
        \bibitem [{\citenamefont {Davidson}\ and\ \citenamefont
          {Polkovnikov}(2015)}]{david15}%
          \BibitemOpen
          \bibfield  {author} {\bibinfo {author} {\bibfnamefont {S.~M.}\ \bibnamefont
          {Davidson}}\ and\ \bibinfo {author} {\bibfnamefont {A.}~\bibnamefont
          {Polkovnikov}},\ }\bibfield  {title} {\bibinfo {title} {$su(3)$ semiclassical
          representation of quantum dynamics of interacting spins},\ }\href
          {https://doi.org/10.1103/PhysRevLett.114.045701} {\bibfield  {journal}
          {\bibinfo  {journal} {Physical Review Letters}\ }\textbf {\bibinfo {volume}
          {114}},\ \bibinfo {pages} {045701} (\bibinfo {year} {2015})}\BibitemShut
          {NoStop}%
        \bibitem [{\citenamefont {Sbaih}\ \emph {et~al.}(2013)\citenamefont {Sbaih},
          \citenamefont {Srour}, \citenamefont {Hamada},\ and\ \citenamefont
          {Fayad}}]{sbaih13}%
          \BibitemOpen
          \bibfield  {author} {\bibinfo {author} {\bibfnamefont {M.~A.~A.}\
          \bibnamefont {Sbaih}}, \bibinfo {author} {\bibfnamefont {M.}~\bibnamefont
          {Srour}}, \bibinfo {author} {\bibfnamefont {M.~S.}\ \bibnamefont {Hamada}},\
          and\ \bibinfo {author} {\bibfnamefont {H.~M.}\ \bibnamefont {Fayad}},\
          }\bibfield  {title} {\bibinfo {title} {{Lie Algebra and Representation of
          SU(4)}},\ }\href@noop {} {\bibfield  {journal} {\bibinfo  {journal}
          {Electronic Journal of Theoretical Physics}\ }\textbf {\bibinfo {volume}
          {10}},\ \bibinfo {pages} {9} (\bibinfo {year} {2013})}\BibitemShut {NoStop}%
    \end{thebibliography}
\end{document}